\documentclass{jfm}

\usepackage{natbib}
\usepackage{amsbsy}
\usepackage{psfrag}
\usepackage{amsmath}

\usepackage{graphicx,color}
\usepackage{amssymb,latexsym}
\usepackage{url}

\newcommand{\mylab}[3]{\raisebox{#2}[0mm][0mm]{%
\makebox[0mm][l]{\hspace*{#1}\textbf{#3}}}}
\def\spacce#1{\hskip #1pt}
\def\drawline#1#2{\raise 2.5pt\vbox{\hrule width #1pt height #2pt}}
\def\solid{\drawline{24}{.5}\nobreak}

\def\bdash{\hbox{\drawline{5.8}{.5}\spacce{2}}}

\def\dashed{\bdash\bdash\bdash\nobreak}

\def\dchndot{\hbox%
{\drawline{4.6}{.5}\spacce{2}\drawline{1}{.5}\spacce{4.6}\drawline{4.6}{.5}\spacce{2}\drawline{1}{.5}}\nobreak }

\def\circle{$\circ$\nobreak }

\def\trian{\raise 1.25pt\hbox{$\scriptstyle\triangle$}\nobreak}

\def\dtrian{\raise 1.25pt\hbox%
{$\scriptscriptstyle\bigtriangledown$}\nobreak}

\def\squar{\raise 1.25pt\hbox{$\scriptstyle\Box$}\nobreak}

\def\diamon{\raise 1.25pt\hbox{$\scriptstyle\diamond$}\nobreak}

\def\solidsquar{$\blacksquare$\nobreak}

\def\beq{\begin{equation}}
\def\eeq{\end{equation}}

\shorttitle{Multiscale analysis of the topological invariants} % for header on odd pages
\shortauthor{A. Lozano-Dur\'an et al.}                         % for header on even pages
\title{Multiscale analysis of the topological invariants in
  the logarithmic region of turbulent channels at $Re_{\tau}=932$}

\author
{
A. Lozano-Dur\'an\aff{1}
\corresp{\email{adrian@torroja.dmt.upm.es}},
M. Holzner\aff{3}
\and
J. Jim\'enez\aff{2}
}

\affiliation
{
  \aff{1}
  Center for Turbulence Research, Stanford University, Stanford, California 94305, USA
  \aff{2}
  School of Aeronautics, U. Polit\'ecnica de Madrid, 28040 Madrid, Spain
  \aff{3}
  ETH Zurich, 8093 Zurich, Switzerland
}

\begin{document}

\maketitle

\begin{abstract}
  The invariants of the velocity gradient tensor, $R$ and $Q$, and
  their enstrophy and strain components are studied in the logarithmic
  layer of an incompressible turbulent channel flow. The velocities
  are filtered in the three spatial directions and the results
  analyzed at different scales. We show that the $R$--$Q$ plane does
  not capture the changes undergone by the flow as the filter width
  increases, and that the enstrophy/enstrophy-production and
  strain/strain-production planes represent better choices.  We also
  show that the conditional mean trajectories may differ significantly
  from the instantaneous behavior of the flow since they are the
  result of an averaging process where the mean is 3-5 times smaller
  than the corresponding standard deviation.  The orbital periods in
  the $R$--$Q$ plane are shown to be independent of the intensity of
  the events, and of the same order of magnitude than those in the
  enstrophy/enstrophy-production and strain/strain-production planes.
  Our final goal is to test whether the dynamics of the flow are
  self-similar in the inertial range, and the answer turns out to be
  no. The mean shear is found to be responsible for the absence of
  self-similarity and progressively controls the dynamics of the
  eddies observed as the filter width increases. However, a
  self-similar behavior emerges when the calculations are repeated for
  the fluctuating velocity gradient tensor. Finally, the turbulent
  cascade in terms of vortex stretching is considered by computing the
  alignment of the vorticity at a given scale with the strain at a
  different one. These results generally support a non-negligible role
  of the phenomenological energy-cascade model formulated in terms of
  vortex stretching.
\end{abstract}

%%%%%%%%%%%%%%%%%%%%%%%%%%%%%%%%%%%%%%%%%%%%%%%%%%%%%%%%%%%%%%%%%%%%%%%%%%%%
\section{Introduction}\label{sec:introduction}
%%%%%%%%%%%%%%%%%%%%%%%%%%%%%%%%%%%%%%%%%%%%%%%%%%%%%%%%%%%%%%%%%%%%%%%%%%%%

% Invariants Q and R
The understanding of the structure of turbulence has been at the core
of turbulence research from its very beginning. It is natural to
investigate the gradients of the velocity, which provide important
information regarding the local behavior of the flow.  The invariants
of the velocity gradient tensor for incompressible flows, $R$ and $Q$,
were first introduced by \citet{Chong1990} and have proved to be a
useful tool to analyze turbulent flows characterized by a wide range
of scales \citep{Chong1990, Chertkov1999, Chacin2000, Tanahashi2004,
  Chevillard2006, Luethi2009, Atkinson2012, Wang2012, Cardesa2013}.
They quantify the relative strength of enstrophy production and strain
self-amplification, and of local enstrophy density and strain density,
respectively. Moving locally with a fluid particle, the velocity
gradient tensor determines the linear approximation to the local
velocity field surrounding the observer. In that frame, invariants can
also be used to classify the local flow topology \citep{Chong1990}.
However, the invariants are gradients of the velocities and hence, are
dominated by the effect of the small scales.  By filtering the
velocity field, we will apply the topological and physical tools
provided by the invariants to scales in the inertial range.

% Vieillefosse tail
Measurements and simulations in different turbulent flows showed that
the joint probability density function of $R$ and $Q$ has a very
particular skewed `tear-drop' shape, e.g. \cite{Soria1994, Ooi1999,
  Chertkov1999, Chacin2000}. That is, there is an increased
probability of points where $R > 0$ and $Q < 0$ along the so-called
Vieillefosse tail, reflecting a dominance of strain self-amplification
over enstrophy production in strain-dominated regions. Such a
signature turns out to be a quite universal feature persistent in many
different turbulent flows, including mixing layers \citep{Soria1994},
channel flows \citep{Blackburn1996}, boundary
layers \citep{Chong1998,Atkinson2012}, isotropic
turbulence \citep{Martin1998,Ooi1999}, etc.

% Decomposition of invariants
A better insight is gained by decomposing the velocity gradient tensor
into its rate-of-rotation and rate-of-strain
tensors \citep{Blackburn1996, Chong1998, Ooi1999} or into their strain
and enstrophy contributions \citep{Gomes2014}. \citet{Luethi2009}
expanded the $R$--$Q$ plane to three dimensions and studied the effect
of enstrophy production and strain self-amplification separately using
direct numerical simulations (DNSs) of isotropic turbulence at a
Taylor-based Reynolds number $Re_{\lambda}=434$.

% CMTs and spirals
\citet{Martin1998} and \citet{Ooi1999} introduced and studied the
conditional mean trajectories of the invariants (henceforth, CMTs) in
DNSs of isotropic turbulence at $Re_{\lambda}=40$-$70$. This involved
the calculation of the mean temporal rate of change along the fluid
particle trajectories of the invariants conditioned on the values of
the invariants themselves, which results in a vector field in the
$R$--$Q$ plane. The resulting conditional vector field can be
integrated to produce trajectories within the space of the invariants.

The analysis of the CMTs provides information about the dynamics of
the small scales of turbulence. Previous results suggest a cyclic and
approximately periodic orbit with a mean clockwise evolution in the
$R$--$Q$ plane and trajectories spiraling towards $(R,Q) =
(0,0)$. Some authors have conjectured about the spurious nature of the
spiraling of the CMTs towards the origin \citep{Martin1998}. However,
others have argued that this effect may be of physical significance
and related to the statistical tendency of the flow to form shear
layers \citep{Elsinga2010}. \citet{loz:hol:jim:2015} showed that the
CMTs describe closed trajectories when the whole domain of a
statistically stationary turbulent flow is considered, but that they
may spiral inwards or outwards if the statistics are restricted to
certain sub-regions of inhomogeneous flows. The dynamical evolution of
the velocity gradient tensor has been also addressed in many
statistical and reduced models \citep{Vieillefosse1982,
  Vieillefosse1984, Cantwell1992, Meneveau2011}.

% Lifetimes
The time-scale associated with the CMTs, i.e., the time to complete
one revolution around the origin, is considered representative of the
times involved in the turbulent dynamics. For isotropic turbulence at
$Re_\lambda \simeq 40$-$70$, \citet{Martin1998} and \cite{Ooi1999}
computed a characteristic time of $3\tau_{eddy}$ to complete one cycle
of the orbit, where $\tau_{eddy}=L/u'$ is the eddy-turnover time, $u'$
is the root-mean-squared velocity and $L$ is the integral length
scale. \citet{Elsinga2010} computed orbital periods using Tomographic
PIV in an experimental boundary layer with a momentum thickness based
Reynolds number $Re_{\theta}=U_{\infty} \theta/\nu \simeq 2460$
($Re_\lambda \simeq 70$ at $x_2=0.5\delta_{99}$), where $U_\infty$ is
the free-stream velocity, $\theta$ the momentum thickness, $\nu$ the
kinematic viscosity, $\delta_{99}$ the boundary layer thickness based
on 99\% of $U_\infty$, and $x_2$ the distance from the wall. The results
were restricted to the logarithmic layer and showed CMTs with a
clockwise evolution similar to those in \citet{Martin1998} and
\citet{Ooi1999}, with a characteristic time close to
$\tau_{eddy}$. \citet{Atkinson2012} computed the CMTs in the
logarithmic and wake region of a DNS of a boundary layer at
$Re_{\theta}=730$-$1954$ ($Re_\lambda \simeq 30$-$50$ at
$x_2=0.4\delta_{99}$), with associated time-scales in inner and outer
units of $658\nu/u_{\tau}^2$ or $2.51\tau_{eddy}$, respectively, where
$u_\tau$ is the friction velocity. \citet{Luethi2009} showed a 3D
pattern in the expanded $R$--$Q$ plane of isotropic turbulence at
$Re_\lambda=434$ more pronounced than the 2D one, and with a
characteristic time of $40$ Kolmogorov units or one $\tau_{eddy}$.

% Coarse-grained and filtering studies
Notably, the invariants of the velocity gradient tensor can be used to
study phenomena at larger scales using the coarse-grained or filtered
velocity gradient tensor \citep{Borue1998, Chertkov1999,
  VanDerBos2002, Naso2005, Naso2006, Naso2007, Luethi2007,
  Meneveau2011}. In both cases, some of the properties mentioned above
are recovered. For example, it has been shown that the characteristic
tear-drop shape in the $R$--$Q$ plane remains visible at scales that
are well in the inertial range \citep{Borue1998, VanDerBos2002,
  Luethi2007}. Using experimental particle tracking,
\citet{Luethi2007} found that the tear-drop shape persisted in the
inertial range of quasi-homogeneous turbulence at $Re_{\lambda} = 150$
even for filter sizes larger than the integral
scale. \citet{Borue1998} showed similar results for isotropic
turbulence using top-hat and Gaussian filters.  On the contrary, in
model calculations of isotropic turbulence, the contours became
increasingly symmetric with growing filter widths, and at scales of
the order of the integral length the results essentially resembled
Gaussian statistics \citep{Chertkov1999, Naso2005, Naso2006, Naso2007,
  Pumir2010}.

% Organization of the paper
In this paper we study the small and inertial scale structure of the
$R$--$Q$ space in a turbulent channel flow at friction Reynolds number
$Re_{\tau}=h u_\tau /\nu=932$, where $h$ is the channel
half-height. The main objective is to improve our knowledge about the
properties and evolution of the filtered velocity gradient tensor. The
study of the inertial scales is of paramount importance, since they
are an intrinsic property of high Reynolds number turbulence. However,
both experimental and numerical limitations have made it difficult to
carry out this task. We would like to improve our insight into the
physics of wall-bounded turbulence, where DNSs with a moderate range
of inertial scales are now available. We decompose $R$ and $Q$ into
their enstrophy and strain components, and analyze their multiscale
dynamics and potential relation with the energy cascade in terms of
vortex stretching. We show that the mean shear is responsible for the
absence of self-similarity and progressively controls the dynamics of
the eddies as the filter width increases. However, a self-similar
behavior emerges when the calculations are repeated for the
fluctuating velocity gradient tensor. Finally, the turbulent cascade
is considered and our results support a non-negligible role of the
phenomenological energy-cascade model in terms of vortex stretching.

The paper is organized as follows.  In the next section, the
topological invariants are revisited and the numerical experiments and
filtering procedure presented. Results are offered in section
\ref{sec:results} which is further divided into five parts. The
dynamics of the $R$--$Q$ plane are studied in \S\ref{subsec:RQ} and
\S\ref{subsec:RQ_CMTs}, their decomposition in strain and enstrophy
components in \S\ref{subsec:QsQwRsRw}, the characteristic orbital
periods in \S\ref{subsec:periods}, the alignment of the vorticity and
the rate-of-strain tensor in \S\ref{subsec:alignment}, and the energy
cascade in terms of vortex stretching in
\S\ref{subsec:alignment_inter}. Finally, we close with the conclusions
in section \ref{sec:conclusions}.

%%%%%%%%%%%%%%%%%%%%%%%%%%%%%%%%%%%%%%%%%%%%%%%%%%%%%%%%%%%%%%%%%%%%%%%%%%%%
\section{Method}\label{sec:method}
%%%%%%%%%%%%%%%%%%%%%%%%%%%%%%%%%%%%%%%%%%%%%%%%%%%%%%%%%%%%%%%%%%%%%%%%%%%%

%-------------------------------------------------------%
\subsection{Topological invariants of the velocity gradient tensor}\label{subsec:invariants}
%-------------------------------------------------------%

% Definitions of R and Q
The second and third invariants of the velocity gradient tensor for an
incompressible flow, $Q$ and $R$, are
\begin{eqnarray}\label{eq:invariants_Q}
Q &=& \frac{1}{4}(\omega_i\omega_i-2s_{ij}s_{ij}), \\
\label{eq:invariants_R}
R &=& -\frac{1}{3}s_{ij}s_{jk}s_{ki}-\frac{1}{4}\omega_i\omega_j s_{ij},
\end{eqnarray}
where summation over repeated indices is implied, $\omega_i$ are the
components of the vorticity vector and $s_{ij}$ of the rate-of-strain
tensor. The first invariant, $P_o=s_{ii}$, is zero due to the
incompressibility of the flow.

The invariants $R$ and $Q$ defined by relations
(\ref{eq:invariants_Q}) and (\ref{eq:invariants_R}) may be interpreted
in two ways. From a physical point of view, $Q$ measures the relative
importance of enstrophy and strain densities.  Enstrophy dominates
over strain for positive values of $Q$, and strain does for negative
ones. The meaning of $R$ depends on the value of $Q$. For $Q>0$, $R<0$
represents vortex stretching and $R>0$ contraction of vorticity (also
refer to as vortex compression in the literature).  For $Q<0$, $R$ is
dominated by the strain self-amplification. The second interpretation
is topological and $R$ and $Q$ characterize the local motion of the
fluid particles for an observer traveling with the fluid. The lines
$D=27/4R^2+Q^3=0$ and $R=0$ divide the $R$--$Q$ plane in four regions
(with $R$ the horizontal axis, as in figure \ref{fig:pdfRQ}a). The
trajectories of the fluid particles are then classified according to
critical point terminology \citep{Chong1990}, as stable
focus/stretching (upper left-hand region), unstable focus/compressing
(upper right-hand region), stable node/saddle/saddle (lower left-hand
region) and unstable node/saddle/saddle (lower right-hand region).

% CMTs
The conditional mean trajectories or CMTs \citep{Martin1998} aim to
study the Lagrangian temporal evolution of the invariants. The method
relies on calculating the average temporal rates of change of the
invariants for the fluid particles, $\mathrm{D}R/\mathrm{D}t$ and
$\mathrm{D}Q/\mathrm{D}t$, conditioned on the values of $R$ and
$Q$. Note that $\mathrm{D}/\mathrm{D}t$ stands for material
derivative. These quantities can be thought of as the components of a
conditionally averaged vector field in the $R$--$Q$ plane,
\begin{eqnarray}\label{eq:v}
  \boldsymbol{V} = (V_R,V_Q) = \left\langle \left ( \frac{
    \mathrm{D}R}{\mathrm{D}t} , \frac{\mathrm{D}Q}{\mathrm{D}t}\right
  ) \right \rangle_{R,Q},
\end{eqnarray}
where $\langle \cdot \rangle_{R,Q}$ denotes conditional average at
point $(R,Q)$. From the vector field $\boldsymbol{V}$, any chosen
initial condition can be integrated resulting in the aforementioned
CMTs.

\citet{loz:hol:jim:2015} argued that CMTs should remain closed when a
statistically stationary wall-bounded or periodic domain is
considered, but in inhomogeneous flows, as in channels, they spiral
outwards and inwards when the statistics are restricted to the buffer
and outer region, respectively.  Since the values of $R'(x_2)$ and
$Q'(x_2)$, where the prime denotes standard deviation with respect to
the mean over homogeneous directions and time, decay several orders of
magnitude from the wall to the center of the channel, it is reasonable
to scale $R$ and $Q$ with a function of $x_2$ which compensates for
the wall-normal inhomogeneity of the channel.  For that reason, we
will use
\begin{eqnarray}\label{eq:normalization}
R/{Q'}^{3/2}, \quad Q/{Q'}.
\end{eqnarray}
Accordingly, we have to compute the vector
field \citep{loz:hol:jim:2015}
\begin{equation}\label{eq:v_norm}
  \boldsymbol{v} = (v_R,v_Q) = \left\langle \left ( \frac{\mathrm{D}}{
    \mathrm{D}t} \left( \frac{R}{{Q'}^{3/2}} \right),
  \frac{\mathrm{D}}{ \mathrm{D}t} \left( \frac{Q}{{Q'}} \right) \right
  ) \right \rangle_{R/{Q'}^{3/2},Q/{Q'}},
\end{equation}
which leads to CMTs consistent with the $(R/{Q'}^{3/2})$--$(Q/{Q'})$
plane. Throughout the paper, we will refer to $R/{Q'}^{3/2}$--$Q/Q'$
as the $R$--$Q$ plane for simplicity. When there is no danger of
ambiguity, the vector $\boldsymbol{v}$ will also denote the
conditionally averaged velocities in other planes.

We focus our study in the logarithmic layer which is chosen to span
from $x_2^+=100$ to $x_2/h=0.4$. All the results shown in the present
manuscript are computed for that region unless otherwise specified.
It was checked that varying these limits within the usual range
\citep{mar:mon:hul:smi:2013} did not significantly alter the results
presented below.

%-------------------------------------------------------%
\subsection{Numerical experiments}\label{subsec:numerical}
%-------------------------------------------------------%

% Numerical data
%----------------------------------------------------------------%
\begin{table}
    \begin{center}
        \begin{tabular}{cccccccccc}
 $Re_{\tau}$ & $L_1/h$ & $L_3/h$ & $\delta_1^+$ & $\delta_3^+$ & $N_1$ & $N_2$ & $N_3$ & $N_f$ & $T u_\tau/h$\\[1ex]
\hline
  932     & $2\pi$  & $\pi$   & 11           & 5.7          & 512   & 385   & 512   & 400   & 20 \\
       \end{tabular}
    \end{center}
\caption{Parameters of the simulation. $L_1$ and $L_3$ are the
  streamwise and spanwise dimensions of the numerical box, and $h$ is
  the channel half-height; $\delta_1$ and $\delta_3$ are the
  resolutions in terms of Fourier modes before dealiasing; $N_1$,
  $N_2$ and $N_3$ are the number of streamwise, wall-normal and
  spanwise modes, respectively; $N_f$ is the number of flow fields
  used to accumulate statistics separated by $0.05h/u_\tau$, and $T
  u_\tau/h$ is the total time simulated in eddy
  turnovers.} \label{table:DNS}
\end{table}
%----------------------------------------------------------------%
%
% DNS database
We use data from a DNS of a turbulent channel flow from
\citet{loz:jim:2014} at a friction Reynolds number $Re_{\tau}=u_\tau
h/\nu=932$.  The superscript $+$ denotes wall units based on $u_\tau$
and $\nu$. The parameters of the simulation are summarized in table
\ref{table:DNS} where $x_1,x_2$ and $x_3$ are the streamwise,
wall-normal and spanwise directions, respectively, with associated
velocities $u_1,u_2$ and $u_3$. The streamwise and spanwise directions
are periodic. The incompressible flow is integrated in the form of
evolution equations for the wall-normal vorticity and for the
Laplacian of the wall-normal velocity \citep{Kim1987}. The spatial
discretization is Fourier in the two wall-parallel directions using
the 3/2 dealiasing rule, and Chebyshev polynomials in the $x_2$
direction. Time stepping is performed with a third-order semi-implicit
Runge-Kutta scheme \citep{Moser1999}. The streamwise and spanwise
lengths of the channel are $L_1=2\pi h$ and $L_3=\pi h$, respectively,
and have been previously shown to be large enough to ensure an
accurate representation of the coherent structures in the logarithmic
layer \citep{Lozano2014,Flores2010}. The DNS was run for 20
eddy-turnover times, $h/u_\tau$, and the fields were stored with a
temporal spacing of $0.05h/u_\tau$ between consecutive snapshots. To
assess the effect of the Reynolds number, an extra DNS at
$Re_\tau=550$ was computed and the results are included in Appendix
\ref{sec:appendix}.

% Computation of invariants
The invariants of the velocity gradient and their material derivatives
are computed from the DNS presented above. All the calculations are
performed in double precision, and the spatial resolution and temporal
numerical schemes are described below. A systematic study of the
numerical effects on the invariants and their CMTs can be found in
\citet{loz:hol:jim:2015}.

The spatial derivatives are computed using spectral methods: Fourier
in $x_1$ and $x_3$, and Chebyshev in $x_2$.  The number of modes of
the velocity field from the DNS in table \ref{table:DNS} is increased
by a factor of three in each direction and padded with zeros before
computing the invariants.

The material derivatives of $R$ and $Q$ (or of their normalized
counterparts) are computed in the form
\begin{equation}\label{eq:material}
  \mathrm{D}/\mathrm{D}t = \partial/\partial t +
  \boldsymbol{u}\cdot\boldsymbol\nabla,
\end{equation}
where $\boldsymbol{u}$ is the flow velocity and $\boldsymbol\nabla$
the gradient operator. For the time derivative, five extra fields are
generated for each flow field advancing in time the DNS with a
constant time step, $\Delta t$. The generated fields are then used to
compute $\partial R/\partial t$ and $\partial Q/\partial t$ with
fourth-order accurate finite differences and $\Delta t^+=4\cdot
10^{-2}$, which corresponds to CFL$=0.075$ on average.

\begin{figure}
\vspace{0.5cm}
\centerline{
\psfrag{X}{ \raisebox{-0.2cm}{$R/[Q'_s]_2^{3/2}$} }\psfrag{Y}{$Q/[Q'_s]_2$}
\includegraphics[width=0.45\textwidth]{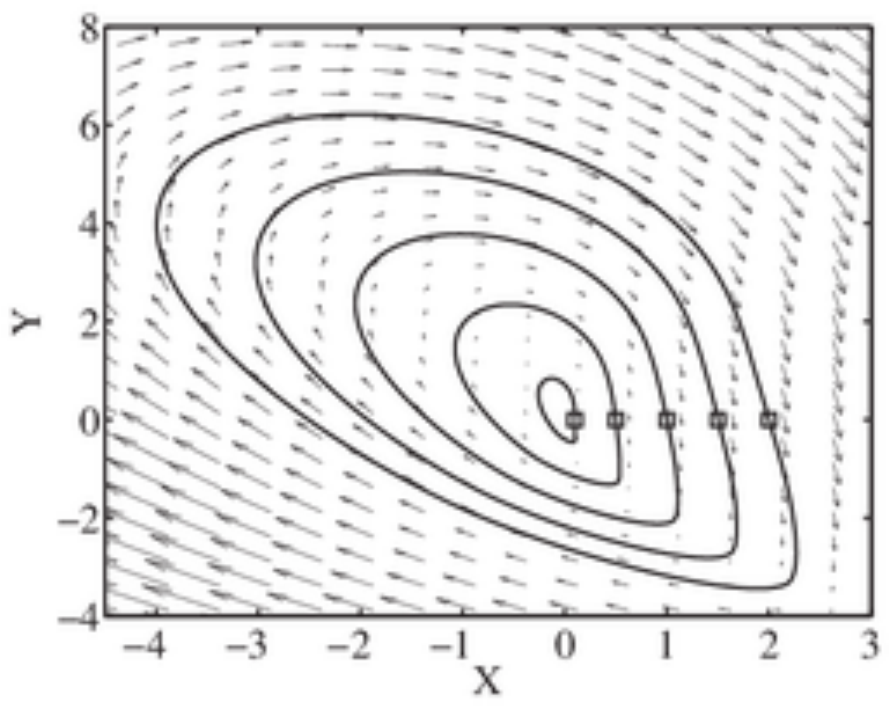}
\mylab{-5.3cm}{4.8cm}{(a)}
\psfrag{X}{ \raisebox{-0.2cm}{$R/{Q'_s}^{3/2}$} }\psfrag{Y}{$Q/{Q'_s}$}
\includegraphics[width=0.45\textwidth]{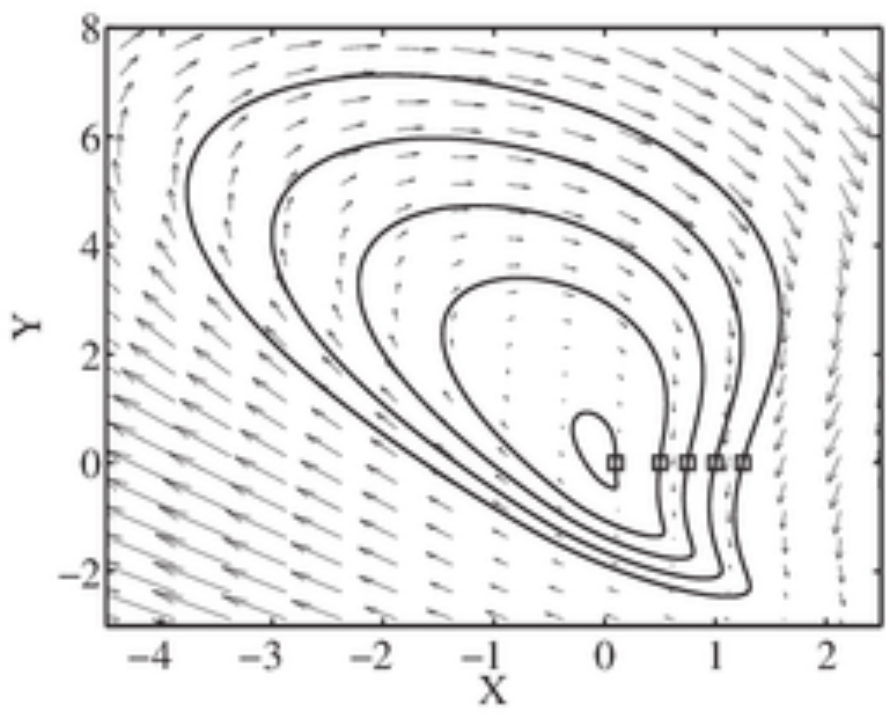}
\mylab{-5.3cm}{4.8cm}{(b)}
}
\caption{ CMTs (solid lines) in (a), the
  $(R/{[Q'_s]_2}^{3/2})$--$(Q/[Q'_s]_2)$ plane; (b), the
  $(R/{Q'_s}^{3/2})$--$(Q/{Q'_s})$ plane. The trajectories are
  integrated from the initial conditions
  $(R/[Q'_s]_2^{3/2},Q/[Q'_s]_2) =(0.1,0), (0.5,0), (1,0),
  (1.5,0)$ and $(2,0)$, and $(R/{Q'_s}^{3/2},Q/{Q'_s})=(0.1,0),
  (0.5,0), (0.75,0), (1,0), (1.25,0)$. The initial positions are
  marked by \squar. The arrows represent the vector fields, (a),
  $(V_R/[Q'_s]_2^2,V_Q/[Q'_s]_2^{3/2})h/u_\tau$ from
  equation (\ref{eq:v}); (b), $(v_R,v_Q)h/u_\tau$ from
  (\ref{eq:v_norm}).
\label{fig:examples}}
\end{figure}

% CMTs
All the CMTs shown in this work are obtained by integrating the
trajectory of a virtual particle in the $R$--$Q$ plane using a
time-marching Runge-Kutta-Fehlberg scheme with a relative error of
$10^{-6}$, and interpolating the vector field $\boldsymbol{v}$ with
cubic splines. As an example, figure \ref{fig:examples}(a) shows the
CMTs computed in the whole turbulent channel normalized with the
constant factor $[Q'_s]_2$, where $[\cdot]_2$ denotes wall-normal
average and $Q_s=-1/2s_{ij}s_{ij}$. Figure \ref{fig:examples}(b) shows
the same result but normalized with the non-uniform function,
$Q'_s(x_2)$, as shown in (\ref{eq:normalization}). In both cases, the
CMTs describe closed trajectories as discussed in
\citet{loz:hol:jim:2015}. The quantity $Q'_s$ was used instead of $Q'$
to avoid dividing $Q$ and $R$ by very small values close to the wall.
%
%================================================================
%/data4/adrian/Q1Q2R1R2/mfiles/cases/plotQR_onfly_cases.m
%================================================================
%

%-------------------------------------------------------%
\subsection{Data filtering}\label{subsec:filter}
%-------------------------------------------------------%

\begin{figure}
\centerline{
\includegraphics[width=0.8\textwidth]{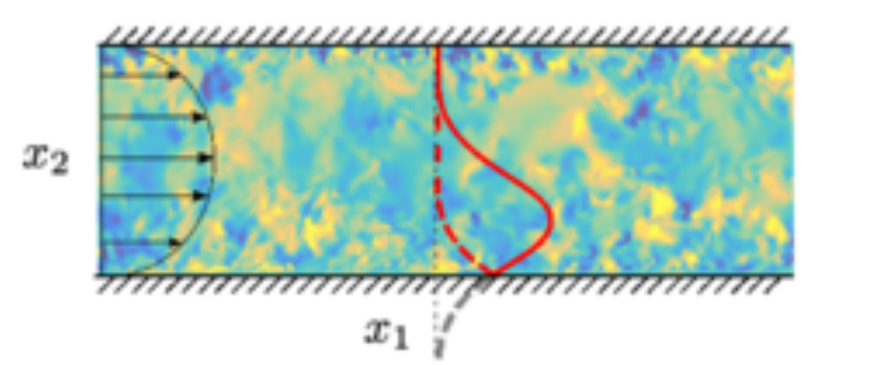}
}
\caption{ Sketch of the filter kernel reflected at the wall. The red
  solid line represents the Gaussian kernel in the $x_2$ direction,
  the gray dashed line its extension out of the domain and the red
  dashed one its reflection. The colormap is the wall-normal
  velocity. \label{fig:filter_sketch}}
\end{figure}

% Filter
The three velocities components $u_i(\boldsymbol{x})$, with
$\boldsymbol{x}=(x_1,x_2,x_3)$, are low-pass filtered with a Gaussian
cut-off,
\begin{equation}\label{eq:filter}
\widetilde{u}_i(\boldsymbol{x})=
\iiint_V a\cdot u_i(\boldsymbol{x}-\boldsymbol{x}') \exp
 \left( -\left(\frac{\pi x_1'}{\Delta_1}\right)^2
        -\left(\frac{\pi x_2'}{\Delta_2}\right)^2
        -\left(\frac{\pi x_3'}{\Delta_3}\right)^2
 \right) \mathrm{d}x_1'\mathrm{d}x_2'\mathrm{d}x_3',
\end{equation}
for $i=1,2$ and $3$, where $\Delta_1$, $\Delta_2$ and $\Delta_3$ are
the filter widths in the streamwise, wall-normal and spanwise
directions, respectively, $V$ is the channel domain extended as
explained below and $a$ a constant such that the integral of the
kernel over $V$ is one. The Gaussian filter is directly applied in the
two homogeneous directions. However, that is not possible in $x_2$ due
to the wall. To overcome this difficulty, the filtering operation is
extended in the wall-normal direction by reflecting the filter at the
walls as if they were a mirror (that is equivalent to copy the
velocity field above the top wall and below the bottom one reversing
the $x_2$ direction) and inverting the sign of $u_2$ (see figure
\ref{fig:filter_sketch}). In this way, the filtered velocity remains
incompressible. To assess the effect of this particular approach, all
the results in the present manuscript were recomputed using the
alternative filter described in Appendix \ref{sec:appendixFil}, and
the differences turned out to be negligible.  Another possible option
was to simply filter in the wall-normal direction and not only in
$x_1$ and $x_3$ as usually done for convenience in the
literature. However, we will show that the two approaches are not
equivalent.  
%
%================================================================
%
%================================================================
%

The aspect ratios of the three filter widths are chosen to be
elongated in the streamwise and spanwise directions, and such that
$\Delta_1/\Delta_2=3$ and $\Delta_3/\Delta_2=1.5$, which is motivated
by the characteristic geometrical shape of eddies attached to the wall
in the logarithmic layer reported by previous works \citep{Lozano2012,
  DelAlamo2006, Jimenez2012}. A homogeneous aspect ratio,
$\Delta_1/\Delta_2=1$ and $\Delta_3/\Delta_2=1$, was also tested and
qualitatively similar results were obtained as discussed in Appendix
\ref{sec:appendix}.

In the present work, we only filter the velocities but quantities
computed from them will be also denoted by $\widetilde{(\cdot)}$. For
instance, $\widetilde{Q}$ is the second invariant of the velocity
gradient tensor computed from the filtered velocities. Consistently,
the material derivative of the quantity $\widetilde{\psi}$ is computed
with respect to the filtered velocity
$\mathrm{D}\widetilde{\psi}/\mathrm{D}t = \partial
\widetilde{\psi}/\partial t + \widetilde{u_i} \partial
\widetilde{\psi}/\partial x_i$.

%----------------------------------------------------------------%
\begin{table}
    \begin{center}
      \begin{minipage}{10cm}
        \begin{tabular}{lccccc}
          Case  & $\Delta_1/h$ & $\Delta_2/h$ & $\Delta_3/h$ & Lines and symbols & Color     \\[1ex]
          \hline
          F0 (unfiltered) &      -       &    -         &  -           &   \solid     & black   \\
          F0.10  &    0.30      &    0.10      &   0.15       &   $-$\circle$-$  & magenta \\
          F0.20  &    0.60      &    0.20      &   0.30       &   $-$\trian$-$   & blue    \\
          F0.25  &    0.75      &    0.25      &   0.38       &   $-$\squar$-$   & green   \\
          F0.30  &    0.90      &    0.30      &   0.45       &   $-\times-$     & red     \\
          F0.40  &    1.20      &    0.40      &   0.60       &   $-\Diamond-$   & yellow  \\
       \end{tabular}
     \end{minipage}
    \end{center}
\caption{Summary of cases. The parameters $\Delta_1$, $\Delta_2$ and
  $\Delta_3$ are the filter widths in the streamwise, wall-normal and
  spanwise direction, respectively.  The velocity field is filtered
  according to (\ref{eq:filter}). The filtered cases are denoted by
  F$\gamma$, where $\gamma$ is the wall-normal filter width,
  $\Delta_2/h$. The symbols and colors in the last two columns are
  used to denote the different cases in the figures unless otherwise
  specified.} \label{table:cases}
\end{table}
%----------------------------------------------------------------%
%
%
%================================================================
% /data4/adrian/Q1Q2R1R2/mfiles/cases/plotQR_onfly_cases.m
\begin{figure}
%%%
\centerline{
\psfrag{X}{$x_2/h$}\psfrag{Y}{${u_{f1}'}^{+}$}
\includegraphics[width=0.45\textwidth]{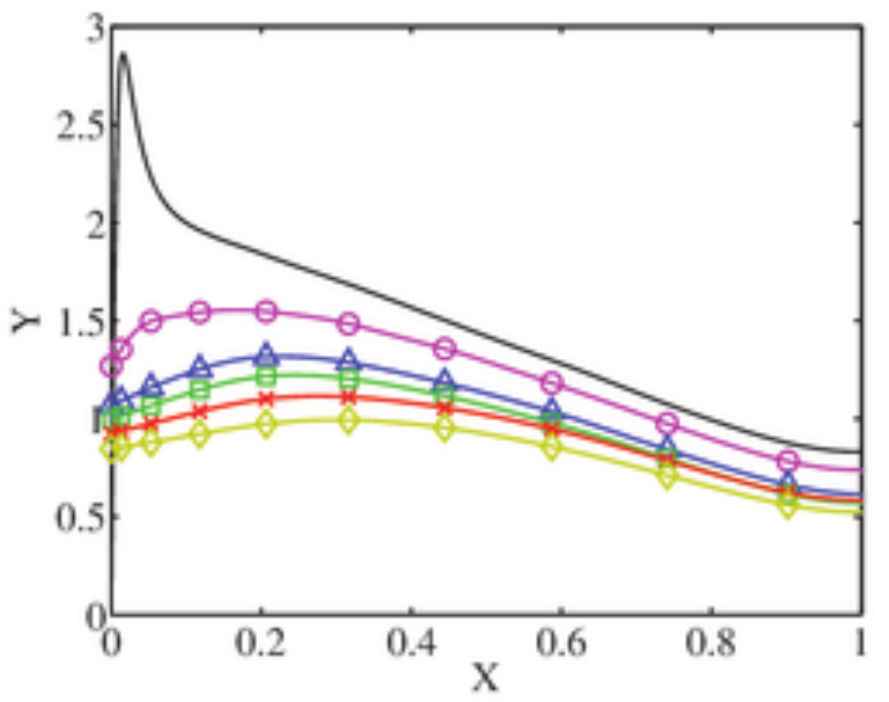}
\mylab{-5.3cm}{4.7cm}{(a)}
\psfrag{X}{$x_2/h$}\psfrag{Y}{$R'/Q^{'3/2}$}
\includegraphics[width=0.45\textwidth]{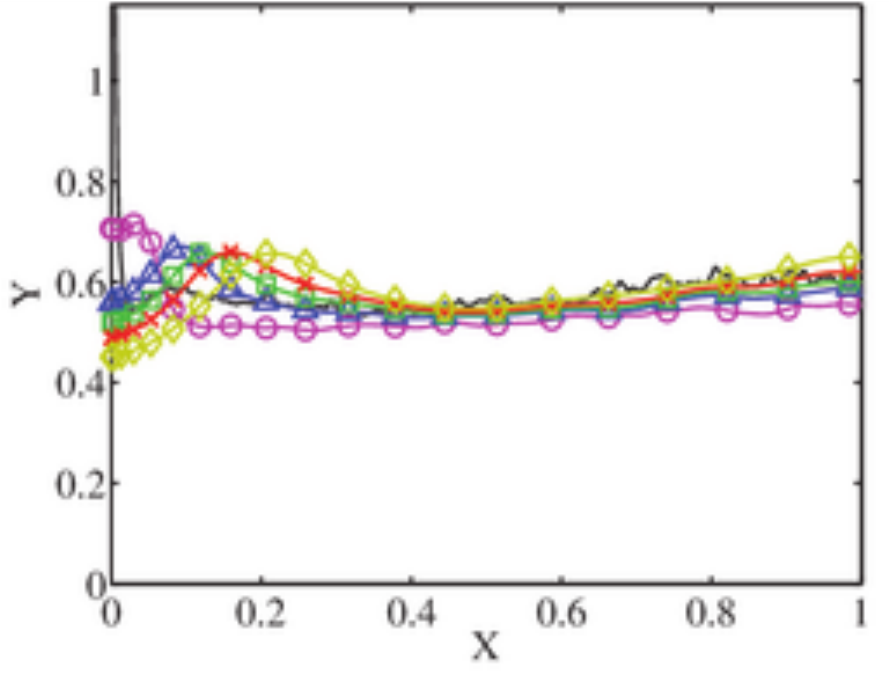}
\mylab{-5.3cm}{4.7cm}{(b)}
}
\caption{ (a) Root-mean-squared streamwise velocity fluctuations,
  normalized with $u_\tau$ from the unfiltered case, as a function of
  the wall-normal distance, $x_2$.  (b) Root-mean-squared third
  invariant, $R'(x_2)$, normalized by $Q'^{3/2}(x_2)$.  Symbols and
  colors are as in table \ref{table:cases}.
\label{fig:vel_scaling}}
\end{figure}
%================================================================
%
% Cases
Based on the filtered velocity, the invariants and their total
derivatives are then calculated as described in
\S\ref{subsec:numerical}.  The results are computed for six filter
widths summarized in Table \ref{table:cases}.  The first case
corresponds to unfiltered data, and the rest are denoted by F$\gamma$,
where $\gamma$ is the wall-normal filter width,
$\Delta_2/h=0.1,0.2,0.25,0.3$ and $0.4$. Note that the largest filter
width is $\Delta_1=1.2h$, which is still far from the streamwise
length of the computational domain $L_1 \approx 6.3h$.  The effects of
the size of the box on the results presented in this manuscript turned
out to be negligible compared to those in larger domains, and are
briefly discussed in Appendix \ref{sec:appendix}.

% Crossed terms problem
As previously mentioned, we compute the invariants of the filtered
velocity components which are the ones usually used in the literature
and perhaps also more useful for comparison with experimental data,
which sometimes suffer from limited spatial resolution and measure
coarse-grained velocity fields. Note that $u_1$ can be expressed as
$u_1 = \widetilde{u}_1+u_{1,r}$, where $u_{1,r}$ is the residual and
$\widetilde{u}_1$ the filtered velocity.  This filtering approach has
the clear physical meaning of removing those scales smaller than the
filter width. However, there is a shortcoming when non-linear terms
are considered (e.g. kinetic energy or the invariants $R$ and $Q$) and
mixed terms of the type $\widetilde{u}_1u_{1,r}$ appear. This poses a
problem in the sense that it is not clear whether they should be added
to the filtered quantity or to the residuals.  Hence, another
possibility for obtaining filtered invariants would be to filter $R$
and $Q$ directly. This approach does not suffer from the ambiguity
above because there are no mixed terms, but the physical meaning is
less clear, e.g. filtered $k$ is not the kinetic energy of a well
defined velocity field. For that reason, this approach is not used in
the present work.

% Scaling
The root-mean-square of the streamwise velocity fluctuations for the
filtered and unfiltered cases is shown in figure
\ref{fig:vel_scaling}(a).  Similar trends are observed in the
wall-normal and spanwise velocity fluctuations (not shown). The
invariants are normalized as shown in (\ref{eq:normalization}), and
for the filtered data, the resulting $\widetilde{Q}'(x_2)$ of each case
is used. The vector field, $\boldsymbol{v}$, from (\ref{eq:v_norm}) is
normalized with the time scale $h/u_\tau$. Figure
\ref{fig:vel_scaling}(b) shows that this normalization provides a very
good collapse of $R'$, at least far from the wall.

%%%%%%%%%%%%%%%%%%%%%%%%%%%%%%%%%%%%%%%%%%%%%%%%%%%%%%%%%%%%%%%%%%%%%%%%%%%%
\section{Results}\label{sec:results}
%%%%%%%%%%%%%%%%%%%%%%%%%%%%%%%%%%%%%%%%%%%%%%%%%%%%%%%%%%%%%%%%%%%%%%%%%%%%

%-------------------------------------------------------%
%-------------------------------------------------------%
\subsection{R and Q joint distributions}\label{subsec:RQ}
%-------------------------------------------------------%
%-------------------------------------------------------%

\begin{figure}
\vspace{0.5cm}
\centerline{
\psfrag{X}{ }\psfrag{Y}{$Q/Q'$}
\includegraphics[width=0.45\textwidth]{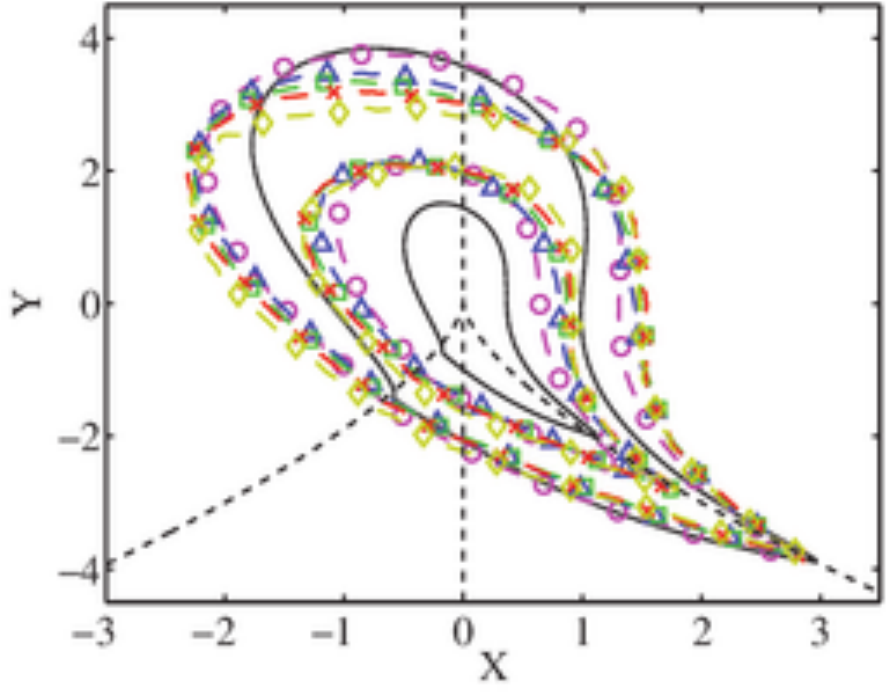}
\mylab{-5.3cm}{4.9cm}{(a)}
\psfrag{X}{ }\psfrag{Y}{ }
\includegraphics[width=0.45\textwidth]{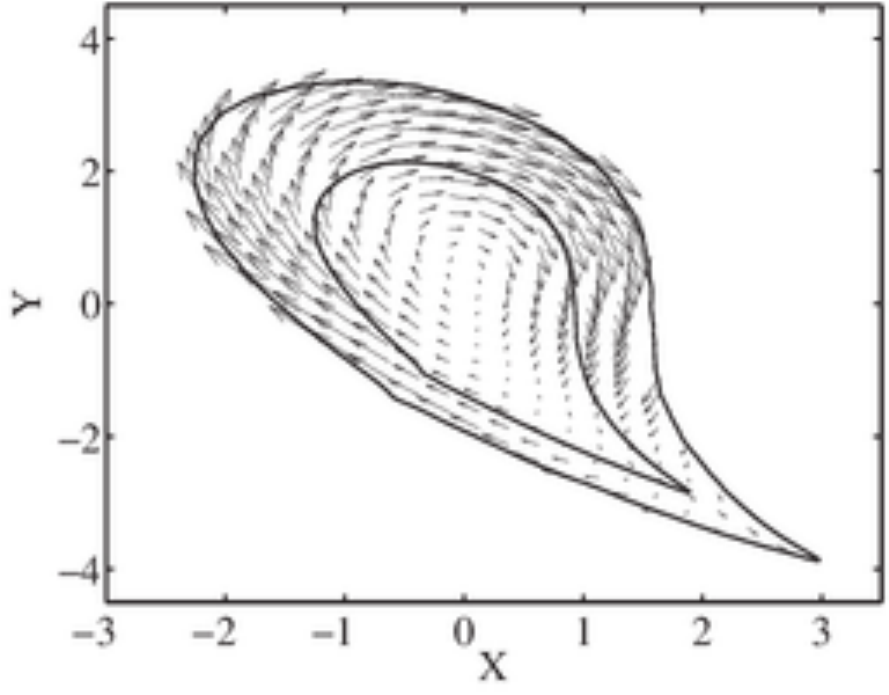}
\mylab{-5.3cm}{5.0cm}{(b)} 
}
\vspace{0.5cm}
\centerline{
\psfrag{X}{ \raisebox{-0.2cm}{$R/Q'^{3/2}$} }\psfrag{Y}{$Q/Q'$}
\includegraphics[width=0.475\textwidth]{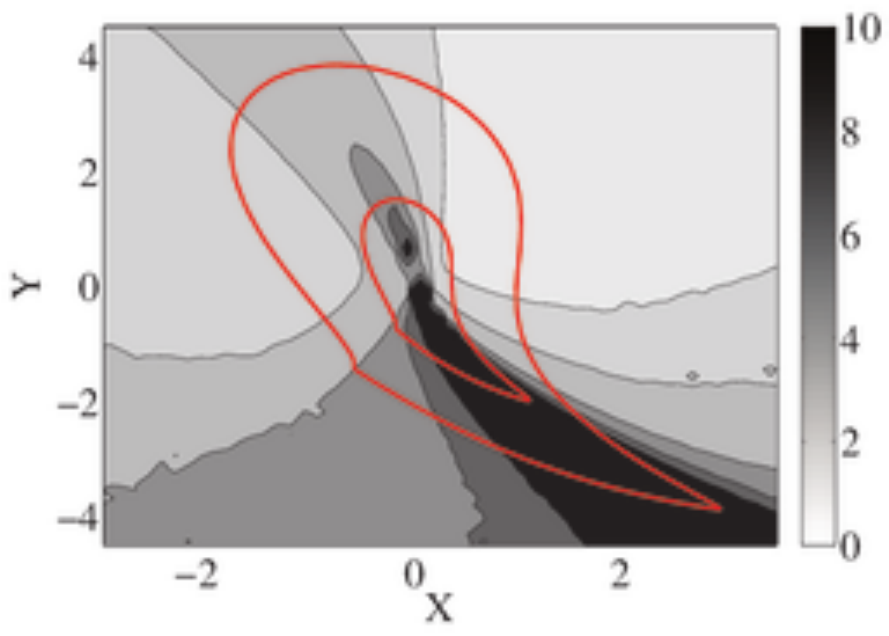}
\mylab{-5.3cm}{4.9cm}{(c)}
\psfrag{X}{ \raisebox{-0.2cm}{$R/Q'^{3/2}$} }\psfrag{Y}{ }
\includegraphics[width=0.485\textwidth]{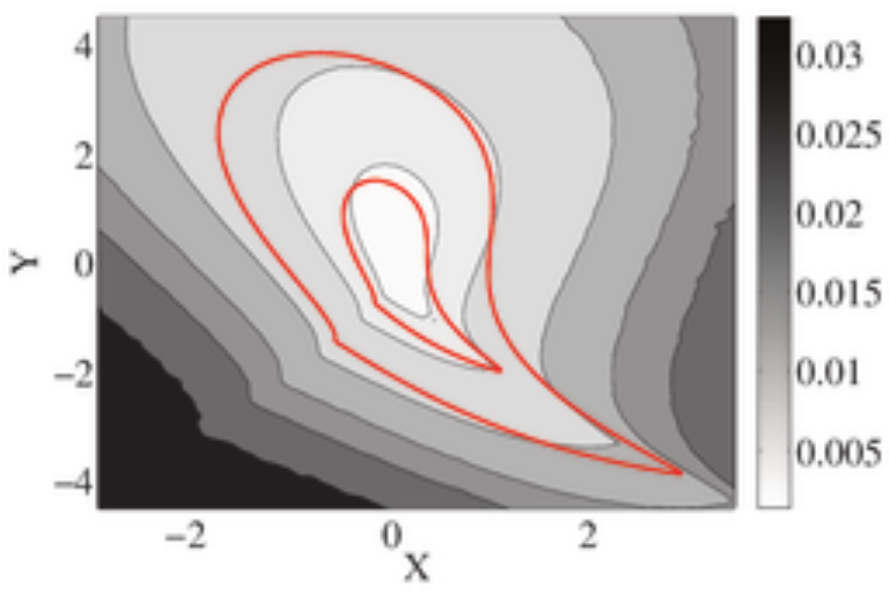}
\mylab{-5.3cm}{5.0cm}{(d)}
}
\caption{ (a) Joint probability density functions of $R$ and $Q$.  The
  iso-probability contours contain 90\% and 98\% of the data (that is,
  90\% and 98\% of the data points are enclosed by each contour
  respectively), corresponding to $1.5\times 10^{-4}$ and $3.3 \times
  10^{-6}$ of the maximum of the p.d.f.s. The dashed lines are $R=0$
  and $D=0$. Symbols and color are as in table \ref{table:cases}.  (b)
  Vector field $\boldsymbol{v}$ for F0.25. The solid lines contain
  90\% and 98\% of the data.  (c) Ratio of the magnitude of the
  conditionally averaged velocity deviation $\boldsymbol{v}_{std}$ and
  the mean $\boldsymbol{v}$, conditioned on the $R$--$Q$ plane.
  Results for the unfiltered case. Although the colorbar ranges from 0
  to 10, values up to 100 are attained close to the Vieillefosse
  tail. (d) Magnitude of the conditionally averaged velocity deviation
  $\boldsymbol{v}_{std}h/u_\tau$. Red solid lines in (c) and (d)
  contain 90\% and 98\% of the data for the probability density
  function of $R$ and $Q$.
\label{fig:pdfRQ}}
\end{figure}

% RQ pdfs
In this section, we compare the invariants computed from the filtered
and unfiltered velocity fields.  The results are presented in figure
\ref{fig:pdfRQ}(a), which shows the joint probability density
functions (p.d.f.s) of $R$ and $Q$ for all the cases in table
\ref{table:cases}. For comparison, figure \ref{fig:pdfRQ}(b) includes
the conditionally averaged velocity $\boldsymbol{v}$ from
(\ref{eq:v_norm}) for case F0.25. Similar vector fields are obtained
for the other cases although they are omitted for the sake of brevity.
Consistent with earlier findings \citep{Borue1998, VanDerBos2002,
  Luethi2007}, the iso-probability lines maintain a tear-drop shape
for the filtered cases, and the strain production dominates over
enstrophy production in strain-dominated regions also in the inertial
scales. This result differs from the more symmetrical p.d.f.s obtained
for the coarse-grained velocity reported in previous works
\citep{Chertkov1999,Naso2005, Naso2007, Pumir2010} and computed
through a stochastic tetrad model based on the evolution of four
tracer points.  Nevertheless, \citet{Luethi2007} showed that the
latter results are aliased and that the tear-drop shape is recovered
when the velocity derivatives are computed using a larger number of
tracers.
%
%================================================================
% /data4/adrian/Q1Q2R1R2/mfiles/plotQ2Q2R1R2_cte.m
% /data4/adrian/Q1Q2R1R2/mfiles/plot_Vmods_cte.m
%================================================================

The five filtered cases collapse well except for some small
differences for large values of $Q$. However, they differ from the
unfiltered case along the horizontal axis where the iso-contours of
the filtered ones tend to broaden, meaning that for a given level of
$Q$, the invariant $R$ is stronger than in the unfiltered case.

% Normalization
The fairly good overlap between the filtered cases also indicates that
the normalization with $\widetilde{Q}'$ is appropriate. Without the
normalization, the values for the filtered cases are strongly reduced
by several orders of magnitude with respect to the unfiltered case
(not shown), since strong and intermittent events are mostly caused by
the small scale structure of turbulence
\citep{bat:tow:1949,jim:2000}. The good collapse for the filtered
cases suggests that the dynamics on the $R$--$Q$ plane are
self-similar in the inertial range, although it will be shown in
section \ref{subsec:QsQwRsRw} that this is not the case for their
enstrophy and strain components.

% Relative variance of velocity in RQ
Note that the vectors in figure \ref{fig:pdfRQ}(b) are statistical
representations of the conditional dynamics of the flow, and only
represent the evolution of individual particles when their standard
deviation is small with respect to the mean. Otherwise, they should be
considered as small residuals of a more complex underlying
evolution. This does not mean that they are irrelevant to the
flow. They represent evolutionary trends, in the same sense as the
bulk flow of a low-Mach-number fluid is a small residue of the much
faster random motion of its molecules. To shed some light on that,
figure \ref{fig:pdfRQ}(c) shows the ratio of the magnitudes of
$\boldsymbol{v}$ and of the deviation vector
\begin{equation}\label{eq:v_deviation}
  \boldsymbol{v}_{std} = \left \langle \left(
\left[ \frac{\mathrm{D}}{\mathrm{D}t} \left( \frac{R}{{Q'}^{3/2}} \right) \right]_{std} ,
\left[ \frac{\mathrm{D}}{\mathrm{D}t} \left( \frac{Q}{{Q'}      } \right) \right]_{std}
\right) \right \rangle_{R/{Q'}^{3/2},Q/{Q'}}, 
\end{equation}
where the subindex $std$ denotes standard deviation at each
$R/{Q'}^{3/2},Q/{Q'}$ point.  The average velocity field is relevant
at those points where $|\boldsymbol{v}_{std}|$ is much smaller than
$|\boldsymbol{v}|$, with $|\cdot|$ the $L^2$-norm. In most of the
$R$--$Q$ plane, $|\boldsymbol{v}_{std}|/|\boldsymbol{v}|$=3--5,
meaning that $\boldsymbol{v}$ is not highly representative of the
trajectories of individual fluid particles but rather a weak trend of
their motion. This is particularly pronounced along the Vieillefosse
tail, where the ratio achieves values up to $100$. Similar results
were reported by \citet{Luethi2009}. This effect is caused by the
small values attained by the mean rather than by the standard
deviation as seen in figure \ref{fig:pdfRQ}(d), which shows
$|\boldsymbol{v}_{std}|$ without dividing by the mean.  Figure
\ref{fig:pdfRQ}(c) corresponds to the unfiltered case, and
qualitatively similar values are obtained for the filtered ones (see
example in Appendix \ref{sec:appendixB}).

%-------------------------------------------------------%
%-------------------------------------------------------%
\subsection{ Conditional mean trajectories }\label{subsec:RQ_CMTs}
%-------------------------------------------------------%
%-------------------------------------------------------%

%
%================================================================
% /data4/adrian/Q1Q2R1R2/mfiles/plotQ2Q2R1R2_cte.m
\begin{figure}
\vspace{0.5cm}
\centerline{
\psfrag{X}{ }\psfrag{Y}{$Q/Q'$}
\includegraphics[width=0.45\textwidth]{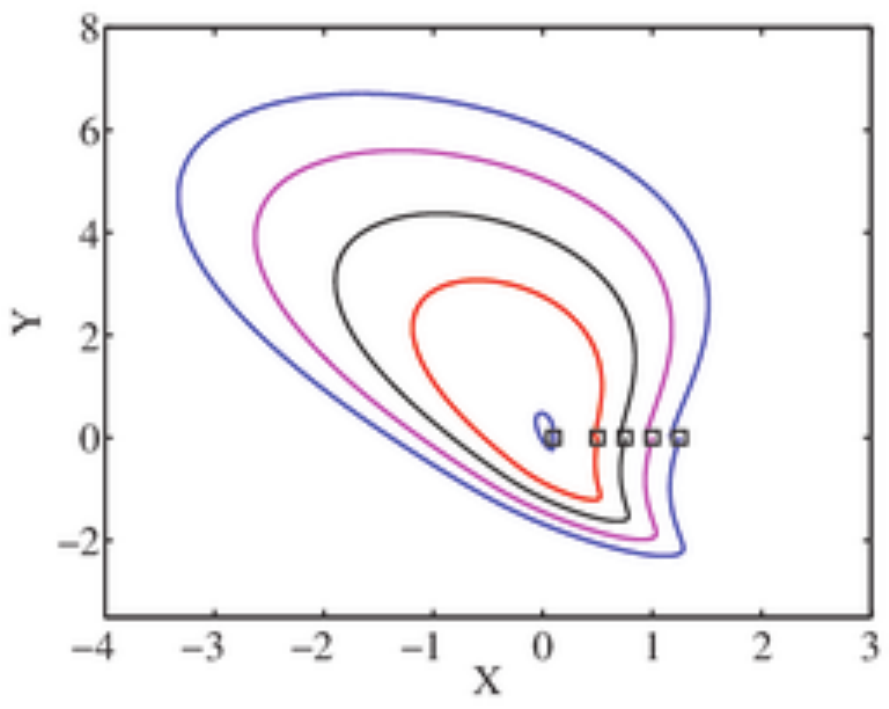}
\mylab{-5.3cm}{4.9cm}{(a)}
\psfrag{X}{ }\psfrag{Y}{ }
\includegraphics[width=0.45\textwidth]{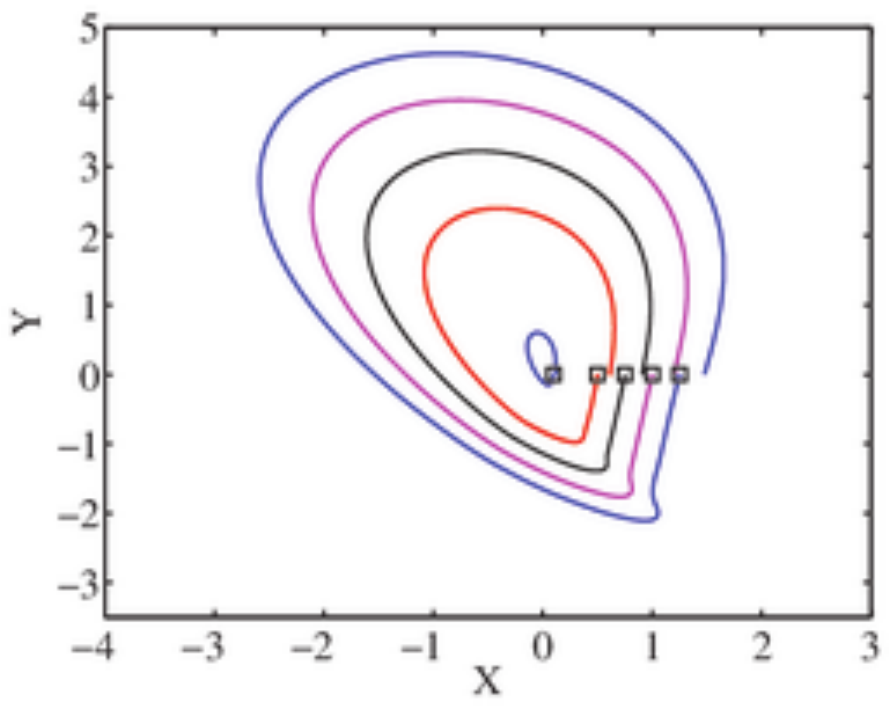}
\mylab{-5.3cm}{4.9cm}{(b)}
}
\centerline{
\psfrag{X}{ }\psfrag{Y}{$Q/Q'$}
\includegraphics[width=0.45\textwidth]{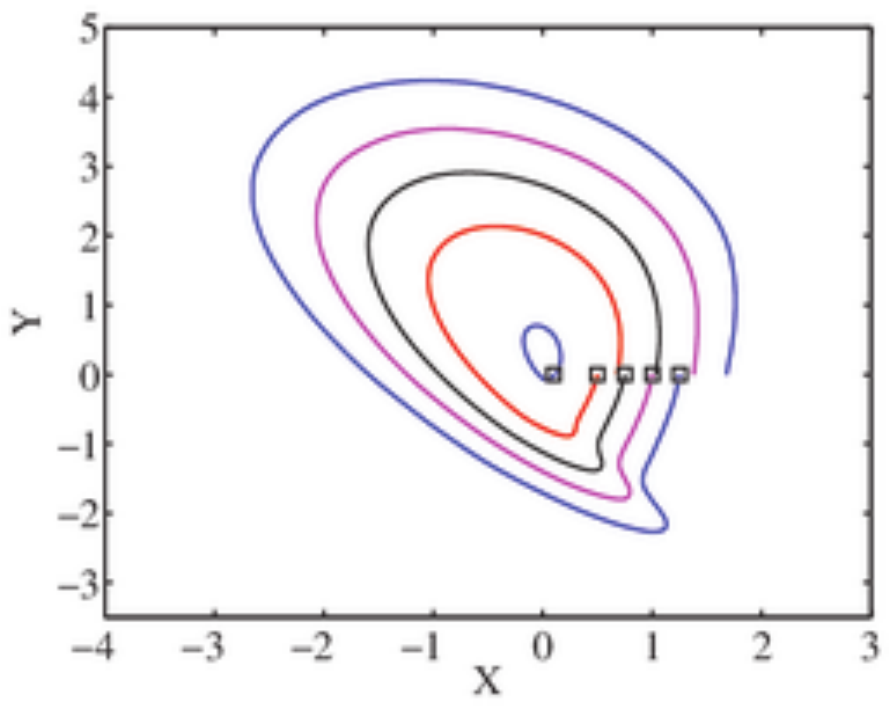}
\mylab{-5.3cm}{4.9cm}{(c)}
\psfrag{X}{ }\psfrag{Y}{ }
\includegraphics[width=0.45\textwidth]{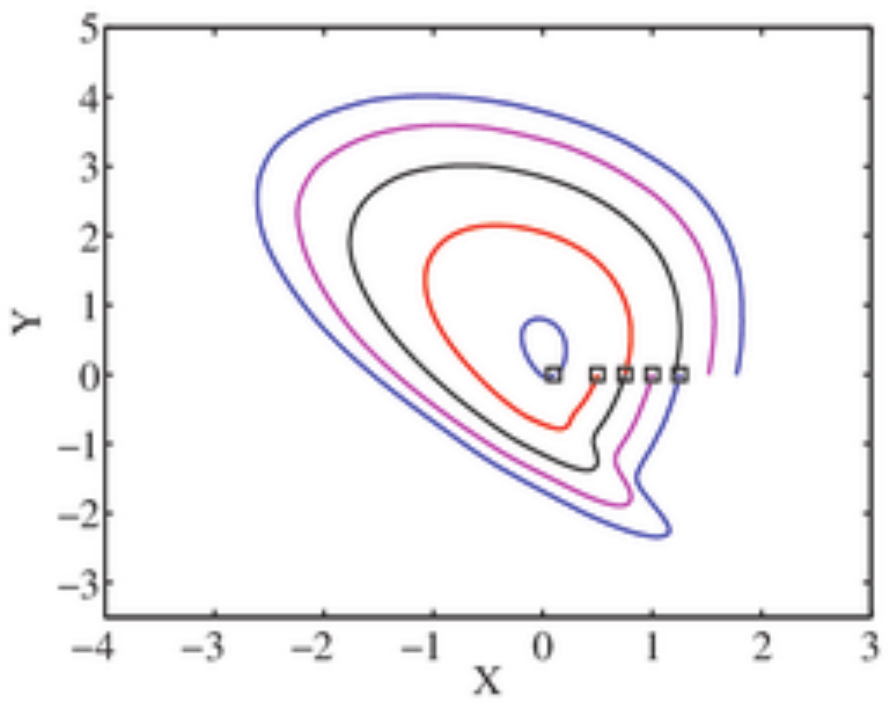}
\mylab{-5.3cm}{4.9cm}{(d)}
}
\centerline{
\psfrag{X}{ \raisebox{-0.2cm}{$R/Q'^{3/2}$} }\psfrag{Y}{$Q/Q'$}
\includegraphics[width=0.45\textwidth]{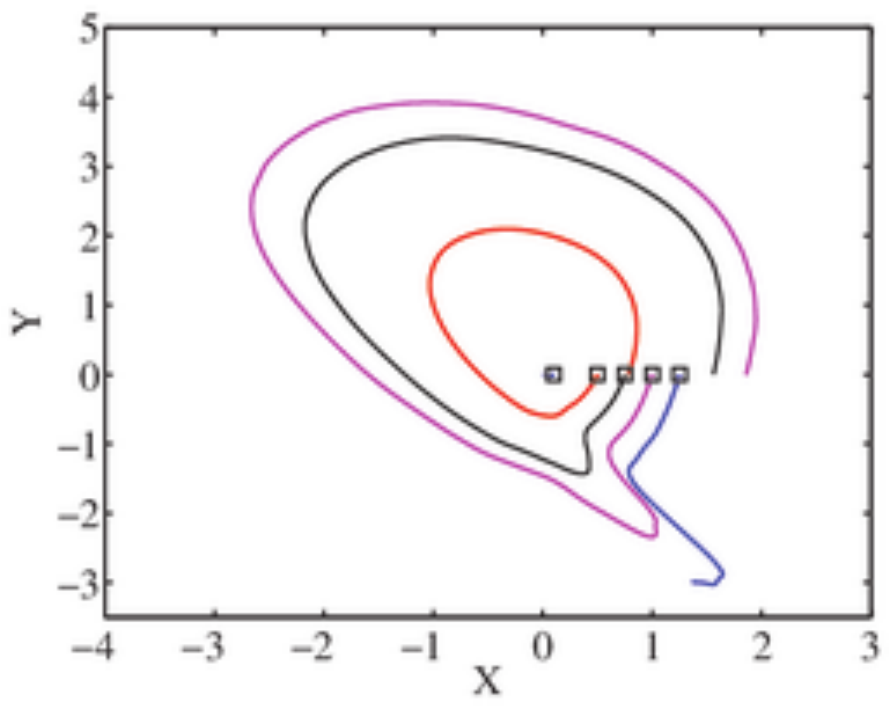}
\mylab{-5.3cm}{4.9cm}{(e)}
\psfrag{X}{ \raisebox{-0.2cm}{$R/Q'^{3/2}$} }\psfrag{Y}{ }
\includegraphics[width=0.45\textwidth]{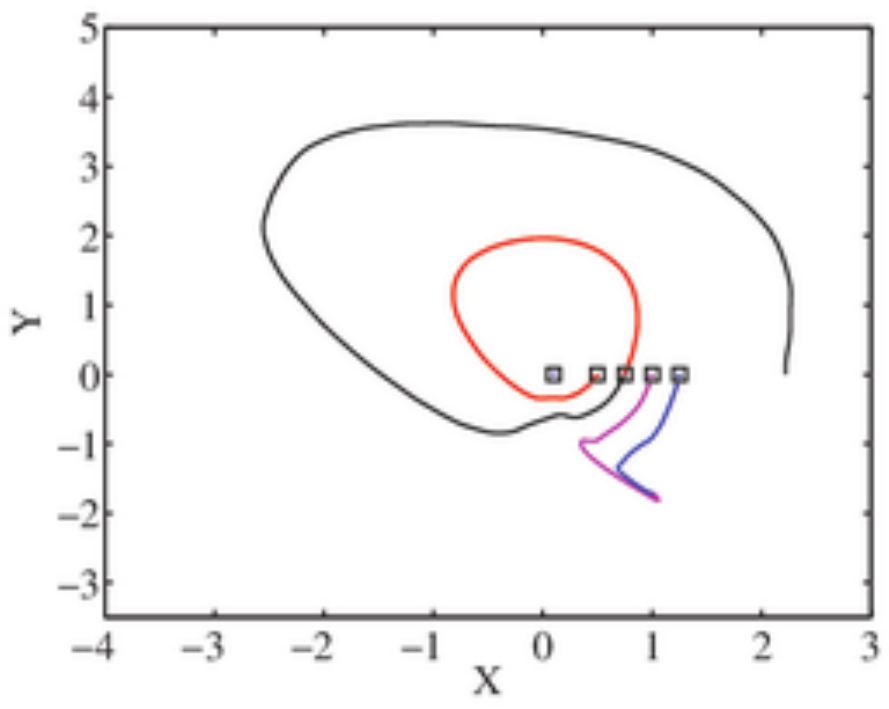}
\mylab{-5.3cm}{4.9cm}{(f)}
}
\caption{ CMTs in the $R$--$Q$ plane computed as shown in
  (\ref{eq:v_norm}). The initial $R$ and $Q$ positions (marked by
  \squar) are the same for all of the cases,
  $(R/{Q'_s}^{3/2},Q/{Q'_s})=(0.1,0), (0.5,0), (0.75,0), (1,0),
  (1.25,0)$.  (a), case F0; (b), case F0.1; (c), case F0.2; (d), case
  F0.25; (e), case F0.3; (f), case F0.4.} \label{fig:pdfRQ_cmts}
\end{figure}
%================================================================
%
% Conditional Mean Trajectories
The CMTs are shown in figure \ref{fig:pdfRQ_cmts}.  It is important to
remark that accurate calculation of CMTs is numerically challenging,
and the inwards spiraling observed in many works is spurious. We have
shown in section \ref{subsec:numerical} that our numerics are good
enough to recover closed CMTs when the full domain is considered.
Results in figure \ref{fig:pdfRQ_cmts} are physical and not the
byproduct of numerical artifacts.

\citet{loz:hol:jim:2015} showed that unfiltered CMTs, scaled as in
(\ref{eq:v_norm}), describe closed trajectories when the whole channel
domain is considered (see figure \ref{fig:examples}), and the same
result is found to be valid here for the filtered cases (not shown).
When the p.d.f. is computed in a subdomain defined by two
wall-parallel planes, as it is done for the logarithmic region in this
paper, the trajectories need not to be closed any more. In that case,
the equation for the conservation of probability of $J=J(R/Q'^{3/2}
,Q/Q')$ for a stationary state is
\begin{equation}\label{eq:prob_conser}
 \boldsymbol{\nabla}_{R,Q}\cdot \left( J \boldsymbol{v} \right) = \psi_t + \psi_b,
\end{equation}
where 
\begin{equation}
\boldsymbol{\nabla}_{R,Q} = \left( \frac{\partial}{\partial R/Q'^{3/2}},\frac{\partial}{\partial Q/Q'}\right),
\end{equation}
$\boldsymbol{v}$ is as defined in
(\ref{eq:v_norm}). $\psi_b$ and $\psi_t$ are the probability fluxes at
the bottom plane ($x_2^+=100$) and at the top plane ($x_2/h=0.4$)
\begin{eqnarray}\label{eq:fluxes1}
\psi_b &=&  \alpha V_b J_b, \\
\psi_t &=& -\alpha V_t J_t,\label{eq:fluxes2}
\end{eqnarray}
where $V_b$, $V_t$, $J_b$ and $J_t$ are functions of $(R/Q'^{3/2} ,
Q/Q')$.  $V_b$ and $V_t$ are the conditional wall-normal velocities on
the $R/Q'^{3/2}$--$Q/Q'$ plane at $x_2=x_{b}$ and $x_2=x_{t}$,
respectively, and $J_b$ and $J_t$ the probability density functions at
those same heights.  $\alpha$ is a scale-factor equal to
$1/(x_t-x_b)$. For $x_{b}$ and $x_{t}$ equal to the bottom and top
walls, fluxes (\ref{eq:fluxes1}) and (\ref{eq:fluxes2}) become zero
and equation (\ref{eq:prob_conser}) is equivalent to the stationary
Fokker-Planck equation used in previous works
\citep{VanDerBos2002,Chevillard2007,Chevillard2011}.  Some guidelines
for deriving equation (\ref{eq:prob_conser}) are provided in Appendix
\ref{sec:appendixProb}.
 
Figure \ref{fig:pdfRQ_cmts}(a) shows that, for the unfiltered case,
the CMTs describe clockwise cycles around the origin in almost closed
trajectories, consistently with the results from
\citet{loz:hol:jim:2015}, who observed a probability flux of strong
$R$--$Q$ leaving the buffer layer and entering the outer zone, but
which canceled when both boundaries were considered.

\begin{figure}
\vspace{0.5cm}
\centerline{
\psfrag{X}{ \raisebox{-0.2cm}{$R/Q'^{3/2}$} }\psfrag{Y}{$Q/Q'$}
\includegraphics[width=0.45\textwidth]{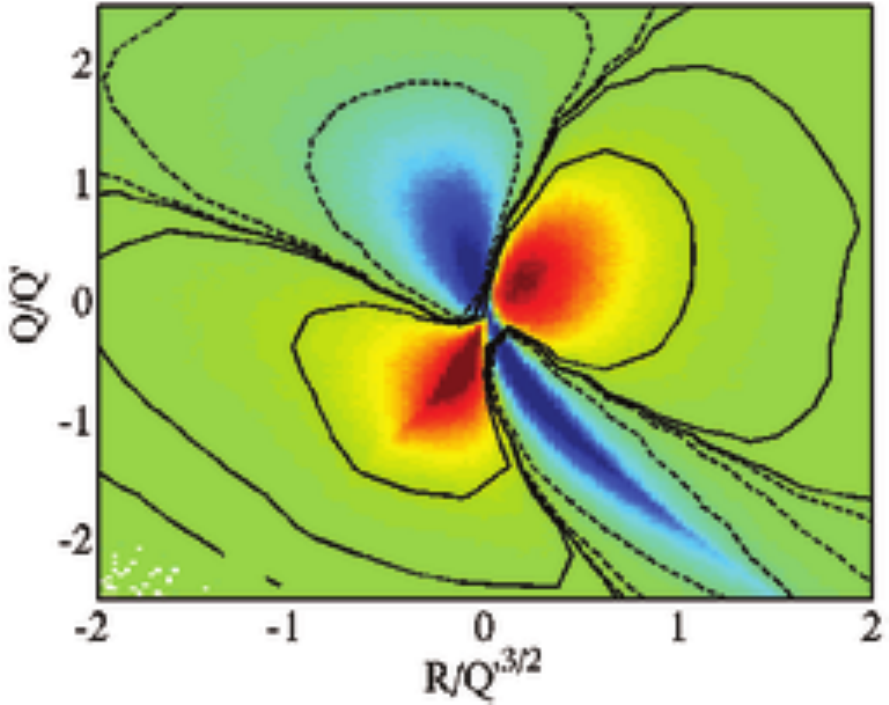}
\mylab{-5.3cm}{4.9cm}{(a)}
\psfrag{X}{ \raisebox{-0.2cm}{$R/Q'^{3/2}$} }\psfrag{Y}{ }
\includegraphics[width=0.45\textwidth]{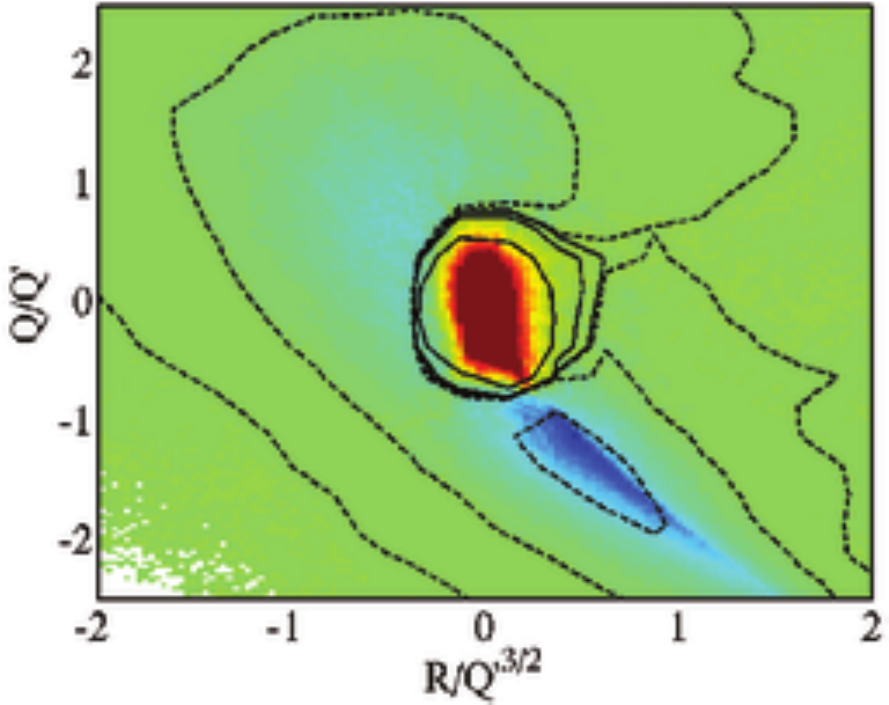}
\mylab{-5.3cm}{4.9cm}{(b)}
}
\centerline{
\psfrag{X}{ \raisebox{-0.2cm}{$R/Q'^{3/2}$} } \psfrag{Y}{$Q/Q'$}
\includegraphics[width=0.45\textwidth]{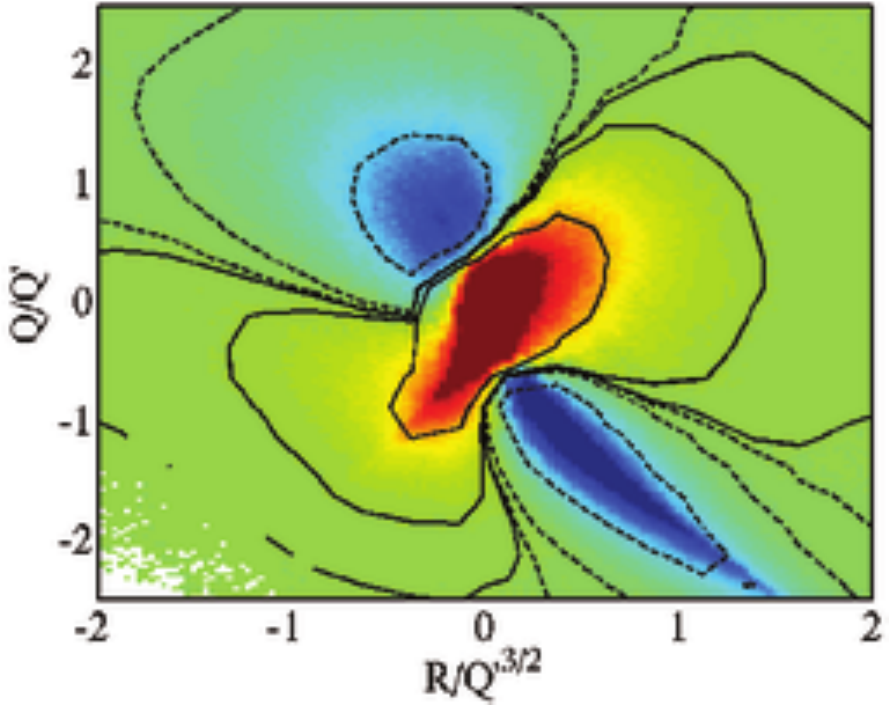}
\mylab{-5.3cm}{4.9cm}{(c)}
}
\caption{ Probability fluxes from equation (\ref{eq:fluxes1}) and
  (\ref{eq:fluxes2}) at plane (a), $x_2^+=100$ ($\psi_b$); (b)
  $x_2/h=0.4$ ($\psi_t$); and (c), net flux through both boundaries
  ($\psi_b+\psi_t$). Positive values are represented by hot colors and
  solid contours. Negative values are cold colors and dashed
  lines. Contours are $\pm 10^{-1}$, $\pm 10^{-2}$ and $\pm 10^{-3}$ of
  the maximum. Data for F0.25.} \label{fig:pdfRQ_fluxes}
\end{figure}

For the filtered cases (figures \ref{fig:pdfRQ_cmts}b to
\ref{fig:pdfRQ_cmts}f), the CMTs spiral outwards.  Intuitively, this
is caused by the unbalanced of $R$ and $Q$ associated with the fluid
leaving and entering the subdomain through the boundaries, and can be
better understood by studying the fluxes $\psi_b$ and $\psi_t$ shown
in figures \ref{fig:pdfRQ_fluxes}(a) and (b) for case F0.25.  At the
bottom boundary, incoming fluxes are located at the first ($R>0$ and
$Q>0$) and third ($R<0$ and $Q<0$) quadrant, whereas at the top one
they concentrate at the centre of the $Q$--$R$ plane.  The resulting
net effect $\psi_b+\psi_t$ (figure \ref{fig:pdfRQ_fluxes}c) is
dominated by incoming flux of weak events into the logarithmic layer,
and a secondary outflow of stronger events distributed in the second
and fourth quadrants. Qualitatively similar results are obtained for
other filter widths (not shown).

In conclusion, the outward spiraling is mainly due to weaker
normalized $R$ and $Q$ transported from the outer region into the
logarithmic layer, where they are amplified. Results from the
remaining cases (not shown) reveal that the increasing spiraling with
wider filter widths is caused by the influx from the lower boundary
being damped by the filter, while fluxes across the upper boundary
remain similar.  If the effect of the viscous terms is considered
negligible for the filtered cases, the outward spiraling of inertial
CMTs may be attributed to the combined effect of self-amplification,
pressure and interscale transfer. Note that the outward spiraling also
implies that the residual CMTs, i.e., those which added to the
filtered cases result in the unfiltered one, must spiral inwards in
order to recover the almost closed CMTs in figure
\ref{fig:pdfRQ_cmts}(a).  This may be caused by viscous and/or
interscale transfer effects (among others), and a term-by-term
analysis of the dynamic equations of $\widetilde{R}$ and
$\widetilde{Q}$ (and of the residual counterparts) would be necessary
to address this question in detail.  This will be tackled in future
studies.
%
%================================================================
% /data4/adrian/Q1Q2R1R2/mfiles/plotQR_fluxes.m
%================================================================
%

%-------------------------------------------------------%
%-------------------------------------------------------%
\subsection{Strain and enstrophy components}\label{subsec:QsQwRsRw}
%-------------------------------------------------------%
%-------------------------------------------------------%

% Decomposition
Despite the similarities found for the joint distributions of $R$ and
$Q$ in the previous section, we show below that decomposing the
invariants in their strain and enstrophy components leads to quite
pronounced differences. Following \citet{Ooi1999}, $Q$ and
$R$ are decomposed as
\begin{eqnarray}
\label{eq:QsQw}
Q_s &=& -\frac{1}{2}s_{ij}s_{ij}, \quad Q_\omega = \frac{1}{4}\omega_i\omega_i,  \\
\label{eq:RsRw}
R_s &=& -\frac{1}{3}s_{ij}s_{jk}s_{ki}, \quad R_\omega = -\frac{1}{4}\omega_i\omega_j s_{ij},
\end{eqnarray}
so that $Q=Q_s+Q_\omega$ and $R=R_s+R_\omega$. Note that this
decomposition differs from that in \cite{Soria1994,Blackburn1996} and
\cite{Davidson2004} who considered the invariants of the
rate-of-strain, $s_{ij}$, and rate-of-rotation, $\Omega_{ij}$,
tensors. In those cases, $Q_s$ and $R_s$ coincide with the second and
third invariants of $s_{ij}$, and so does $Q_\omega$ with the second
invariant of $\Omega_{ij}$. However, the third invariant of
$\Omega_{ij}$ is zero and $R_\omega$ is not.

% Interpretation of Qs, Qw, Rs, Rw
Relations (\ref{eq:QsQw}) and (\ref{eq:RsRw}) show that $Q_\omega$ is
proportional to the enstrophy density whose intense values tend to
concentrate in tube-like structures \citep{jim:wra:saf:rog:93}. $Q_s$
is proportional to the strain, which is proportional to the local rate
of viscous dissipation of kinetic energy, $\varepsilon=-4\nu Q_s$,
with high values organized in sheets or ribbons \citep{Moisy2004}.
The meaning of $R_s$ and $R_\omega$ is closely connected to the
evolution equations for the strain and enstrophy densities, that for
the filtered cases are (see \cite{Ooi1999} for the original equations
for the unfiltered case)
\begin{eqnarray}
\label{eq:DQw}
\frac{D}{Dt}\left( \frac{\widetilde{\omega}_i\widetilde{\omega}_i}{2} \right) &=&
\widetilde{\omega}_i\widetilde{\omega}_j \widetilde{s}_{ij} +
\nu \widetilde{\omega}_i \frac{\partial^2 \widetilde{\omega}_i}{\partial x_k \partial x_k} +
\epsilon_{ilm} \widetilde{\omega}_i \frac{\partial^2 \tau_{mk}}{\partial x_l \partial x_k},\\
\label{eq:DQs}
\frac{D}{Dt}\left( \frac{\widetilde{s}_{ij}\widetilde{s}_{ij}}{2} \right) &=&
-\widetilde{s}_{ik}\widetilde{s}_{kj}\widetilde{s}_{ij} -
\frac{1}{4}\widetilde{\omega}_i\widetilde{\omega}_j\widetilde{s}_{ij} -
\widetilde{s}_{ij}\frac{\partial^2 \widetilde{p}}{\partial x_i \partial x_j} +
\nu \widetilde{s}_{ij} \frac{\partial^2 \widetilde{s}_{ij}}{\partial x_k \partial x_k}+
\widetilde{s}_{ij}\frac{\partial B_{ijk}}{\partial x_k},
\end{eqnarray}
where $\tau_{mk}=\widetilde{u}_m \widetilde{u}_k-\widetilde{u_m u_k}$
and $B_{ijk}= 1/2(\partial \tau_{ik} /\partial x_j + \partial
\tau_{jk} /\partial x_i)$ are responsible for the interscale transfer
of strain and enstrophy.  The corresponding relations for the
unfiltered case are recovered by taking $\tau_{ik}=0$.  Relations
(\ref{eq:DQw}) and (\ref{eq:DQs}) show that $R_s$ and $R_\omega$ are
proportional to the strain self-amplification and enstrophy
production, respectively. We will use $Q'_s(x_2)$ to
non-dimensionalize quantities related to the strain (such as $Q_s$ and
$R_s$) and $Q'_\omega(x_2)$ for those related to the enstrophy (such
as $Q_\omega$ and $R_\omega$). The material derivatives are computed
for the normalized quantities as in (\ref{eq:v_norm}).

%================================================================
% /data4/adrian/Q1Q2R1R2/mfiles/plotQ1Q2R1R2_cte.m
\begin{figure}%[h!tb]
\centerline{
\psfrag{X}{$R_\omega$}\psfrag{Y}{$Q_\omega$}
\includegraphics[width=0.432\textwidth]{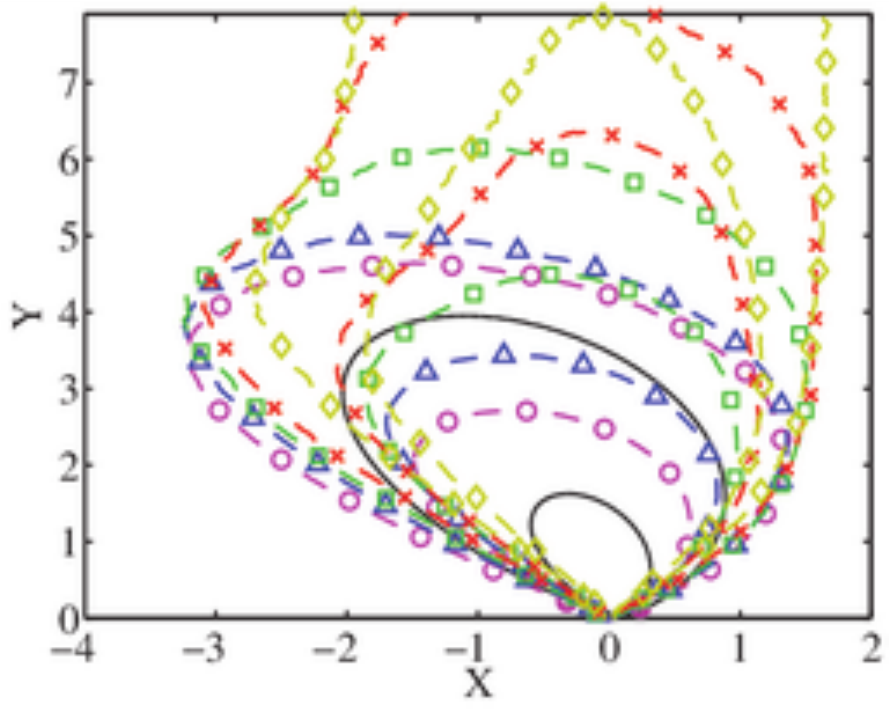}
\mylab{-1cm}{4.0cm}{(a)}
\psfrag{X}{$R_\omega$}\psfrag{Y}{$Q_\omega$}
\includegraphics[width=0.432\textwidth]{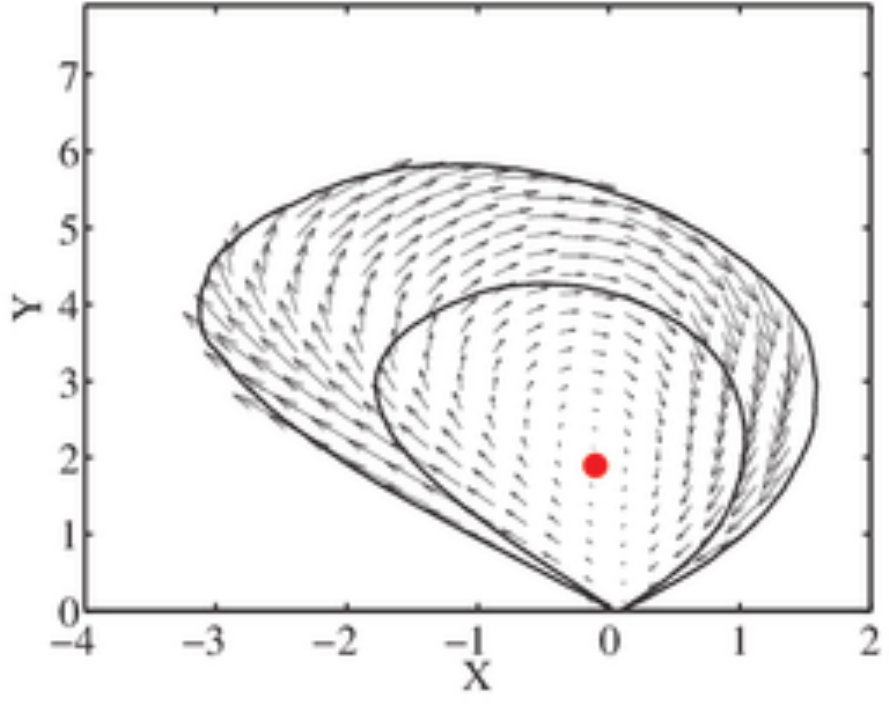}
\mylab{-1cm}{4.0cm}{(b)}
}
\centerline{
\psfrag{X}{$R_s$}\psfrag{Y}{$Q_s$}
\includegraphics[width=0.435\textwidth]{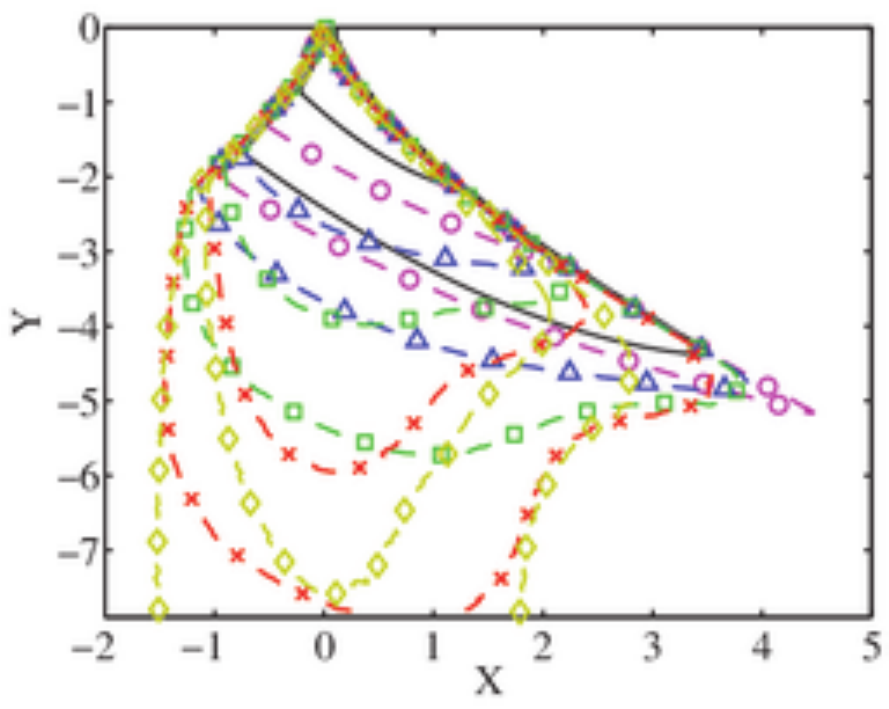}
\mylab{-1.0cm}{4.0cm}{(c)}
\psfrag{X}{$R_s$}\psfrag{Y}{$Q_s$}
\includegraphics[width=0.435\textwidth]{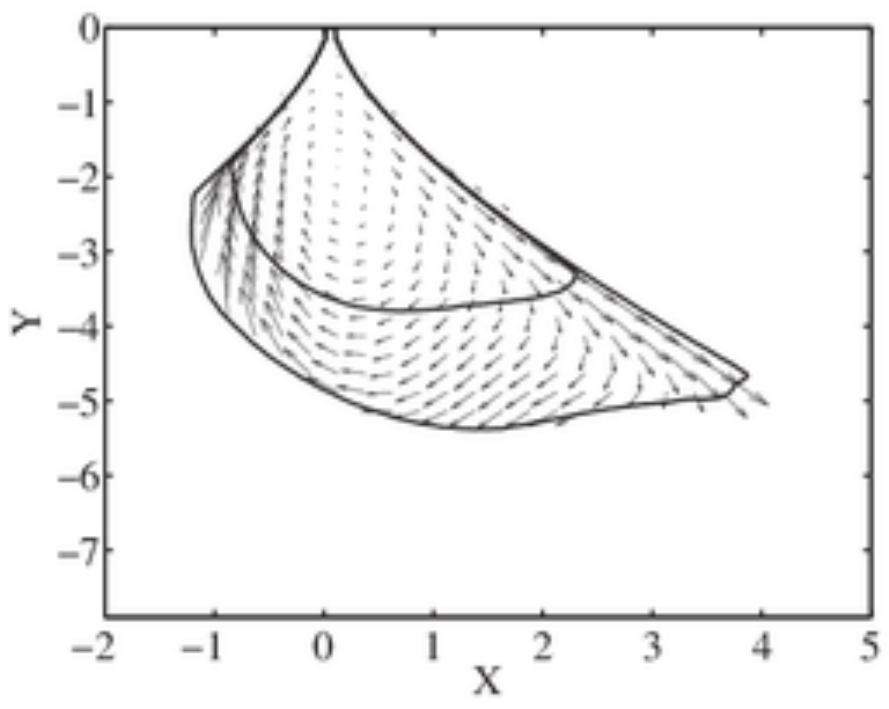}
\mylab{-1.0cm}{4.0cm}{(d)}
}
\centerline{
\psfrag{X}{$Q_s$}\psfrag{Y}{$Q_\omega$}
\includegraphics[width=0.435\textwidth]{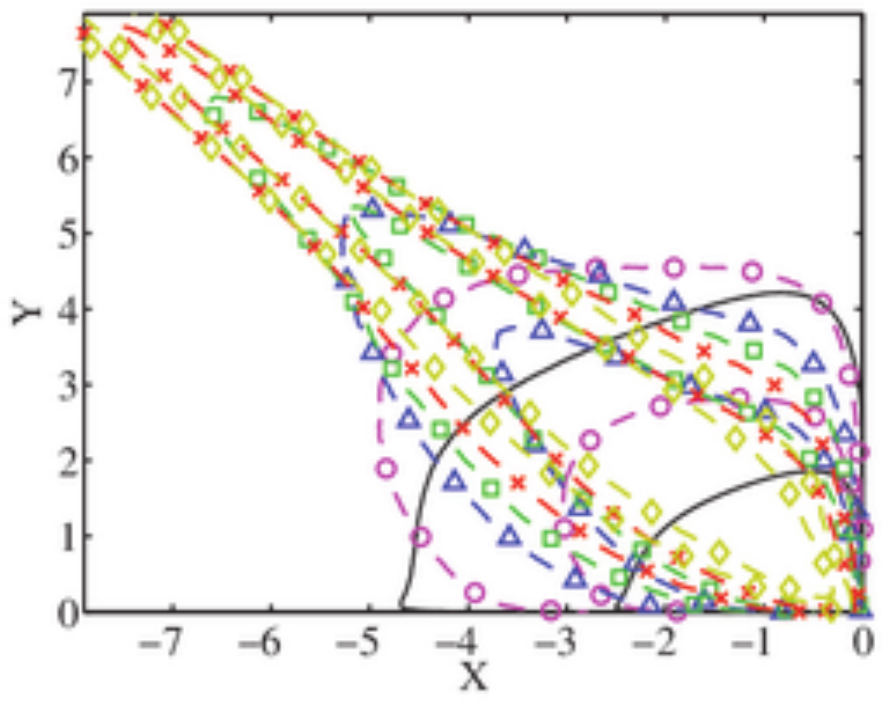}
\mylab{-1.0cm}{4.0cm}{(e)}
\psfrag{X}{$Q_s$}\psfrag{Y}{$Q_\omega$}
\includegraphics[width=0.435\textwidth]{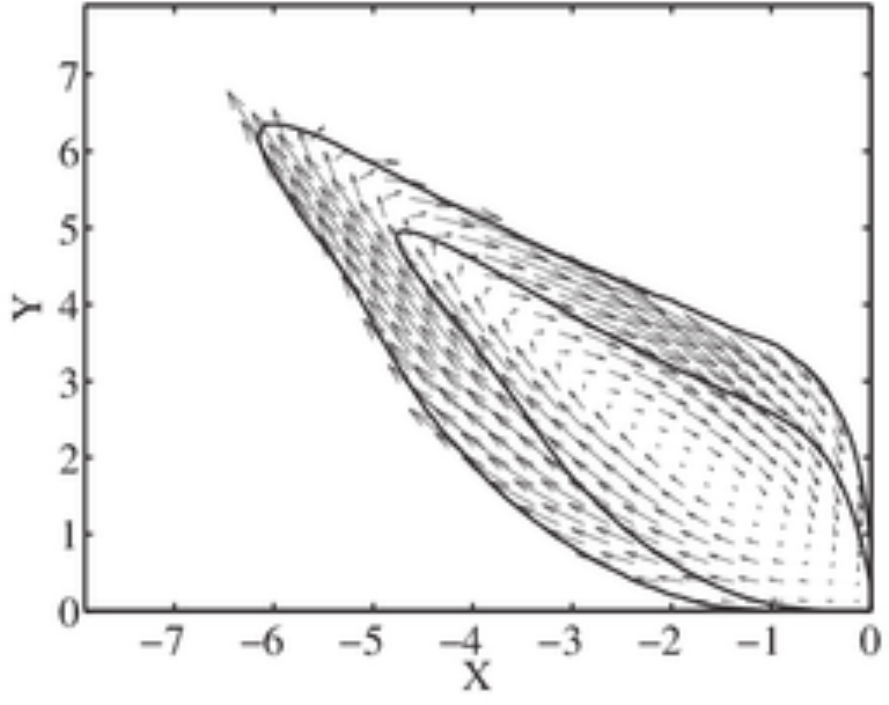}
\mylab{-1.0cm}{4.0cm}{(f)}
}
\centerline{
\psfrag{X}{$R_s$}\psfrag{Y}{$R_\omega$}
\includegraphics[width=0.435\textwidth]{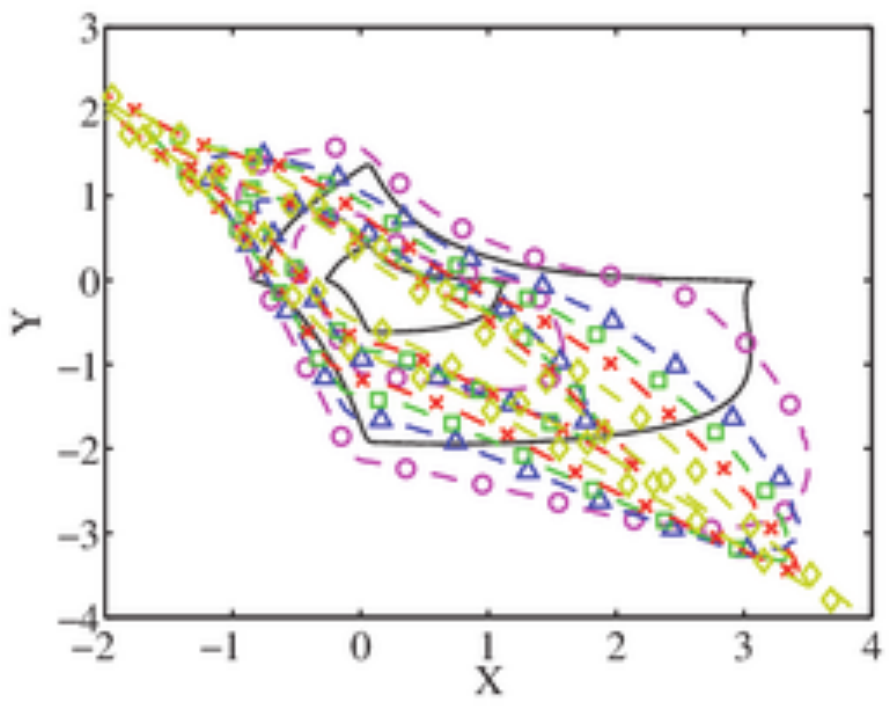}
\mylab{-1.0cm}{4.0cm}{(g)}
\psfrag{X}{$R_s$}\psfrag{Y}{$R_\omega$}
\includegraphics[width=0.435\textwidth]{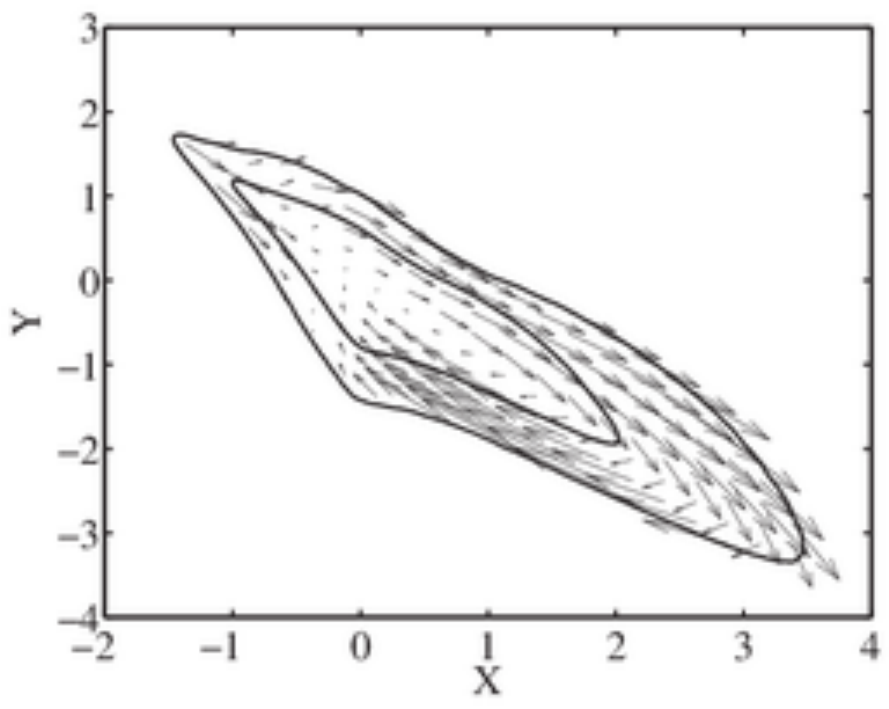}
\mylab{-1.0cm}{4.0cm}{(h)}
}
\caption{ Joint probability density functions of (a) and (b),
  $R_\omega$--$Q_\omega$; (c) and (d), $R_s$--$Q_s$; (e) and (f),
  $Q_s$--$Q_\omega$; (g) and (h), $R_s$--$R_\omega$.  The left column
  includes the conditionally averaged velocity $\boldsymbol{v}$ for
  case F0. Symbols and colors as in table \ref{table:cases}. The right
  column includes the joint p.d.f. and $\boldsymbol{v}$ for F0.25. The
  contours contain 90\% and 98\% of the data. The red dot in (b) is
  $\boldsymbol{v}=\boldsymbol{0}$. See text for details about the
  normalization.
\label{fig:QsQwRsRw}}
\end{figure}
%================================================================
%
%
% Explain the figures
The joint p.d.f.s for several combinations of (\ref{eq:QsQw}) and
(\ref{eq:RsRw}) are plotted in figure \ref{fig:QsQwRsRw}.  Figures
\ref{fig:QsQwRsRw}(a,c,e,g) show the iso-probability contours for all
of the cases in table \ref{table:cases}, and the corresponding
conditionally averaged velocity for the unfiltered case.  For
comparison, figures \ref{fig:QsQwRsRw}(b,d,f,h) show the contours and
velocities for F0.25.  In all of them, the corresponding conditional
velocity $\boldsymbol{v}$ reveals a cyclical behavior of the CMTs that
is consistent with previous literature,
e.g. \cite{Ooi1999,Luethi2009}. Although not shown for the
distributions in figure \ref{fig:QsQwRsRw},
$|\boldsymbol{v}_{std}|/|\boldsymbol{v}|$ attains values similar to
those reported above in the $R$--$Q$ plane (see examples in Appendix
\ref{sec:appendixB}).

One of the most remarkable results is the lack of collapse of the
p.d.f.s for the different filter widths.  The distributions in the
$Q_\omega$--$R_\omega$ and $Q_s$--$R_s$ planes lose their skewed
shape, at least partially compared to the unfiltered case, and become
more symmetric, specially for $\Delta_2\ge0.3h$.  On the contrary, the
enstrophy and strain densities become increasingly anti-correlated as
the filter width grows, and follow the relation,
$Q_s\approx-Q_\omega$. The same result applies to the enstrophy
production and strain self-amplification distributions which exhibit a
strong anti-correlation, $R_s\approx-R_\omega$, although in this case
the trend saturates above $\Delta_2=0.2h$. Note that $Q_s=-Q_\omega$,
i.e. $Q=0$, represents a degenerate flow topology \citep{Chong1990}
that can be associated with pure shear, and that $R_s=-R_\omega$
implies a linear relation between enstrophy production and strain
self-amplification.

% Surprising the dynamics in RQ planes change so little
The results above reveal that a lot of information is hidden in the
$R$--$Q$ plane, and that decomposing the invariants in their strain
and enstrophy contributions offers a more comprehensive view to study
the dynamics of the flow.  The resulting dynamics in the $R$--$Q$
plane are obtained by adding the quantities $Q_s+Q_\omega$ and
$R_s+R_\omega$ which have similar magnitude but opposite signs most of
time, making it difficult to predict the final shape of the $R$--$Q$
iso-contour in figure \ref{fig:pdfRQ}(a) from those in figure
\ref{fig:QsQwRsRw}(a). It is still intriguing how the tear-drop shape
persists at different scales despite the changes undergone by the
strain and enstrophy components of $R$ and $Q$.

\begin{figure}
\vspace{0.5cm}
\centerline{
\psfrag{X}{$\Delta_2/h$}\psfrag{Y}{Ratio}
\includegraphics[width=0.45\textwidth]{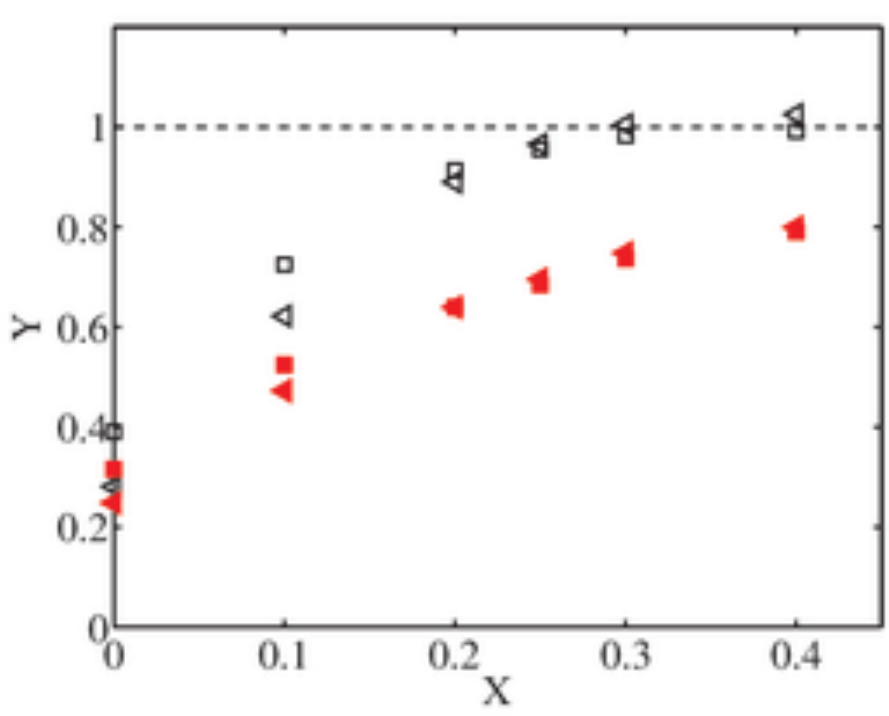}
\mylab{-5.3cm}{4.9cm}{(a)}
\psfrag{X}{$\Delta_2/h$}\psfrag{Y}{skewness}
\includegraphics[width=0.45\textwidth]{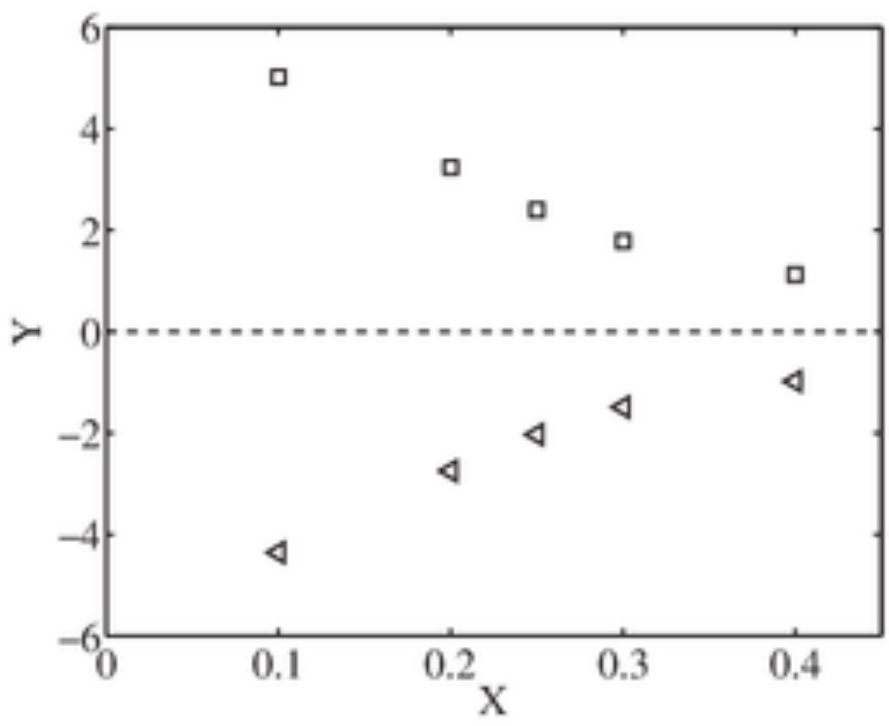}
\mylab{-5.3cm}{4.9cm}{(b)}
}
\caption{ (a) Ratios of the standard deviation of the terms defined by
  (\ref{eq:Q_S}) and (\ref{eq:R_S}), and those in (\ref{eq:QsQw}) and
  (\ref{eq:RsRw}), averaged along the log-layer and as a function of
  the filter width, $\Delta_2$. \squar, $\overline{Q'^S_s/Q'_s}$;
  $\triangleleft$, $\overline{Q'^S_\omega/Q'_\omega}$;
  \textcolor{red}{\solidsquar}, $\overline{R'^S_s/R'_s}$;
  \textcolor{red}{$\blacktriangleleft$}, $\overline{R'^S_\omega/R'_\omega}$. (b)
  Skewness of the probability density functions of \squar,
  $R_s/{Q'_s}^{3/2}$; \trian, $R_\omega/{Q'_\omega}^{3/2}$, as a
  function of the wall-normal filter width, $\Delta_2$. The skewnesses
  for the unfiltered cases are omitted and roughly equal to $\pm40$.
\label{fig:QR_S}}
\end{figure}

% Shear responsible
The lack of collapse in the previous results may be explained taking
into account the increasing contribution of $\partial
\widetilde{u}_1/\partial x_2$ with the filter width. We can write the
relations for $\widetilde{Q}_s$, $\widetilde{Q}_\omega$,
$\widetilde{R}_s$ and $\widetilde{R}_\omega$ in the limiting case in
which the wall-normal derivative of $\widetilde{u}_1$ is the most
important gradient,
\begin{eqnarray}
\label{eq:Q_S}
\widetilde{Q}^S_s &=& -\frac{1}{4} \left( \frac{\partial \widetilde{u}_1}{\partial x_2} \right)^2, \quad
\widetilde{Q}^S_\omega = \frac{1}{4} \left( \frac{\partial \widetilde{u}_1}{\partial x_2} \right)^2, \\
\label{eq:R_S}
\widetilde{R}^S_s &=&
\frac{1}{4} \frac{\partial \widetilde{u}_3}{\partial x_3} \left( \frac{\partial \widetilde{u}_1}{\partial x_2} \right)^2, \quad
\widetilde{R}^S_\omega =
-\frac{1}{4} \frac{\partial \widetilde{u}_3}{\partial x_3} \left( \frac{\partial \widetilde{u}_1}{\partial x_2} \right)^2.
\end{eqnarray}
The superscript $S$ is used to distinguish them from the regular
definitions in (\ref{eq:QsQw}) and (\ref{eq:RsRw}).  Relations
(\ref{eq:Q_S}) and (\ref{eq:R_S}) show that $\widetilde{Q}^S_s =
-\widetilde{Q}^S_\omega$ and $\widetilde{R}^S_s =
-\widetilde{R}^S_\omega$, which is consistent with the trends observed
in figures \ref{fig:QsQwRsRw}(e,g). In order to test whether the
decomposed invariants are dominated by the contribution of $\partial
\widetilde{u}_1/\partial x_2$ as the filter width increases, figure
\ref{fig:QR_S}(a) shows the ratios of the standard deviations,
$Q'^S_s/Q'_s$, $Q'^S_\omega/Q'_\omega$, $R'^S_s/R'_s$ and
$R'^S_\omega/R'_\omega$ averaged in $x_2$ along the log-layer, denoted
by $\overline{(\cdot)}$, and as a function of the filter width. The
results suggest that the dynamics of the eddies are progressively
controlled by $\partial \widetilde{u}_1 / \partial x_2$ as their scale
increases. Note that $\partial \widetilde{u}_1 / \partial x_2$ is the
instantaneous gradient but is related to the mean shear, $S$, by
averaging in the homogeneous directions and in time. This is in
agreement with Corrsin's argument \citep{Corrsin1958} whereby the
dynamics of the eddies with sizes comparable or larger than the
Corrsin scale, $l_C=(\varepsilon/S^3)^{1/2}$, are dominated by the
effect of the mean shear. It is also consistent with $l_C$ being on
average $\approx 0.08h$ in the range considered for the logarithmic
layer, which is well below the filter widths used (see table
\ref{table:cases}). The key role of $S$ in the dynamics of
wall-attached eddies in the logarithmic layer have also been
highlighted in previous works \citep{jim:2013,Lozano2014}.  At this
point, it is interesting to add that the trends shown above are much
weaker if the filter is only performed in the homogeneous but not in
$x_2$, and the reader is referred to Appendix \ref{sec:appendixFxz}
for more details and some examples.

Another conclusion from (\ref{eq:Q_S}) and (\ref{eq:R_S}) is that the
distributions $\widetilde{Q}_s$--$\widetilde{R}_s$ and
$\widetilde{Q}_\omega$--$\widetilde{R}_\omega$ should become mirror
images of each other as the filter width increases, since their
variables may be interchanged as $\widetilde{Q}_s\rightarrow
-\widetilde{Q}_\omega$ and $\widetilde{R}_s\rightarrow
-\widetilde{R}_\omega$. This is clearly visible in figures
\ref{fig:QsQwRsRw}(a,c) for F0.4.  Figure \ref{fig:QR_S}(b) shows that
the skewness of $\widetilde{R}_s$ and $\widetilde{R}_\omega$
decreases, and justifies the increasingly symmetrical shape of the
$\widetilde{Q}_s$--$\widetilde{R}_s$ and
$\widetilde{Q}_\omega$--$\widetilde{R}_\omega$ distributions with the
$\Delta_2$ (figures \ref{fig:QsQwRsRw}(a) and (c) respectively). Exact
zero average enstrophy production is not expected for any filter width,
since averaging (\ref{eq:DQw}) yields to
\begin{equation}\label{eq:DQw_balance}
\langle \widetilde{\omega}_i\widetilde{\omega}_j \widetilde{s}_{ij}  \rangle =
- \langle \nu \widetilde{\omega}_i \frac{\partial^2 \widetilde{\omega}_i}{\partial x_k \partial x_k} \rangle
- \langle \epsilon_{jil} \widetilde{\omega}_l \frac{\partial^2 \tau_{ik}}{\partial x_k \partial x_j} \rangle,
\end{equation}
where $\langle \cdot \rangle$ denotes ensemble average.  Assuming that
the viscous effects are negligible at the inertial scales,
(\ref{eq:DQw_balance}) implies that the average filtered enstrophy
production is not zero but balanced by the interscale transfer of
enstrophy density.
%
%================================================================
% /data4/adrian/Q1Q2R1R2/mfiles/old/plotLinear.m
%================================================================
%

%
%----------------------------------------------------------------%
\begin{table}
    \begin{center}
      \begin{minipage}{10cm}
        \begin{tabular}{lccccc}
          Case  & $\Delta_1/h$ & $\Delta_2/h$ & $\Delta_3/h$ & Lines and symbols & Color   \\[1ex]
          \hline
          S0 (unfiltered)    &      -       &      -       &     -        &   \dashed        & black   \\
          S0.10  &    0.30      &    0.10      &   0.15       &   $-$\circle$-$  & magenta \\
          S0.20  &    0.60      &    0.20      &   0.30       &   $-$\trian$-$   & blue    \\
          S0.25  &    0.75      &    0.25      &   0.38       &   $-$\squar$-$   & green   \\
          S0.30  &    0.90      &    0.30      &   0.45       &   $-*-$          & yellow  \\
          S0.40  &    1.20      &    0.40      &   0.60       &   $-\Diamond-$   & black   \\
       \end{tabular}
     \end{minipage}
    \end{center}
\caption{Summary of cases computed for the velocity fluctuations. The
  parameters $\Delta_1$, $\Delta_2$ and $\Delta_3$ are the filter
  widths in streamwise, wall-normal and spanwise directions,
  respectively.  The fluctuating velocity field is filtered according
  to (\ref{eq:filter}). The cases are denoted by S$\gamma$, where
  $\gamma$ is the wall-normal filter width, $\Delta_2/h$. The symbols
  and colors in the last columns are used to denote the different
  cases in the figures.} \label{table:cases_flu}
\end{table}
%----------------------------------------------------------------%
%
%
%================================================================
% /data4/adrian/Q1Q2R1R2/mfiles/plotQ2Q2R1R2_cte.m
\begin{figure}%[h!tb]
\centerline{
\psfrag{X}{$R$}\psfrag{Y}{$Q$}
\includegraphics[width=0.425\textwidth]{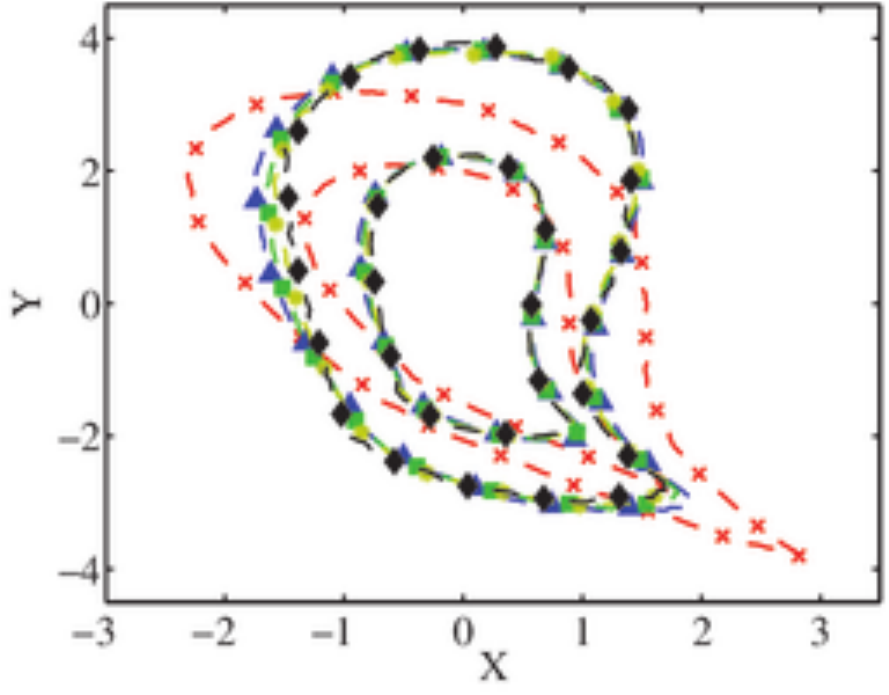}
\mylab{-1cm}{4.0cm}{(a)}
\psfrag{X}{$Q_s$}\psfrag{Y}{$Q_\omega$}
\includegraphics[width=0.425\textwidth]{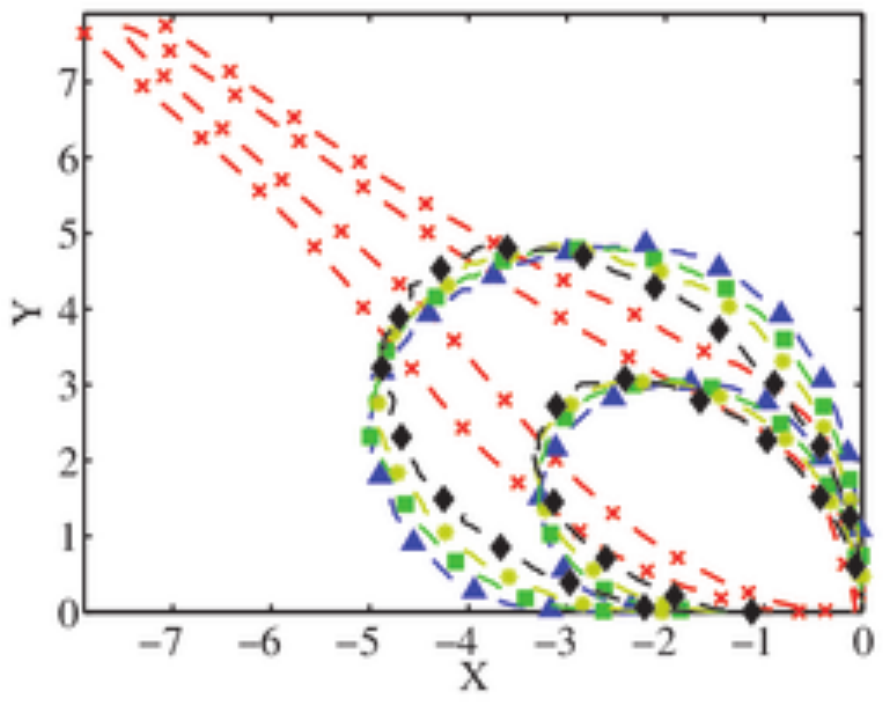}
\mylab{-1.0cm}{4.0cm}{(b)}
}
\centerline{
\psfrag{X}{$R_s$}\psfrag{Y}{$Q_s$}
\includegraphics[width=0.425\textwidth]{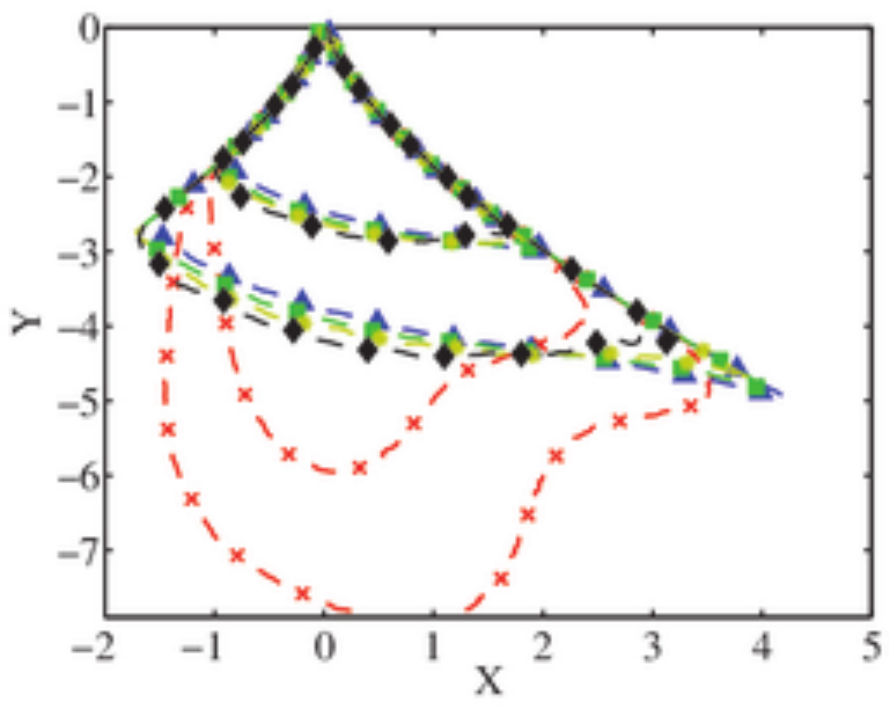}
\mylab{-1.0cm}{4.0cm}{(c)}
\psfrag{X}{$R_\omega$}\psfrag{Y}{$Q_\omega$}
\includegraphics[width=0.425\textwidth]{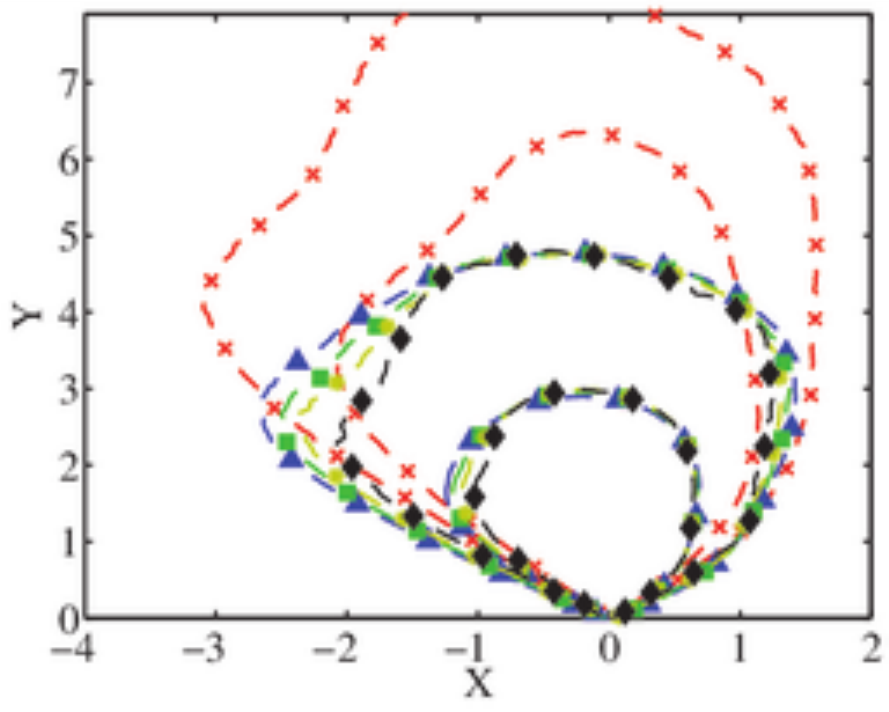}
\mylab{-1.0cm}{4.0cm}{(d)}
}
%%%%%%%%%
\centerline{
\psfrag{X}{$R$}\psfrag{Y}{$Q$}
\includegraphics[width=0.425\textwidth]{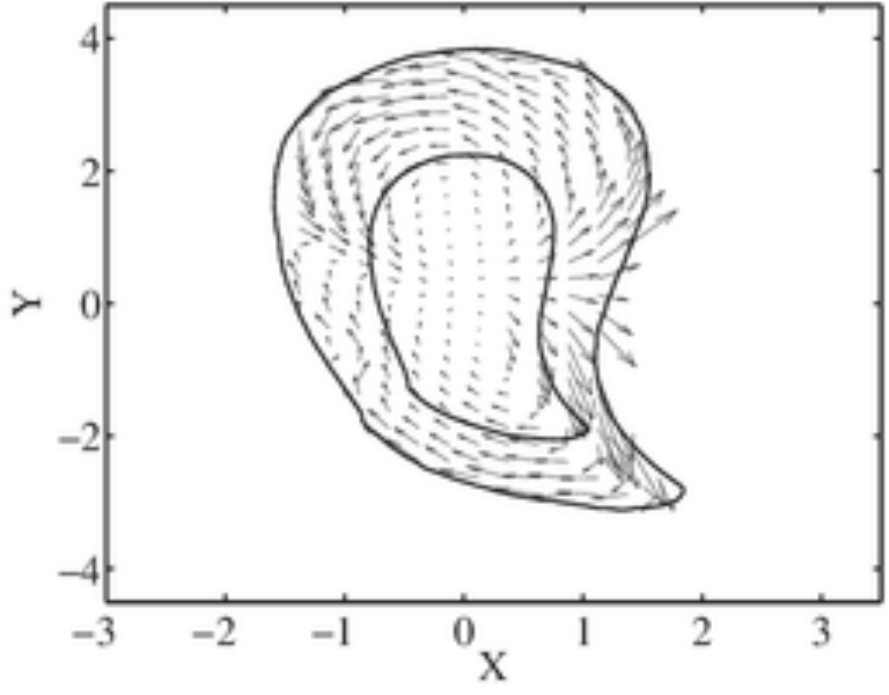}
\mylab{-1cm}{4cm}{(e)}
\psfrag{X}{$Q_s$}\psfrag{Y}{$Q_\omega$}
\includegraphics[width=0.425\textwidth]{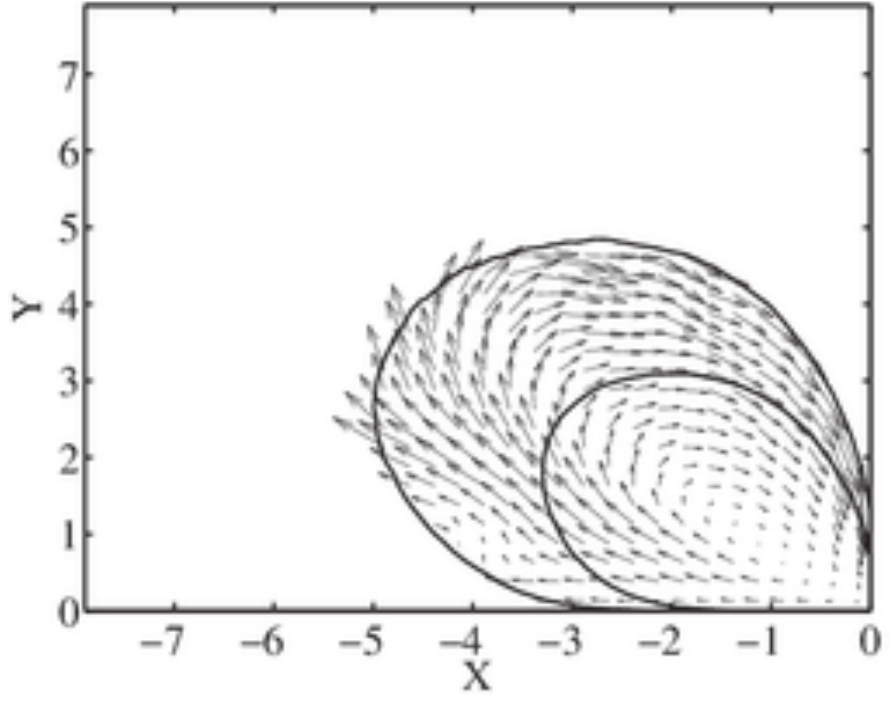}
\mylab{-1.0cm}{4cm}{(f)}
}
\centerline{
\psfrag{X}{$R_s$}\psfrag{Y}{$Q_s$}
\includegraphics[width=0.425\textwidth]{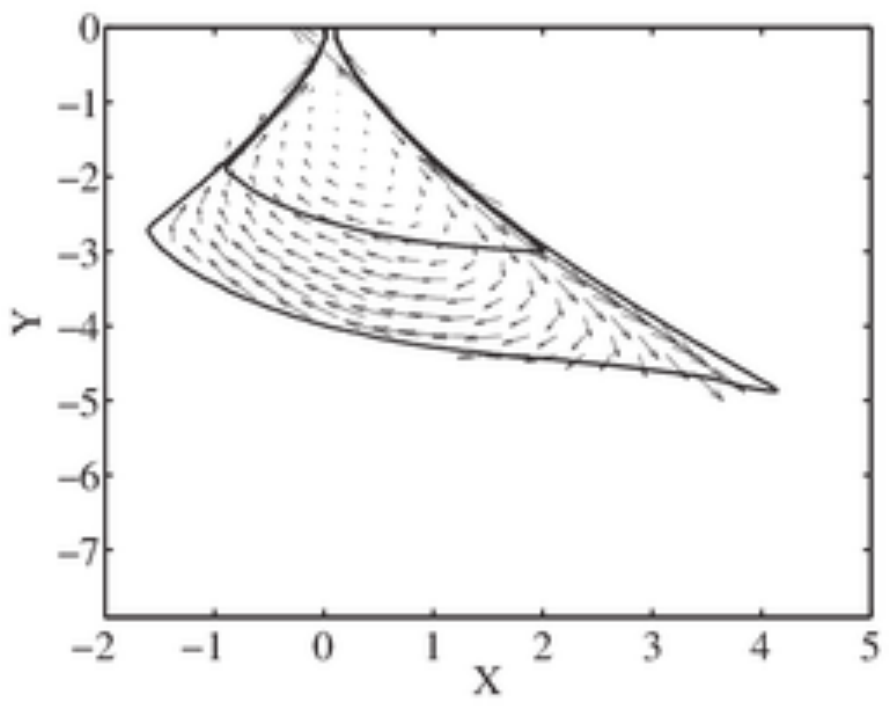}
\mylab{-1.0cm}{4cm}{(g)}
\psfrag{X}{$R_\omega$}\psfrag{Y}{$Q_\omega$}
\includegraphics[width=0.425\textwidth]{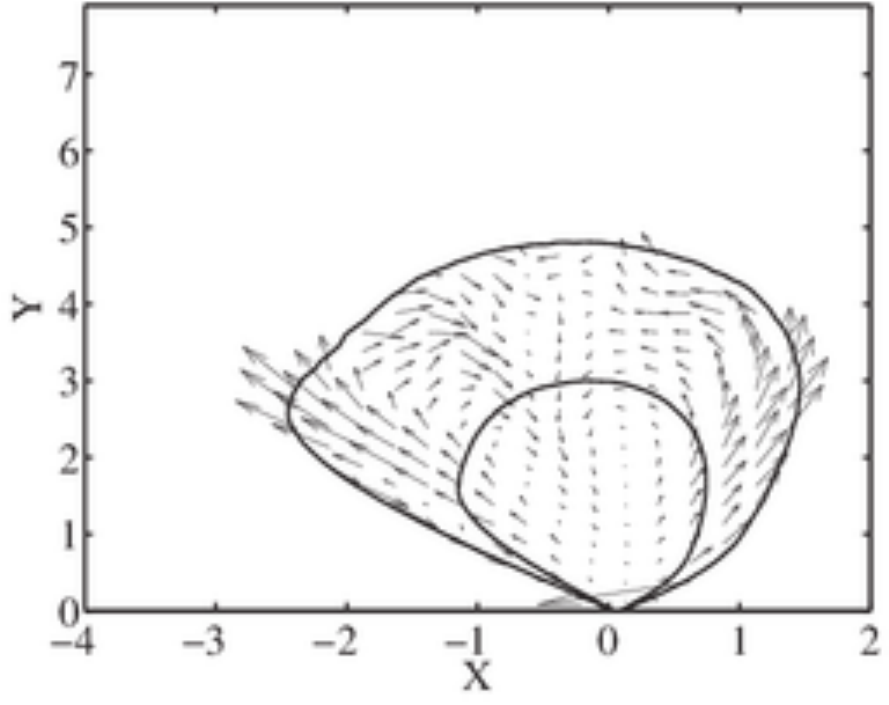}
\mylab{-1.0cm}{4cm}{(h)}
}
\caption{ Results computed for the velocity fluctuations. Joint
  probability density functions of (a), $R$--$Q$; (b),
  $Q_s$--$Q_\omega$; (c), $R_s$--$Q_s$; (d), $R_\omega$--$Q_\omega$.
  The iso-probability contours contain 90\% and 98\% of the
  data. Symbols and colors are as in table \ref{table:cases_flu}. For
  comparison, one more case is included: \textcolor{red}{$\times$},
  case F0.3. (e), (f), (g), and (h), conditionally averaged velocity
  field computed as in (\ref{eq:v_norm}) in the planes (e),
  $R$--$Q$; (f), $Q_s$--$Q_\omega$; (g), $R_s$--$Q_s$; (h),
  $R_\omega$--$Q_\omega$.  The contours are 90\% and 98\% of the data
  of the corresponding p.d.f..  Results for case S0.25. See text for
  details about the normalization.
\label{fig:fluctuations}}
\end{figure}
%================================================================
%
% Results of the fluctuations
The direct effect of the mean shear in the previous results may be
removed by computing the invariants of the velocity fluctuations
instead of those of the total velocity.  Hence, six more cases were
computed using the fluctuating velocities, and their parameters are
summarized in table \ref{table:cases_flu}. The new cases are denoted
by S$\gamma$, where $\gamma$ is the wall-normal filter width,
$\Delta_2/h$, with the same aspect ratios $\Delta_1/\Delta_2$ and
$\Delta_3/\Delta_2$ as in F$\gamma$. Although not shown, the
differences in the probability distributions in figure
\ref{fig:QsQwRsRw} computed with and without the mean shear are
insignificant for the unfiltered case, meaning that the small scales
are barely affected by $S$.

The results for S$\gamma$ are shown in figure
\ref{fig:fluctuations}. One noteworthy difference is the change in the
joint p.d.f.s of $\widetilde{R}$ and $\widetilde{Q}$, which become
more symmetric and lose a large portion of the Vieillefosse tail
(figure \ref{fig:fluctuations}a). The strong correlation between
$\widetilde{Q}_s$ and $\widetilde{Q}_\omega$ is also lost (figure
\ref{fig:fluctuations}b) and the good collapse of the iso-probability
contours reinforces the conclusion that the mean shear is responsible
for the trends observed in figures \ref{fig:QsQwRsRw}. The
distribution of $\widetilde{R}_s$--$\widetilde{R}_\omega$ behaves in a
similar manner (not shown).  The joint p.d.f.s of
$\widetilde{Q}_s$--$\widetilde{R}_s$ and
$\widetilde{Q}_\omega$--$\widetilde{R}_\omega$ become more symmetric
too (figures \ref{fig:fluctuations}c,d) and, contrary to the results
observed for cases F$\gamma$, the contours collapse quite well for
$\Delta_2>0.1h$. Case S0.1 lies in between S0 and S0.2, probably
because it is an intermediate stage between the small and inertial
scales, and was omitted from figure \ref{fig:fluctuations} for the
sake of clarity.

% CMTs for fluctuations
Interestingly, the conditionally averaged velocities for S$\gamma$,
computed from $\mathrm{D}/\mathrm{D}t$ based on the fluctuating
filtered velocity, do not always rotate around one center as in
F$\gamma$ (see figure \ref{fig:QsQwRsRw}). As an example, figures
\ref{fig:fluctuations}(e-h) contain the conditional velocity
$\boldsymbol{v}$ for S0.25, and show that the CMTs in the $R$--$Q$ and
$R_\omega$--$Q_\omega$ planes may be classified into two families
according to their clockwise/counter-clockwise rotation. It is
remarkable that the trajectories in the upper quadrant of the $R$--$Q$
plane now cycle counter-clockwise, in contrast to the result for the
invariants of the total velocity gradient showed in figures
\ref{fig:pdfRQ}(c)-(d).  The difference comes mostly from the behavior
of $R_\omega$ and $Q_\omega$, since $R_s$ and $Q_s$ remain similar to
those observed in F$\gamma$ (cf. figures \ref{fig:QsQwRsRw} and
\ref{fig:fluctuations}). The fluctuating $R$--$Q$ plane may be
explained noting that intense contraction of vorticity ($R_\omega>0$)
is now associated with increasing $Q_\omega$ (figure
\ref{fig:fluctuations}h), which is responsible for the
counter-clockwise part in the first quadrant of figure
\ref{fig:fluctuations}(e). Also, the strongest $Q_\omega$ at negative
$R_\omega$ (vortex stretching) is associated with decreasing
$Q_\omega$, which explains the counter clockwise part of the second
quadrant in figure \ref{fig:fluctuations}(e). This very interesting
behavior comes from the increase of fluctuating enstrophy ($Q_\omega$)
in the presence of contraction of vorticity ($R_\omega>0$), in
contrast to the usual decrease observed for the total velocity. It
also implies that $R_\omega$ is not the dominant term for the budget
of $Q_\omega$ (nor $Q$) in those regions, and that the interactions
between mean and fluctuating gradients at the inertial scales or the
interscale transfer are presumably responsible for these trends.
Double cycles similar to those shown in figures
\ref{fig:fluctuations}(e) and (h) for S0.25 appear in the remaining
filtered cases too (not shown).

% Comparison with HIT
%
%================================================================
% /data4/adrian/Q1Q2R1R2/mfiles/cases/plotQR_onfly_cases.m
\begin{figure}
\vspace{0.5cm}
\centerline{
\psfrag{X}{$R$}\psfrag{Y}{$Q$}
\includegraphics[width=0.42\textwidth]{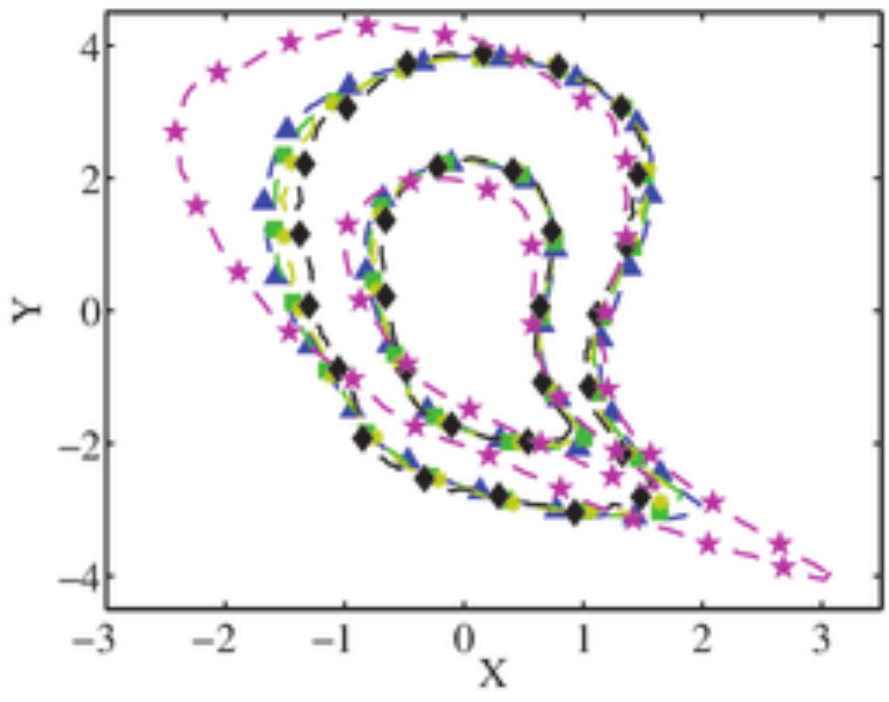}
\mylab{-5.1cm}{4.6cm}{(a)}
\psfrag{X}{$Q_s$}\psfrag{Y}{$Q_\omega$}
\includegraphics[width=0.42\textwidth]{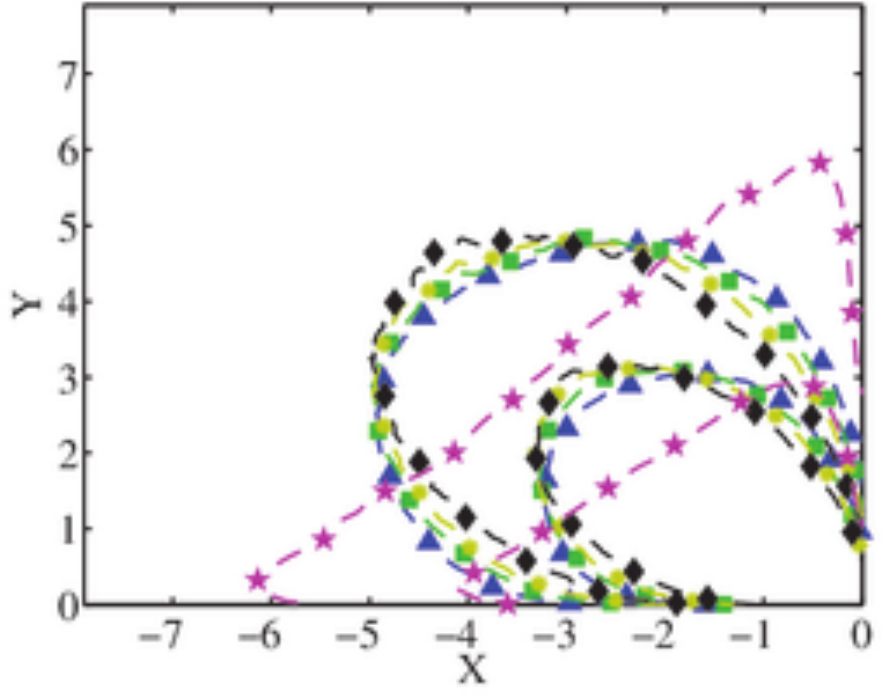}
\mylab{-5.1cm}{4.6cm}{(b)}
}
\vspace{0.5cm}
\centerline{
\psfrag{X}{$R_s$}\psfrag{Y}{$Q_s$}
\includegraphics[width=0.42\textwidth]{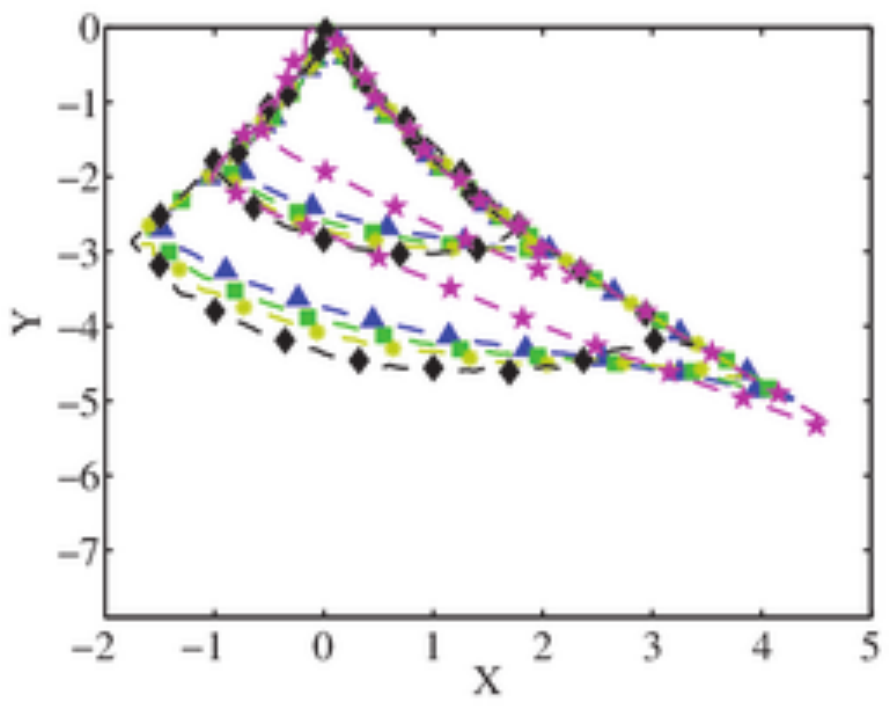}
\mylab{-5.1cm}{4.6cm}{(c)}
\psfrag{X}{$R_\omega$}\psfrag{Y}{$Q_\omega$}
\includegraphics[width=0.42\textwidth]{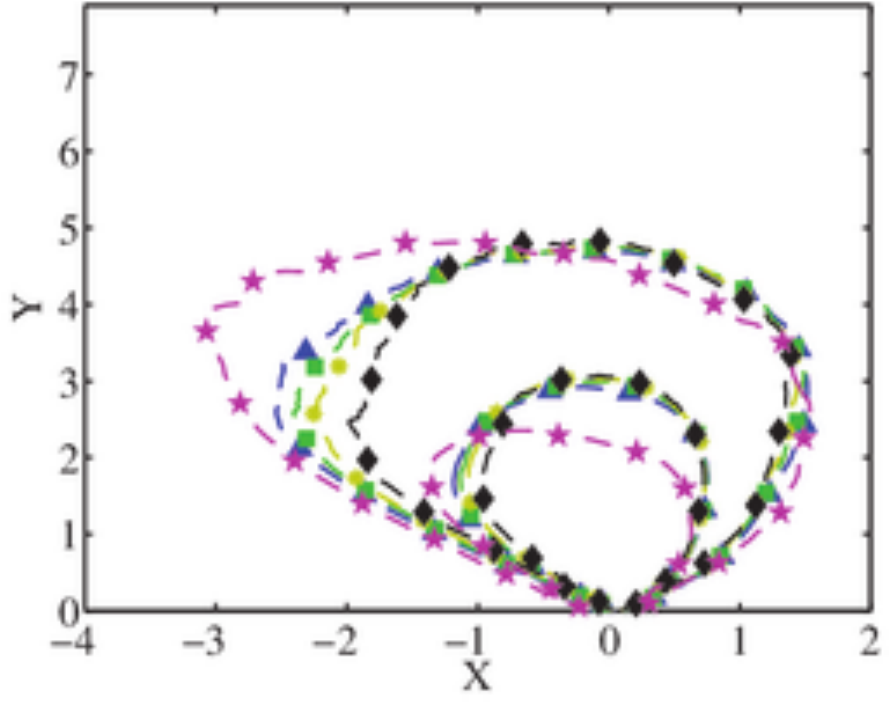}
\mylab{-5.1cm}{4.6cm}{(d)}
}
\caption{ Results computed for the velocity fluctuations. Joint
  probability density functions of (a), $R$--$Q$; (b),
  $Q_s$--$Q_\omega$; (c), $R_s$--$Q_s$; (d), $R_\omega$--$Q_\omega$.
  The iso-probability contours contain 90\% and 98\% of the
  data. Symbols and colors are as in table \ref{table:cases_flu}. For
  comparison, one more case is included,
  \textcolor{magenta}{$\bigstar$}, isotropic turbulence from JHU
  turbulence database \citep{JHU2008} filtered with
  $\Delta_1=\Delta_2=\Delta_3=90\eta$. See text for details about the
  normalization.
\label{fig:fluctuations_HIT}}
\end{figure}
%================================================================
%
Another interesting question is whether the results for the velocity
fluctuations presented above resemble those of isotropic turbulence
(HIT). The idea is tested in figures \ref{fig:fluctuations_HIT}(a-d)
and turns out not to be the case. The data used is HIT from JHU
turbulence database \citep{JHU2008} filtered using (\ref{eq:filter})
with $\Delta_1=\Delta_2=\Delta_3=90\eta$, which is comparable to the
filter widths from table \ref{table:cases_flu}, where
$\Delta_2=0.3h\approx90\eta$.  \citet{Borue1998} reported
distributions consistent with ours from their analysis of HIT using a
top-hat filter. The distributions for HIT are more skewed than those
for the fluctuations in the channel, specially in the $R$--$Q$ and
$Q_s$--$R_s$ maps.  The enstrophy/enstrophy-production p.d.f.  for HIT
is the only one close to the results obtained for cases
S$\gamma$. However, the trends in the $Q_s$--$Q_\omega$ plane are
opposite, and the CMTs from HIT resemble the clockwise cycling of
figure \ref{fig:QsQwRsRw} rather than the ones observed in figures
\ref{fig:fluctuations}(e) and (h). These results suggest that
considering only fluctuating velocities removes the direct effect of
the shear on the dynamics but there still remains the indirect effect,
which is expected since the mean shear and the fluctuations are
coupled by the non-linear dynamics of the Navier--Stokes equations.

%-------------------------------------------------------%
%-------------------------------------------------------%
\subsection{Orbital periods}\label{subsec:periods}
%-------------------------------------------------------%
%-------------------------------------------------------%

\begin{figure}
\vspace{0.5cm}
\centerline{
\psfrag{X}{$r_0$}\psfrag{Y}{$\tau \overline{Q'}^{1/2}$}
\includegraphics[width=0.45\textwidth]{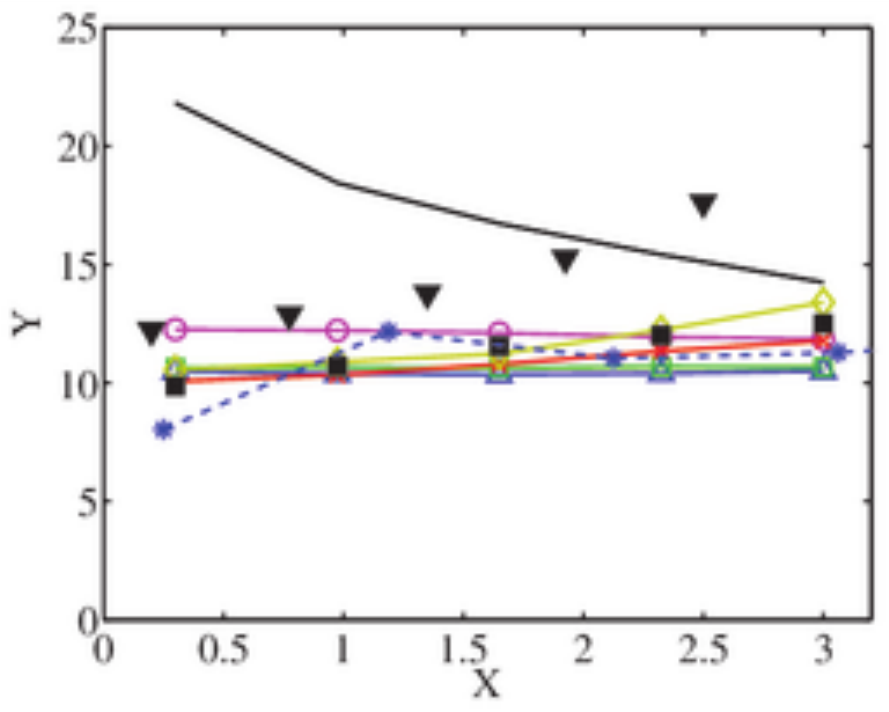}
\mylab{-5.3cm}{4.9cm}{(a)}
\psfrag{X}{$\Delta_2/h$}\psfrag{Y}{$\tau u_\tau/h$}
\includegraphics[width=0.44\textwidth]{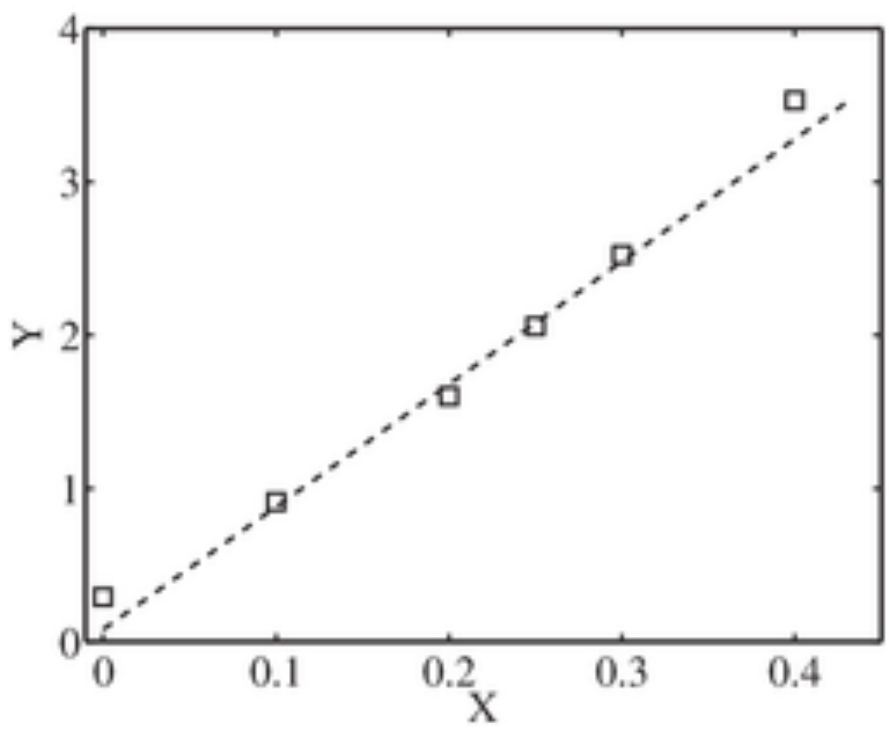}
\mylab{-5.3cm}{4.9cm}{(b)}
}
\caption{ (a) Orbital periods of the CMTs, $\tau$, as a function of
  the distance of their initial condition to the stagnation point of
  $\boldsymbol{v}$ in the corresponding plane, $r_0$.  \solidsquar,
  periods computed in the $R$--$Q$ plane for case F0.2;
  $\blacktriangledown$, periods computed in the $Q_s$--$R_s$ plane for
  case F0.2; \textcolor{blue}{*}, periods computed in the $R$--$Q$
  plane for CMTs in the regions $Q>0$ and case S0.25. The rest of the
  symbols are as in table \ref{table:cases} and correspond to periods
  extracted from the $Q_\omega$--$R_\omega$ plane and filter widths
  $\Delta_2=0,0.1,0.2,0.25,0.3$ and $0.4$.  (b) Orbital periods of the
  CMTs in the $Q_\omega$--$R_\omega$ averaged in $r_0$ as a function
  of the wall-normal filter width, $\Delta_2$.  The dashed line is
  $\tau u_\tau/h = 8 \Delta_2/h + 0.08$.
\label{fig:cmt_T}}
\end{figure}

% CMTs to compute lifetimes
The CMTs are useful to compute orbital periods, i.e., the time $\tau$
elapsed to complete one full revolution around the stagnation point,
$\boldsymbol{v}=\boldsymbol{0}$.  This can be done for all the
distributions in figure \ref{fig:QsQwRsRw}, and the results will be
shown to be of the same order throughout this section. We focus on the
periods in the $Q_\omega$--$R_\omega$ plane and some examples are
discussed for other cases.  Figure \ref{fig:cmt_T}(a) shows the values
of $\tau$ as a function of the initial distance to the stagnation
point, $r_0$ (see figure \ref{fig:QsQwRsRw}b). Many initial conditions
randomly distributed in each plane ($Q_\omega$--$R_\omega$, $R$--$Q$,
etc) were used to compute $\tau$, and its average value is shown in
figure \ref{fig:cmt_T}(a).

The periods collapse reasonable well for all the filter widths when
normalized by $\overline{\widetilde{Q}'}^{1/2}$ (where the bar
represents average along the logarithmic layer) and do not vary much
with $r_0$, i.e., weak and strong events show similar characteristic
time-scales. This is the consequence of the larger velocities sampled
by the CMTs as $r_0$ increases, which compensates for the longer paths
traveled, leading to $\tau\approx$ constant.

For the unfiltered case, $\tau$ decreases $~30$\% by the time $r_0$
has tripled, but it remains always above the filtered cases. The
dependence of $\tau$ with $r_0$ suggest that weak small-scale events
take longer time to complete one dynamic cycle, although the
underlying physical meaning is unclear. On average, $\tau
\overline{Q'}^{1/2} \approx 17$ for the unfiltered case, and $\tau
\overline{\widetilde{Q}'}^{1/2} \approx 10$ for the filtered
ones. Results of the same order but slightly larger are obtained in
the $Q_s$--$R_s$ plane, and one example is included in figure
\ref{fig:cmt_T}(a).

% Lifetimes scaling with Q'
A Kolmogorov-scale normalization, $\tau_{\eta} \sim
\overline{Q'}^{-1/2}$, where $\tau_\eta$ is the Kolmogorov time-scale,
is the natural choice for the unfiltered case since $\tau_{\eta}$ is
the typical decorrelation time-scale during Lagrangian evolution
\citep{Meneveau2011}. The fact that such a normalization works with
$\widetilde{Q}'$ for the filtered cases suggests that the dominating
eddies in the filtered flow are those with characteristic size
$\Delta_2$ and lifetimes proportional to their local eddy-turnover
time $\overline{\widetilde{Q}'}^{-1/2}$.

% More examples
Figure \ref{fig:cmt_T}(a) also includes the orbital periods in the
$R$--$Q$ plane for case F0.2 and the trend is similar to those
computed for other joint distributions. However, our experience shows
that the conditionally averaged velocity on the $R$--$Q$ plane changes
direction very fast close to the Vieillefosse tail, specially for
large filter widths, which poses some numerical issues and makes the
CMTs to follow wrong paths. For instance, it was not possible to
complete a single orbit for case F0.4 for large $r_0$ (figure
\ref{fig:pdfRQ_cmts}d). The $Q_\omega$--$R_\omega$ and $Q_s$--$R_s$
planes are more reliable than the $R$--$Q$ space to compute periods
for large filter widths.

The last period computed corresponds to the CMTs from case S0.25 in
the $R$--$Q$ plane restricted to $Q>0$ (see figure
\ref{fig:fluctuations}e). The results are included in figure
\ref{fig:cmt_T}(a) and remain within the scatter of previous orbital
times. Many other periods can be computed and show different degrees
of agreement with those shown before. These are not discussed here
since we do not pretend to perform an exhaustive analysis of all the
possibilities, but just to remark that they are of the same order.
%
%================================================================
% /data4/adrian/Q1Q2R1R2/mfiles/cases/plotQR_onfly_cases.m
%================================================================
%

% Lifetimes in eddy turnovers
The value $\tau \overline{\widetilde{Q}'}^{1/2} \approx 10$ extracted
from figure \ref{fig:cmt_T}(a) for cases F$\gamma$ implies that the
absolute orbital period increases with $\Delta_2$, since the magnitude
of $\overline{Q'}^{1/2}$ decreases with the filter width.  Figure
\ref{fig:cmt_T}(b) shows that the relation between orbital periods and
filter widths follows approximately the linear trend, $\tau u_\tau/h
\approx 8 \Delta_2/h + 0.08$, when the time is normalized by the
eddy-turnover time, $h/u_\tau$.  The first and last points,
$\Delta_2=0$ and $\Delta_2=0.4$, were excluded from the previous
fitting, the former for being dominated by the viscous effects, and
the latter for exceeding or being at the edge of the usual range
considered for the logarithmic layer \citep{mar:mon:hul:smi:2013}. The
dependence of $\overline{\widetilde{Q}'}^{1/2}$ with $\Delta_2$ can be
estimated analytically for a known velocity gradient spectrum and, for
the range of filter widths considered here, turns out to be almost
linear, which explains the trend in figure \ref{fig:cmt_T}(b). The
constant factor based on the filter width and friction velocity,
$\Delta_2/u_{\tau}$, is a suitable normalization factor for the
orbital period.

%In fact, $\Delta_2=0.4$ is slightly off the trend, which could be
%indicative of the end of the self-similar eddies in the logarithmic
%layer or of the decrease of their influence in favor of larger
%non-self-similar scales. Other fittings are also reasonable, even if
%$\Delta_2=0.4h$ is included, at the expense of higher error in the
%fit.

% Compare lifetimes with other people
To close this section we compare our periods with those in the
literature.  The orbital periods discussed above for the unfiltered
case, $\tau \overline{Q'}^{1/2} \approx 17$, correspond to
$\tau^+=280$ and $\tau/\tau_\eta=26$ in wall and Kolmogorov units,
respectively, with $\tau_\eta = \overline{ (\nu/\varepsilon)^{1/2}
}$. The value obtained by \citet{Atkinson2012} in a turbulent boundary
layer flow at a comparable Reynolds number is $\tau^+=658$, which
exceeds ours, although in their case the periods are computed for the
logarithmic and wake region. \citet{Elsinga2010} reported a smaller
value, $\tau^+=470$, which is still larger than ours, although they
used experimental data filtered over $50$ wall units in each
direction.  For isotropic turbulence, \citet{Luethi2009},
\citet{Martin1998} and \citet{Ooi1999} obtained
$\tau/\tau_{\eta}\approx 40$ and $\tau/\tau_{\eta}\approx 30$,
respectively, which are not so far from the value of $26$ obtained
here. Nevertheless, some differences are expected since we compute the
Lagrangian time derivative of the normalized invariants, as in
(\ref{eq:v_norm}), and not of the invariants themselves, as it is the
case of previous works. It is difficult to find in the literature
orbital periods for the filtered invariants, although values of the
same order to those shown in figure \ref{fig:cmt_T}(b) were obtained
for the bursting periods, $T_b$, of minimal log-layer channels by
\citet{Flores2010}, $T_b u_\tau/h\approx 6 x_2/h$, if we take
$\Delta_2=x_2$. However, it is not simple to establish a link between
orbital and bursting periods, and it is unclear whether they are
related or not.  A linear relation between lifetime and scale has also
been observed in previous works \citep{DelAlamo2006, Lozano2014,
  leh:gua:mck:2013}, and is explained in the context of self-similar
log-layer eddies with lifetimes proportional to their
size. \citet{Lozano2014} found $T u_\tau/h \approx\Delta_2/h$, with
$T$ the lifetimes of wall-attached eddies, and $\Delta_2$ their size,
that will be considered equivalent to the filter width. However, these
lifetimes are much shorter than the orbital periods reported here,
$\tau u_\tau/h\approx 8 \Delta_2/h + 0.08$. This discrepancy may be
related to the ratio $|\boldsymbol{v}_{std}|/|\boldsymbol{v}|$ shown
in figure \ref{fig:pdfRQ}(b).  The average
$|\boldsymbol{v}_{std}|/|\boldsymbol{v}|$ along a CMT is $6$--$12$ for
all filter widths, and its inverse value can be interpreted as the
fraction of time the fluid particles travel in the `correct' direction
to complete a cycle instead of drifting in other directions. This may
responsible for the factor of 8 between orbital periods and lifetimes
of individual eddies mentioned above.

%-------------------------------------------------------%
%-------------------------------------------------------%
\subsection{Alignment of the vorticity and the rate-of-strain tensor}\label{subsec:alignment}
%-------------------------------------------------------%
%-------------------------------------------------------%

% Strain-vorticity alignment
In order to gain a better insight into the dynamics at different
scales, we analyze the alignment of the vorticity vector,
$\boldsymbol{\omega}$, with the eigenvectors of the rate-of-strain
tensor, $ \boldsymbol{ \lambda_1}$, $ \boldsymbol{\lambda_2}$, and
$\boldsymbol{\lambda_3}$, whose associated eigenvalues are
$\lambda_1>\lambda_2>\lambda_3$ (note that in this case the subindex
refers to the sorting of the eigenvalues, in contrast to the
convection used for other quantities, $s_{ij}$, $u_{i}$,... where the
index denotes the spatial direction $x_i$). This alignment is of
interest since the enstrophy production may be expressed as
\citep{bet:1956}
\begin{equation}\label{eq:p1p2p3}
\omega_i \omega_j s_{ij} =
\omega^2 \left( p_1 + p_2 + p_3\right),
\end{equation}
where $p_i= \lambda_i \cos^2(\boldsymbol{\omega},
\boldsymbol{\lambda_i})$ and $\omega$ is the $L^2$-norm of
$\boldsymbol{\omega}$. An equivalent expression applies to the
filtered cases.

\begin{figure}
\vspace{0.5cm}
\centerline{
\psfrag{X}{ \raisebox{-0.2cm}{$|\cos( \boldsymbol{\omega}, \boldsymbol{\lambda_2})|$} }\psfrag{Y}{p.d.f.s}
\includegraphics[width=0.45\textwidth]{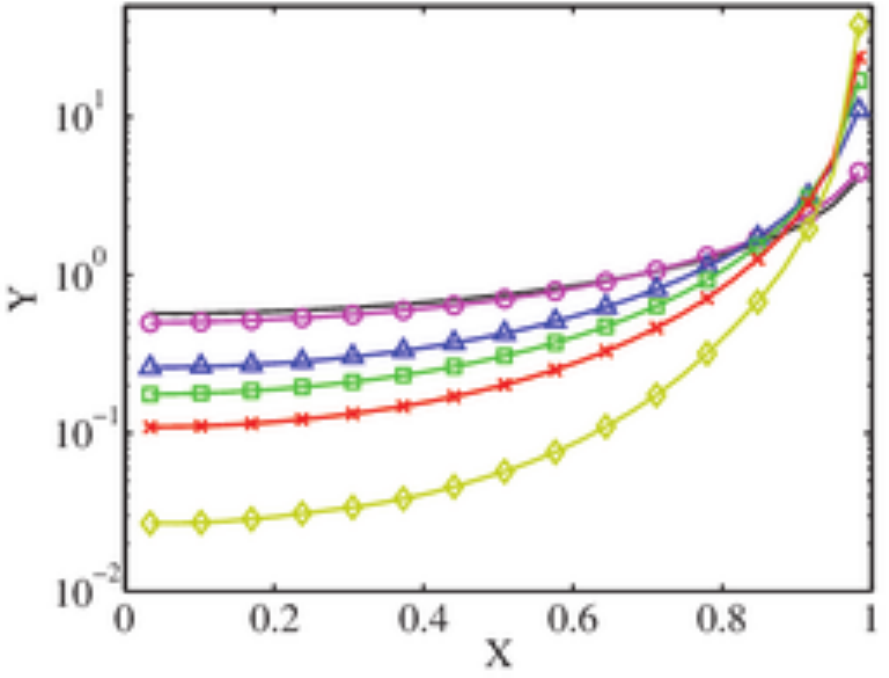}
\mylab{-5.3cm}{4.9cm}{(a)}
\psfrag{X}{ \raisebox{-0.2cm}{$\lambda_{1,3}/{Q'_s}^{1/2}$} }\psfrag{Y}{p.d.f.s}
\includegraphics[width=0.45\textwidth]{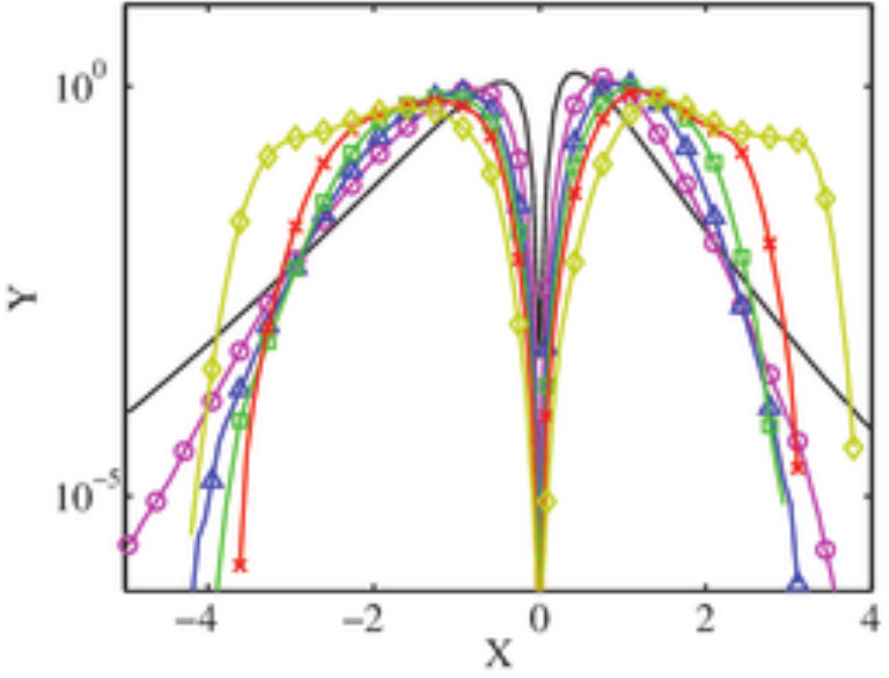}
\mylab{-5.3cm}{4.9cm}{(b)}
}
\vspace{0.5cm}
\centerline{
\psfrag{X}{ \raisebox{-0.2cm}{$\lambda_2/{Q'_s}^{1/2}$} }\psfrag{Y}{p.d.f.s}
\includegraphics[width=0.45\textwidth]{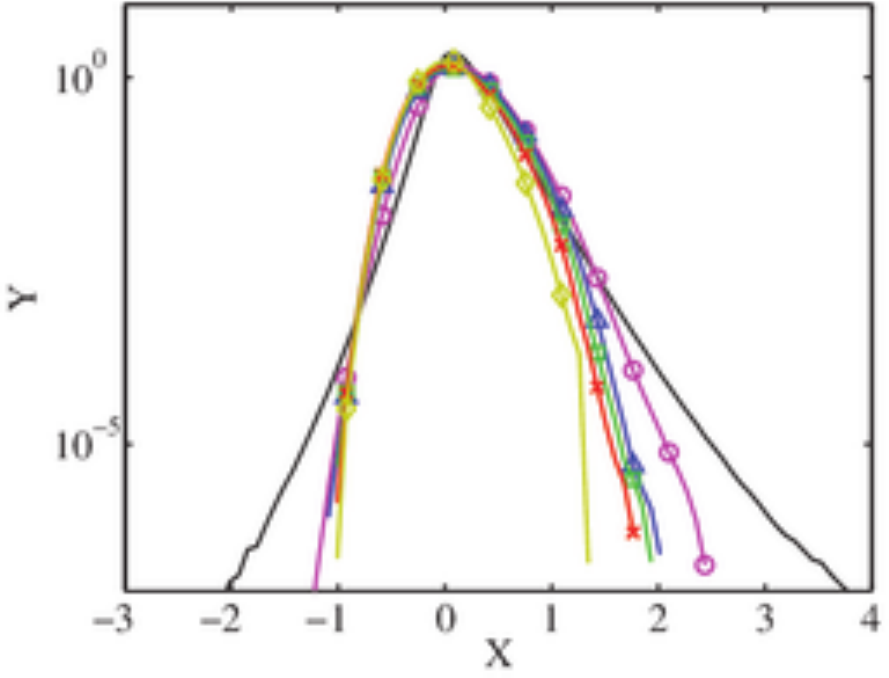}
\mylab{-5.3cm}{4.9cm}{(c)}
\psfrag{X}{ \raisebox{-0.2cm}{$\Delta_2/h$} }\psfrag{Y}{$p'_i/p'_1$}\psfrag{Z}{$\langle p_i \rangle/\langle p_1 \rangle$}
\includegraphics[width=0.43\textwidth]{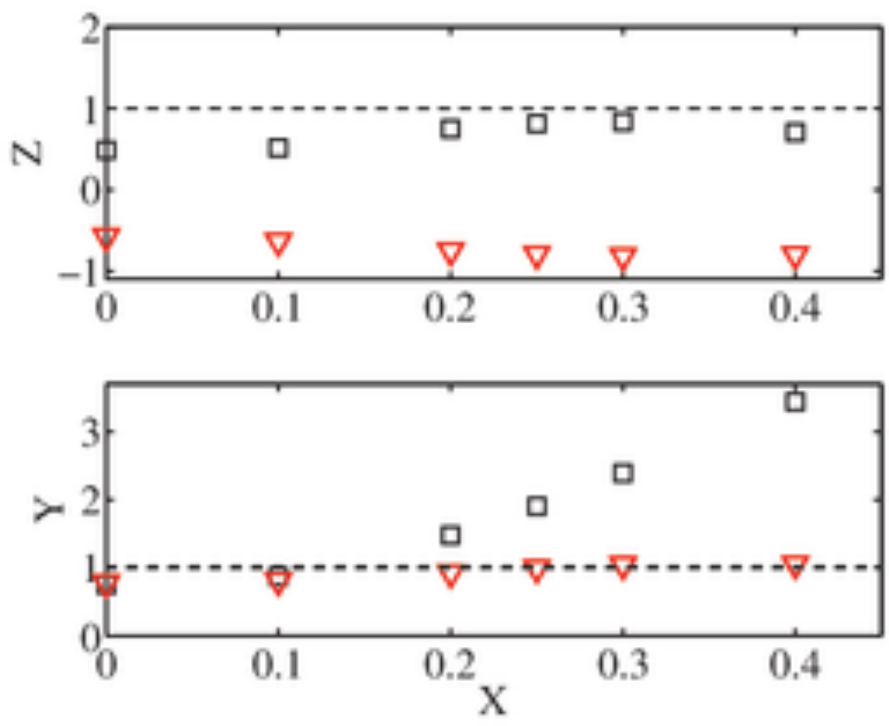}
\mylab{-5.3cm}{4.9cm}{(d)}
}
\caption{ Results computed for the total velocities. (a) Probability
  density functions of the cosine of the angle of the vorticity,
  $\boldsymbol{\omega}$, and the second eigenvector of the
  rate-of-strain tensor, $\boldsymbol{\lambda_2}$.  (b) and (c) are
  the probability density functions of the eigenvalues of the
  rate-of-strain tensor, (b), $\lambda_1$ and $\lambda_3$; (c),
  $\lambda_2$. For (a), (b) and (c) the symbols and colors are as in
  table \ref{table:cases}. (d) Relative contributions to the enstrophy
  production of $p_1 = \lambda_1 \cos^2( \boldsymbol{\omega},
  \boldsymbol{\lambda_1})$, $p_2 = \lambda_2 \cos^2(
  \boldsymbol{\omega},\boldsymbol{\lambda_2})$ and $p_3 = \lambda_3
  \cos^2( \boldsymbol{\omega},\boldsymbol{\lambda_3})$ as a function
  of the wall-normal filter width, $\Delta_2$. The top panel shows the
  ratio of the means, \squar, $\langle p_2 \rangle/\langle p_1
  \rangle$; \trian, $\langle p_3 \rangle/\langle p_1 \rangle$.  The
  bottom panel shows the ratio of the standard deviations, \squar,
  $p'_2/p'_1$; \dtrian, $p'_3/p'_1$.
\label{fig:align_c123}}
\end{figure}

% Alignment
Since the trace of $s_{ij}$ must be zero owing to incompressibility,
it is satisfied that $\lambda_1>0$ and $\lambda_3<0$.  The former
eigenvalue represents stretching in the direction of
$\boldsymbol{\lambda_1}$, whereas the latter is a contraction of
vorticity along $\boldsymbol{\lambda_3}$. The second eigenvalue,
$\lambda_2$, takes both positive and negative values and it is
well-known that, for the small scales, $\boldsymbol{\omega}$ aligns
preferentially with $\boldsymbol{\lambda_2}$ in isotropic turbulence
\citep{Ashurst1987}, free shear flows \citep{mul:2006} and turbulent
channels \citep{Blackburn1996}. Similar results hold for the filtered
cases as shown in figure \ref{fig:align_c123}(a). However, the
alignment of $\boldsymbol{\omega}$ and $\boldsymbol{\lambda_2}$
intensifies for larger scales and, as a result, the angles of
$\boldsymbol{\lambda_1}$ and $\boldsymbol{\lambda_3}$ with
$\boldsymbol{\omega}$ tend to zero, i.e., perpendicular to
$\boldsymbol{\omega}$ (not shown).
%
%================================================================
% /data4/adrian/Q1Q2R1R2/mfiles/cases/plotQR_onfly_cases.m
%================================================================
%

% lambda pdfs
Figures \ref{fig:align_c123}(b,c) show the p.d.f.s of $\lambda_i$
scaled by the corresponding ${Q'_s}^{1/2}$ of each case.  The
distribution for the unfiltered $\lambda_2$ is skewed towards positive
values as already reported in previous works \citep{Ashurst1987,
  Vincent1991, Blackburn1996}. The p.d.f.s of $\lambda_1$ do not
collapse for the different filter widths, nor they do for $\lambda_2$
and $\lambda_3$, respectively, although the distributions become more
symmetric as $\Delta_2$ increases, consistent with the results for the
skewness of $R_\omega$ in figure \ref{fig:QR_S}(b).

% Explanation
The preferential alignment of $\boldsymbol{\omega}$ with
$\boldsymbol{\lambda_2}$ is predicted by angular momentum conservation
in the Restricted Euler Model \citep{Cantwell1992}, and was explained
from a kinematic point of view in \citet{jim:1992}.  Such an
alignment, or at least part of it, may also be expected in the context
of eddies controlled by the mean shear as proposed in section
\ref{subsec:QsQwRsRw}. In fact, in the very simple scenario in which
$\partial u_1/\partial x_2$ is the most important term, the flow
behaves like a pure shear, and $\boldsymbol{\omega}$ and
$\boldsymbol{\lambda_2}$ align \citep{Tennekes1972}. In this case, the
associated $\lambda_2$ tends to zero, which is a consequence of the
over-simplifications made and could be solved by retaining higher
order terms.

% Contribution to enstrophy production
The preferential alignment of $\boldsymbol{\omega}$ with
$\boldsymbol{\lambda_2}$ does not imply that most of the contribution
to the enstrophy production is due to $\omega^2 p_2$ \citep{Tsi:1997}.
Figure \ref{fig:align_c123}(d) shows the relative importance of $p_1$,
$p_2$ and $p_3$ in terms of its mean $\langle\cdot\rangle$ (top
panel), and its standard deviation $(\cdot)'$ (bottom panel), for
different filter widths. For the unfiltered case, the contributions of
$\langle p_2\rangle$ and $\langle p_3\rangle$ to the mean enstrophy
production/destruction are, in magnitude, roughly one half of the
contribution of $\langle p_1\rangle$, which turns out to be the most
important term \citep{Vincent1994,Tsinober1998}. When filtering, the
ratios $\langle p_2\rangle/\langle p_1\rangle$ and $\langle
p_3\rangle/\langle p_1\rangle$ approach to $\pm 1$, respectively.
This implies that the small-scale vortices are mostly dominated by
vortex stretching, but large-scale vorticity is equally influenced by
the three terms from (\ref{eq:p1p2p3}). From a geometric point of
view, the results above suggest that the tube-like structures are
favored at the small scales but that sheet-like objects dominate at
larger ones as in the geometrical analysis of turbulent structures in
\citet{Moisy2004}.  The scale-dependence of the ratios is even more
pronounced for the standard deviations of $p_i$, and $p'_2/p'_1$
increases steadily, attaining values up to three times those of
$p'_3/p'_1$.
%This is caused by the slower decrease of $p'_2$ with $\Delta_2$ (not
%shown).

%
% note: unfiltered case (<lambda_1>/<lambda_2>=4,1,<lambda_3>/<lambda_2>)=(4,1,-5)
%                       vs. isotropic turbulence (3,1,-4)

%
%================================================================
% /data4/adrian/Q1Q2R1R2/mfiles/cases/plotQR_onfly_cases.m
\begin{figure}
\vspace{0.5cm}
\centerline{
\psfrag{X}{ \raisebox{-0.2cm}{$|\cos( \boldsymbol{\omega}, \boldsymbol{\lambda_2})|$} }\psfrag{Y}{p.d.f.s}
\includegraphics[width=0.45\textwidth]{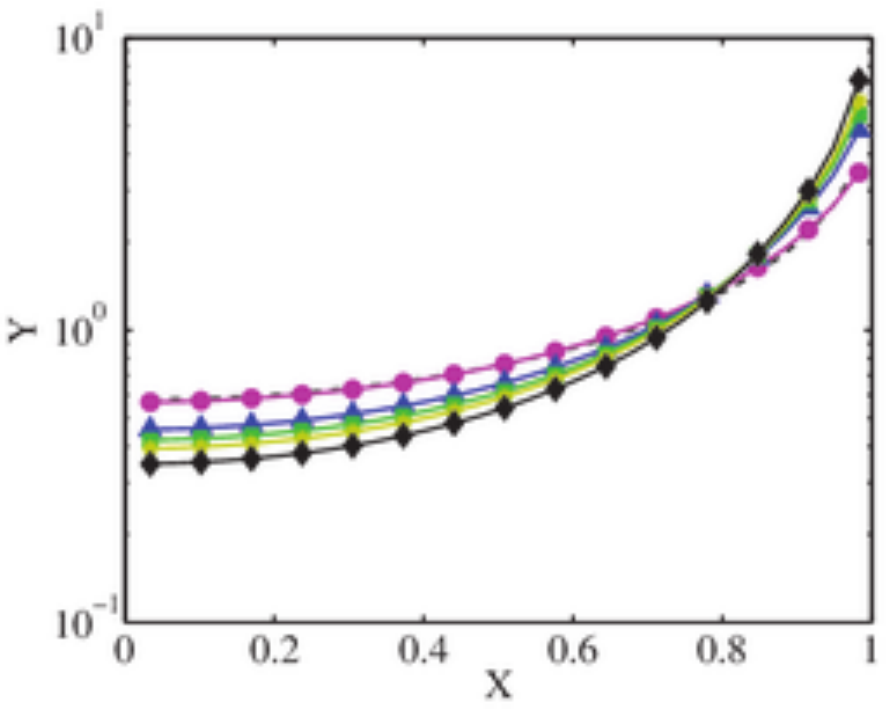}
\mylab{-5.3cm}{4.9cm}{(a)}
\psfrag{X}{ \raisebox{-0.2cm}{$\lambda_{1,3}/{Q'_{s}}^{1/2}$} }\psfrag{Y}{p.d.f.s}
\includegraphics[width=0.45\textwidth]{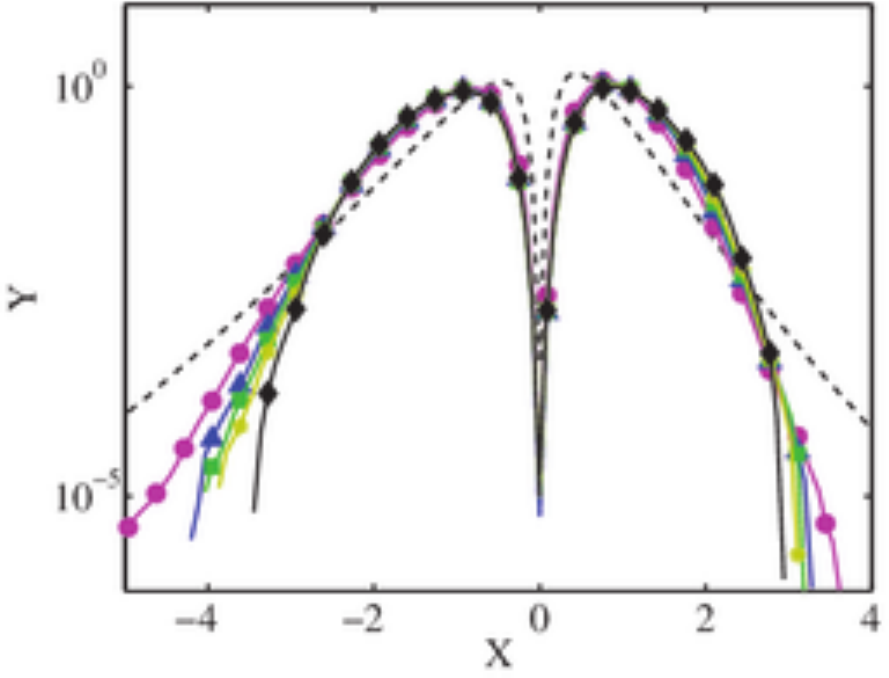}
\mylab{-5.3cm}{4.9cm}{(b)}
}
\vspace{0.5cm}
\centerline{
\psfrag{X}{ \raisebox{-0.2cm}{$\lambda_{2}/{Q'_{s}}^{1/2}$} }\psfrag{Y}{p.d.f.s}
\includegraphics[width=0.45\textwidth]{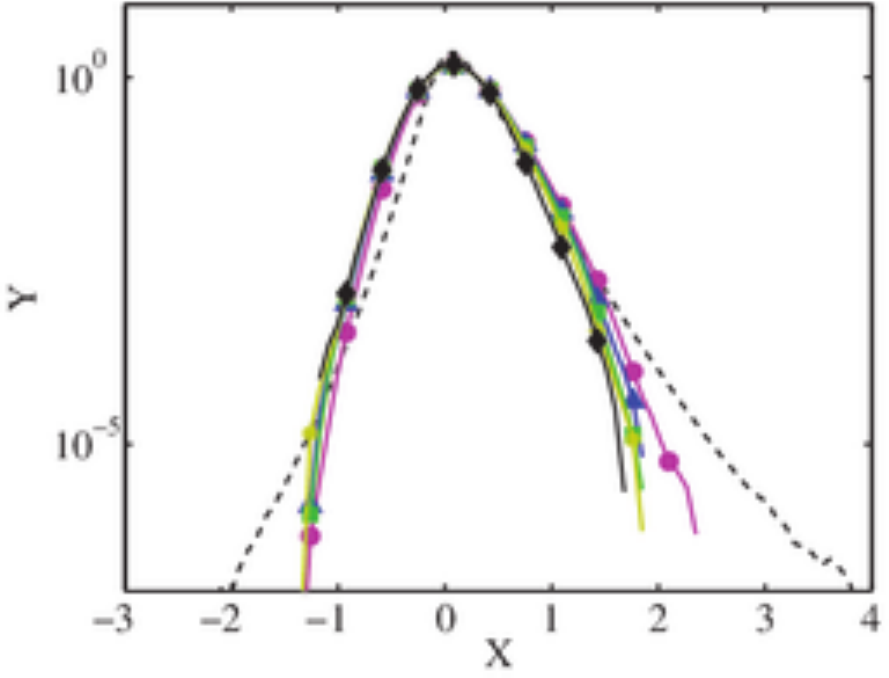}
\mylab{-5.3cm}{4.9cm}{(c)}
\psfrag{X}{ \raisebox{-0.2cm}{$\Delta_2/h$} }\psfrag{Y}{$p'_{i}/p'_{1}$}\psfrag{Z}{$\langle p_{i} \rangle/\langle p_{1} \rangle$}
\includegraphics[width=0.43\textwidth]{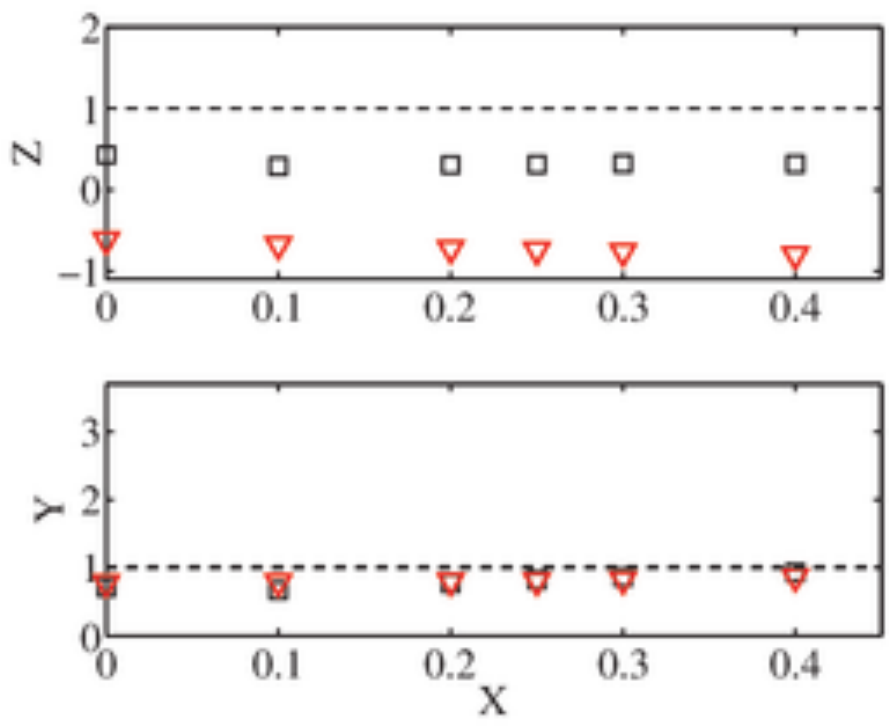}
\mylab{-5.3cm}{4.9cm}{(d)}
}
\caption{ Same as figure \ref{fig:align_c123_flu} but for the
  fluctuating velocities (cases S$\gamma$ table
  \ref{table:cases_flu}). For (a), (b) and (c) the symbols and colors
  are as in table \ref{table:cases_flu}.
\label{fig:align_c123_flu}
}
\end{figure}
%================================================================
%
% Fluctuations
The results above suggest that the dynamics of the flow differ at
different scales. However, it was noticed in section
\ref{subsec:QsQwRsRw} that this may be caused by the effect of the
mean shear. For that reason, the calculations were repeated for cases
S$\gamma$, where only the velocity fluctuations are considered, and
the results are shown in figure \ref{fig:align_c123_flu}. The
fluctuating vorticity also aligns predominantly with the second
eigenvector of the fluctuating rate-of-strain tensor (figure
\ref{fig:align_c123_flu}a), implying that the alignment shown in
figure \ref{fig:align_c123}(a) is not entirely caused by the effect of
the mean shear, and the distributions of the three eigenvalues become
more symmetric, specially for $\Delta_2>0.1h$ (figures
\ref{fig:align_c123_flu}b,c). The main difference compared to the
results computed for the total velocities is the improved collapse of
the p.d.f.s for both the alignments and eigenvalues. Figure
\ref{fig:align_c123_flu}(d) is equivalent to figure
\ref{fig:align_c123}(d), but the ratios remain roughly constant and
independent from the filter width, suggesting that vortices defined
through the fluctuating velocities are better candidates than the
total vorticity to study the multiscale dynamics of the flow.  This
scale-independent behavior also suggest that these large-scale
vortices are geometrically self-similar as opposed to those described
above for the total velocities, although that should be tested in more
detail with a geometrical analysis of the structures that is out of
the scope of the present paper. From the results above, we can
conclude that once the direct effect of the mean shear is removed, the
multiscale dynamics of the enstrophy production become roughly
self-similar, consistent with the results shown before for the joint
distributions of $Q_s$, $Q_\omega$, $R_s$ and $R_\omega$ associated
with the fluctuating velocity.

%-------------------------------------------------------%
%-------------------------------------------------------%
\subsection{The energy cascade in terms of vortex stretching}\label{subsec:alignment_inter}
%-------------------------------------------------------%
%-------------------------------------------------------%

% Analysis between different scales and motivation
In the previous section we have studied the alignment of the vorticity
and the eigenvectors of the rate-of-strain tensor at the same
scale. Here we analyze the alignment of vorticity and strain at
different ones, i.e., the vorticity associated with the filtered
velocity at scale $\Delta^\omega$, whose quantities will be denoted by
$\widetilde{(\cdot)}$, and the rate-of-strain associated with the
filtered velocity at a scale $\Delta^s$, represented by
$\widehat{(\cdot)}$. The study is motivated by the classical energy
cascade in terms of vortex stretching, where the strain at a given
scale stretches the vortices at a smaller one and induces higher
velocities by the conservation of angular momentum.  This scenario
provides a mechanism for the interscale energy transfer required by
the energy cascade, but is presumably in contradiction with previous
studies \citep{Ashurst1987,She1991,Vincent1994} and with section
\ref{subsec:alignment}, where the vorticity aligns most probably with
the intermediate strain eigenvector. This may be caused by the fact
that $\boldsymbol{\omega}$ and $s_{ij}$ are both studied at the same
scale, and it has been noted before that the alignment of vorticity
with the intermediate strain eigenvector decreases when the local
strain induced by the vortices is eliminated
\citep{jim:1992,Hamlington2008}. This suggests that such an alignment
could change for vorticity fields and strain tensors calculated each
at a different scale.

%================================================================
% /data4/adrian/Q1Q2R1R2/mfiles/cases/plot_Omega_S_scales.m
\begin{figure}
\vspace{0.5cm}
\centerline{
\psfrag{X}{ \raisebox{-0.2cm}{$|\cos(\widetilde{\boldsymbol{\omega}},\widehat{\boldsymbol{\lambda}}_i)|$} } \psfrag{Y}{p.d.f.s}
\includegraphics[width=0.45\textwidth]{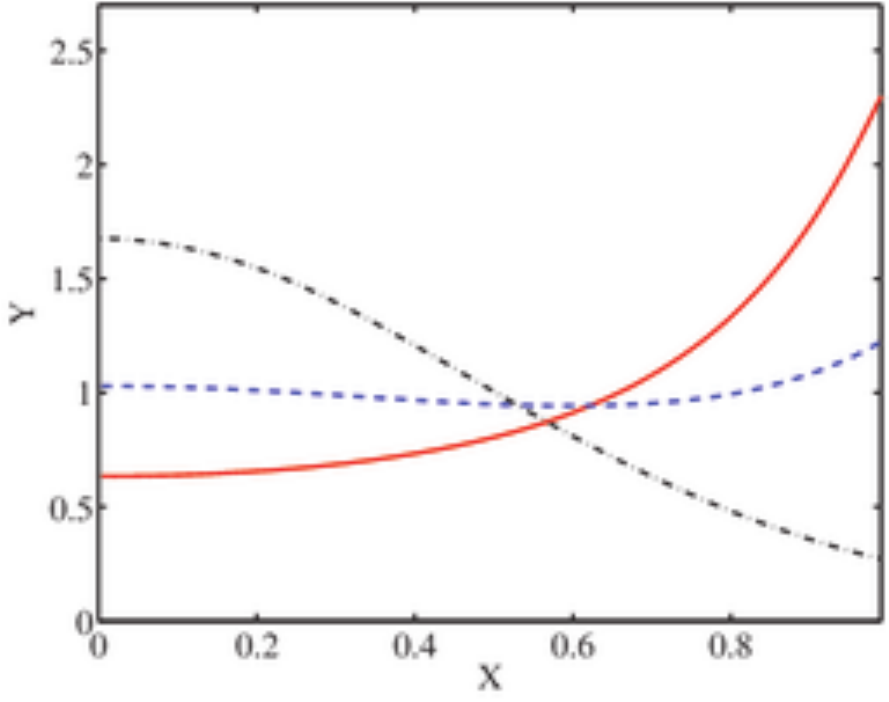}
\mylab{-5.3cm}{5.cm}{(a)}
\psfrag{X}{ \raisebox{-0.2cm}{$|\cos(\widetilde{\boldsymbol{\omega}},\widehat{\boldsymbol{\lambda}}_i)|$} } \psfrag{Y}{ }
\includegraphics[width=0.45\textwidth]{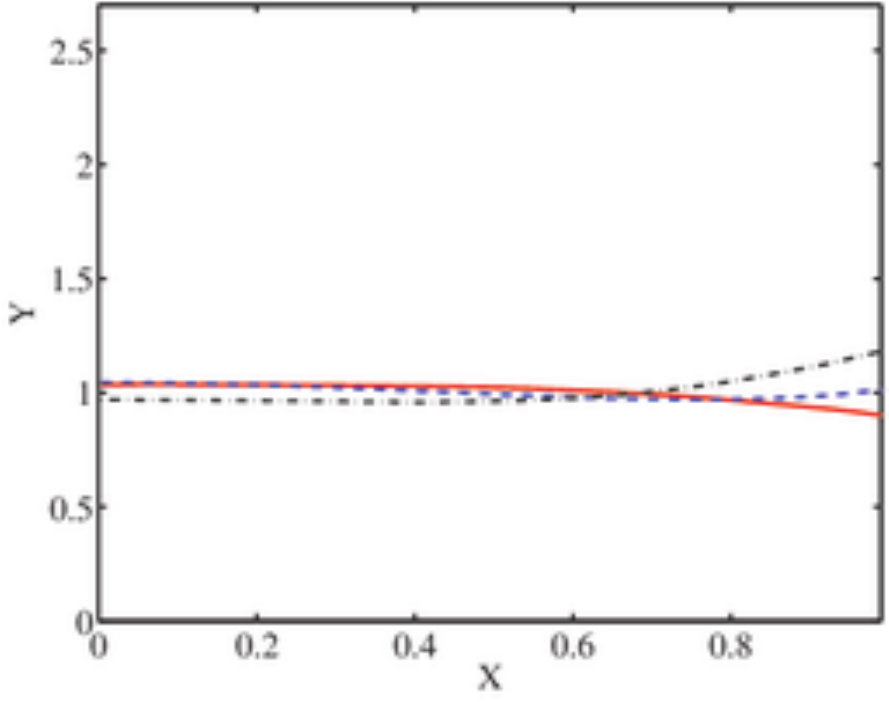}
\mylab{-5.3cm}{5.cm}{(b)}
}
\caption{ Probability density functions of the cosine of the angle of
  the vorticity, $\widetilde{\boldsymbol{\omega}}$, filtered with
  $\Delta^\omega_2$, and the eigenvectors of the rate-of-strain
  tensor, \solid, $\widehat{\boldsymbol{\lambda}}_1$; \dashed,
  $\widehat{\boldsymbol{\lambda}}_2$; \dchndot,
  $\widehat{\boldsymbol{\lambda}}_3$, filtered with $\Delta^s_2$. (a)
  $\Delta^\omega_2 = 0.05h$ and $\Delta^s_2 = 0.55h$, (b)
  $\Delta^\omega_2 = 0.55h$ and $\Delta^s_2 = 0.05h$. Results computed
  for the total velocity.
  \label{fig:align_inter_examples} }
\end{figure}
%================================================================
%
% Particular example
The previous idea was tested in \citet{leung2012} for isotropic
turbulence and a few scales. Here we extend the study to the
logarithmic layer of a turbulent channel and a wider range of
scales. We will denote the streamwise, wall-normal and spanwise filter
widths by $\Delta^j_1$, $\Delta^j_2$ and $\Delta^j_3$ with
$j=\omega,s$, and ratios as described in \S\ref{subsec:filter}. The
interest of this analysis is further motivated by figure
\ref{fig:align_inter_examples}(a), which shows that the roles of
$\widehat{\boldsymbol{\lambda}}_1$ and
$\widehat{\boldsymbol{\lambda}}_2$ change completely for
$\Delta^\omega_2=0.05h$ and $\Delta^s_2=0.55h$ compared to those
reported for the vorticity and strain at the same scale. In this
particular case, the smaller-scale vorticity aligns predominantly with
the most extensional eigenvector and is stretched by the larger-scale
strain.  On the other hand, figure \ref{fig:align_inter_examples}(b)
shows that the p.d.f.s of $\cos(\widetilde{\boldsymbol{\omega}},
\widehat{\boldsymbol{\lambda}}_i)$ follow a uniform distribution with
no preferential alignment when computed for $\Delta^\omega_2=0.55h$
and $\Delta^s_2=0.05h$, i.e., large-scale vorticity versus
smaller-scale strain.

% causality
It is difficult to infer the causal relation between scales from
instantaneous flow fields without a more detailed time-resolved
information of the flow. However, it is reasonable to suppose that, on
average, the causality is from larger scales to smaller ones, since
the characteristic times of the former are longer than those of the
latter. Thus, for $\Delta^s_2>\Delta^\omega_2$, we will assume that
the vorticity is stretched/compressed by the larger-scale strain, and
vice versa for $\Delta^\omega_2>\Delta^s_2$.

% Analysis of alignment:
%        - systematic analysis
%        - new filters
%        - definition of dominant peak
%        - results: align with c1 for large strain, c3 large vorticity, off-diagonal elements
%        - interpretation of the plot, scales, causal relation (time scales)
% Notes: - flat pdf distribution in cosine implies a sine distribution in angles
%        - Better way of measure dominance of vortex stretching?

To perform a more systematic analysis of the dominant alignment of
$\widetilde{\boldsymbol{\omega}}$ and
$\widehat{\boldsymbol{\lambda}}_{i}$, we expand the number of filters
previously used in order to sample the scale-space in more detail. The
new wall-normal filter widths range from $\Delta^j_2=0.05h$ to
$\Delta^j_2=0.55h$ in increments of $0.05h$ with $j=\omega,s$, and all
the possible combinations of $(\Delta^\omega_2,\Delta^s_2)$ are
considered, which yields a total number of 121 cases.  Alignments are
measured by the ratio of probabilities of
$|\cos(\widetilde{\boldsymbol{\omega}},\widehat{\boldsymbol{\lambda}}_i)|>0.9$,
with $i=1,2,3$, for each $(\Delta^\omega_2,\Delta^s_2)$ pair,
\begin{equation}\label{eq:ratios_align}
r_i(\Delta^\omega_2,\Delta^s_2) =
\frac{P(|\cos(\widetilde{\boldsymbol{\omega}},\widehat{\boldsymbol{\lambda}}_i)|>0.9)}
     {P(|\cos(\widetilde{\boldsymbol{\omega}},\widehat{\boldsymbol{\lambda}}_2)|>0.9)}, \ i=1,3
\end{equation}
where $P$ stands for probability, and
$|\cos(\widetilde{\boldsymbol{\omega}},\widehat{\boldsymbol{\lambda}}_2)|>0.9$
is used as a reference value. $\cos(\alpha)=0.9$ corresponds to an
angle close to $\pi/7$ radians or $25^{\circ}$, and $r_i>1$ implies a
dominant alignment of $\widetilde{\boldsymbol{\omega}}$ and
$\widehat{\boldsymbol{\lambda}}_i$. The discussion below is valid for
values equal to 0.7 and 0.8 are used instead of 0.9.

\begin{figure}
\vspace{0.5cm}
\centerline{
\psfrag{X}{ \raisebox{-0.2cm}{$\Delta^\omega_2$} } \psfrag{Y}{ $\Delta^s_2$ }
\includegraphics[width=0.47\textwidth]{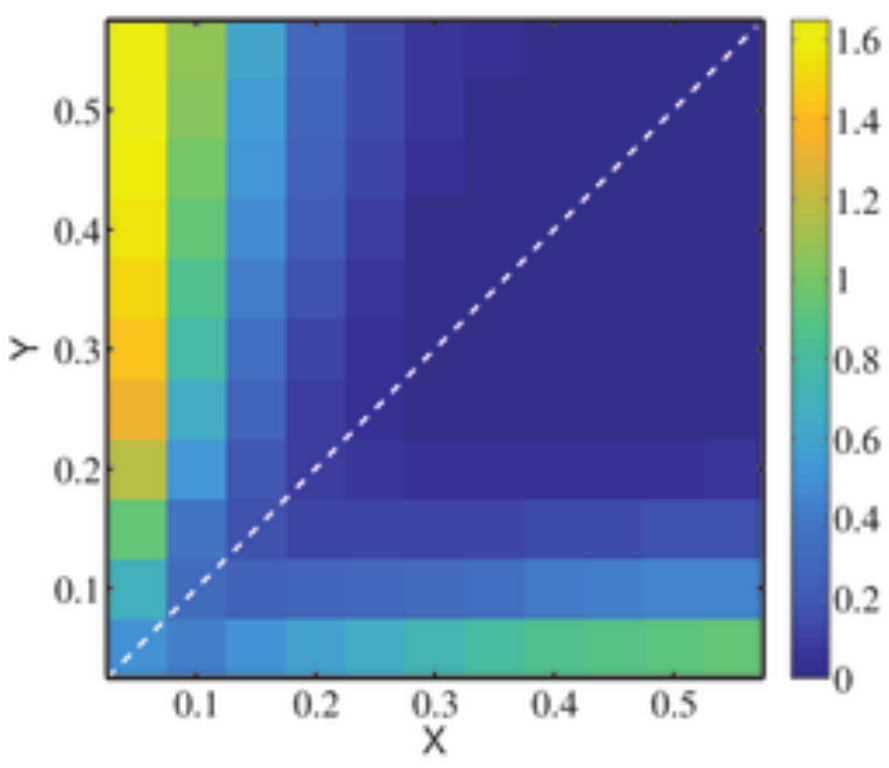}
\mylab{-5.3cm}{5.6cm}{(a)}
\psfrag{X}{ \raisebox{-0.2cm}{$\Delta^\omega_2$} } \psfrag{Y}{ }
\includegraphics[width=0.47\textwidth]{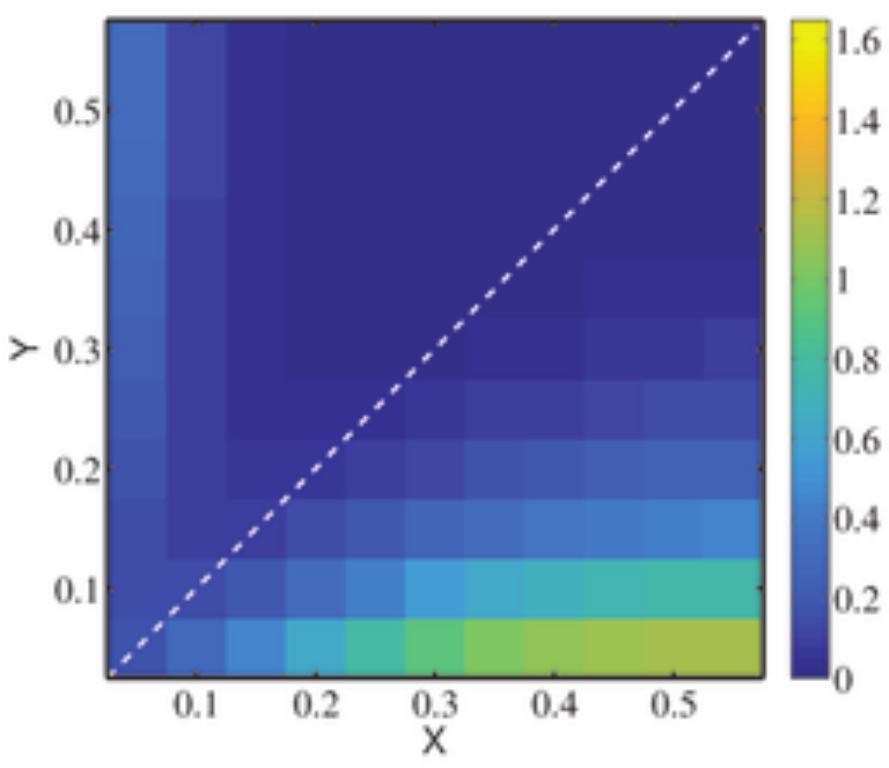}
\mylab{-5.3cm}{5.6cm}{(b)}
}
\caption{ (a) Ratio of the probability of
  $|\cos(\widetilde{\boldsymbol{\omega}},\widehat{\boldsymbol{\lambda}}_1)|>0.9$
  and
  $|\cos(\widetilde{\boldsymbol{\omega}},\widehat{\boldsymbol{\lambda}}_2)|>0.9$
  as a function of the vorticity and strain wall-normal filter widths,
  $\Delta^\omega_2$ and $\Delta^s_2$, respectively. Values equal to
  one imply equal probability of finding
  $|\cos(\widetilde{\boldsymbol{\omega}},\widehat{\boldsymbol{\lambda}}_1)|>0.9$
  and
  $|\cos(\widetilde{\boldsymbol{\omega}},\widehat{\boldsymbol{\lambda}}_2)|>0.9$. (b)
  Same as (a) but for the ratio of the probability of
  $|\cos(\widetilde{\boldsymbol{\omega}},\widehat{\boldsymbol{\lambda}}_3)|>0.9$
  and
  $|\cos(\widetilde{\boldsymbol{\omega}},\widehat{\boldsymbol{\lambda}}_2)|>0.9$. The
  dashed line is $\Delta^\omega_2=\Delta^s_2$. Results computed for
  the total velocity.
  \label{fig:align_inter_align} }
\end{figure}

Results for $r_i$ are shown in figure \ref{fig:align_inter_align},
where low values of the $r_1$ and $r_3$ appear along the diagonal
$\Delta^s_2=\Delta^\omega_2$, in agreement with the dominant alignment
of $\widetilde{\boldsymbol{\boldsymbol{\omega}}}$ and
$\widetilde{\boldsymbol{\boldsymbol{\lambda}}}_2$ discussed in
\S\ref{subsec:alignment}. For $\Delta^s_2>\Delta^\omega_2$ (upper
diagonal), the alignment of $\widetilde{\boldsymbol{\omega}}$ and
$\widehat{\boldsymbol{\lambda}}_1$ increases with the distance to the
diagonal, consistent with the energy cascade framework described
above. This is specially the case for low values of $\Delta^\omega_2$,
around $0.05-0.10h$, and $\Delta^s_2>0.2h$, where $r_1\approx
1.6$. For $\Delta^s_2<\Delta^\omega_2$ (lower diagonal), the alignment
of $\boldsymbol{\omega}$ and $\widehat{\boldsymbol{\lambda}}_3$
increases (compression of the flow field by large-scale vorticity)
although the effect is weaker than the one observed in the upper
diagonal for $r_1$. It is important to remark that the probability
distributions flatten in this region, as shown in figure
\ref{fig:align_inter_examples}(b), and the peaks at
$|\cos(\widetilde{\boldsymbol{\omega}},\widehat{\boldsymbol{\lambda}}_i)|=1$
are less pronounced than those found in the upper diagonal. It was
checked that the kurtosis coefficients of the p.d.f.s in the lower
diagonal are closer to the theoretical value of a perfectly flat
distribution than those in the upper part (not shown). This suggests
that the alignment of $\widetilde{\boldsymbol{\omega}}$ and
$\widehat{\boldsymbol{\lambda}}_3$ are less relevant than the one of
$\widetilde{\boldsymbol{\omega}}$ and
$\widehat{\boldsymbol{\lambda}}_1$ for $\Delta^s_2>\Delta^\omega_2$.
%
%================================================================
% /data4/adrian/Q1Q2R1R2/mfiles/cases/plot_Omega_S_scales.m
%================================================================
%

\begin{figure}
\vspace{0.5cm}
\centerline{
\psfrag{X}{ \raisebox{-0.2cm}{$\Delta^\omega_2$} } \psfrag{Y}{ $\Delta^s_2$ }
\includegraphics[width=0.47\textwidth]{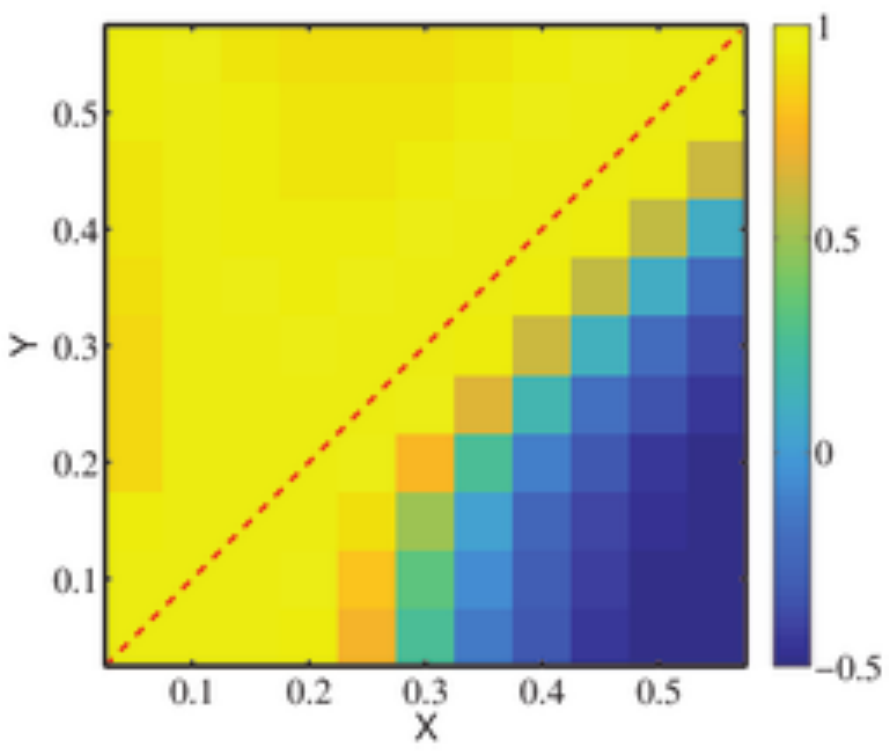}
\mylab{-5.3cm}{5.4cm}{(a)}
\psfrag{X}{ \raisebox{-0.2cm}{$\Delta^\omega_2$} } \psfrag{Y}{ }
\includegraphics[width=0.47\textwidth]{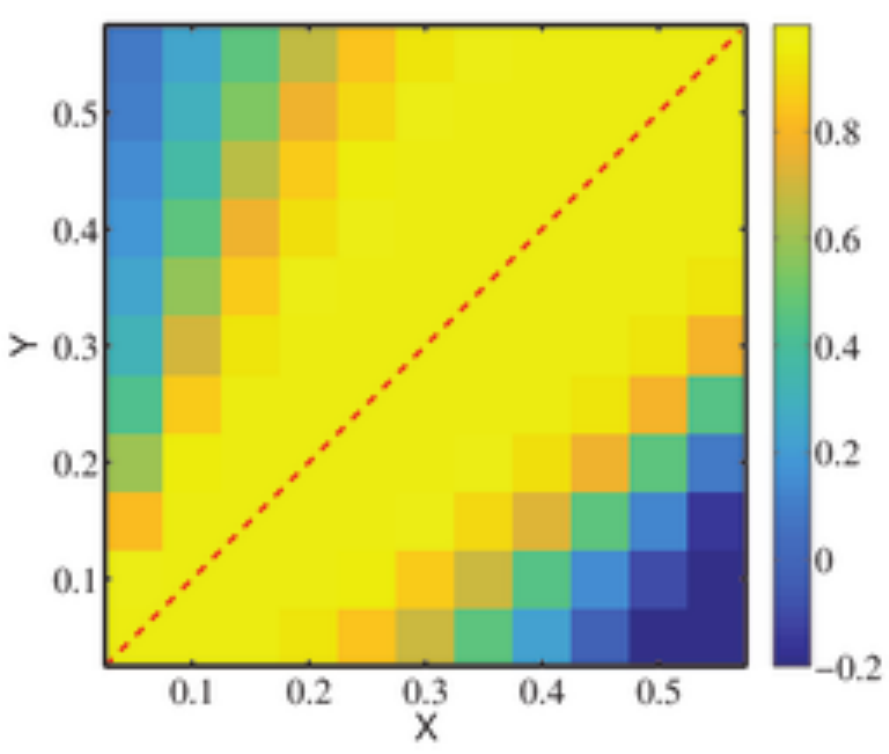}
\mylab{-5.3cm}{5.4cm}{(b)}
}
\caption{Ratio of the average enstrophy productions, $r^p$, as a
  function of the wall-normal filter width for the vorticity,
  $\Delta^\omega_2$ denoted by $\widetilde{(\cdot)}$ and the strain,
  $\Delta^s_2$ denoted by $\widehat{(\cdot)}$. Results computed from
  (a), total velocity; (b), fluctuating velocity.
  \label{fig:align_inter_production} }
\end{figure}

\begin{figure}
\vspace{0.45cm}
\centerline{
\includegraphics[width=0.9\textwidth]{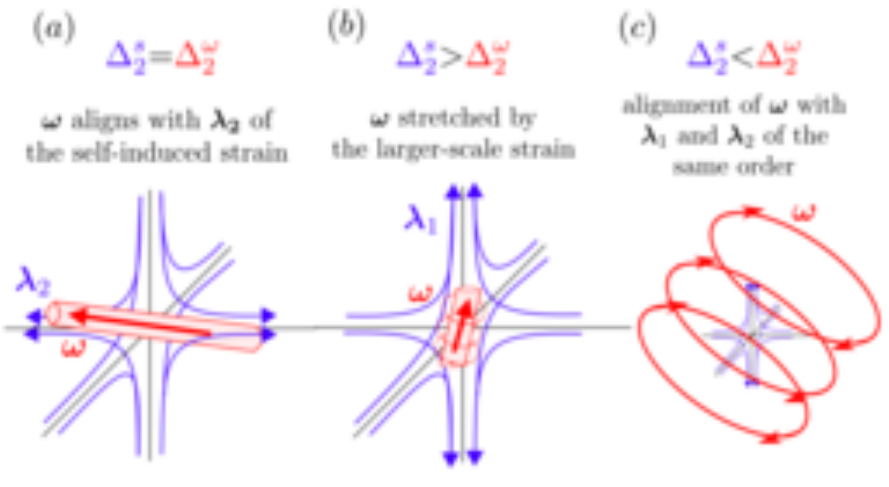}
}
\caption{ Sketches of the vorticity-strain interaction. (a) At the
  same scale; (b), strain-scale larger than vorticity-scale; (c),
  strain-scale smaller than vorticity-scale. \label{fig:cas_ske}}
\end{figure}

% Analysis of production:
%     - production equation
%     - production map
%     - results: non-negligible contribution of production
The
$\widetilde{\boldsymbol{\omega}}$--$\widehat{\boldsymbol{\lambda}}_i$
alignments presented in figure \ref{fig:align_inter_align} are of little
value if its contribution to the enstrophy production of
$\widetilde{\omega}^2$ is negligible. The dynamical equation for the
average enstrophy from (\ref{eq:DQw_balance}) can be re-written
neglecting the viscous term as 
\begin{equation}\label{eq:DQw_inter}
  \langle \widetilde{\omega}_i\widetilde{\omega}_j \widetilde{s}_{ij} \rangle =
  \langle \widetilde{\omega}_i\widetilde{\omega}_j \widehat{s}_{ij} \rangle +
  \langle \widetilde{\omega}_i\widetilde{\omega}_j         s^r_{ij} \rangle \approx
- \langle \epsilon_{jil} \widetilde{\omega}_l \frac{\partial^2 \tau_{ik}}{\partial x_k \partial x_j} \rangle,
\end{equation}
where $\widetilde{s}_{ij}$ has been decomposed into its contribution
from scale $\Delta^s_2$ and a residual term such that
$\widetilde{s}_{ij} = \widehat{s}_{ij} + s^r_{ij}$. Relation
(\ref{eq:DQw_inter}) shows that the enstrophy production may be
expressed as the interaction of $\widetilde{\omega}_i$ and
$\widehat{s}_{ij}$ plus a residual, and that the sum of both is
balanced by the interscale transfer term on the right-hand-side.  Note
that the
$\widetilde{\boldsymbol{\omega}}$--$\widehat{\boldsymbol{\lambda}}_i$
alignment calculated above is directly related to the enstrophy
production $\widetilde{\omega}_i\widetilde{\omega}_j
\widehat{s}_{ij}$.  The importance of this term is quantified in
figure \ref{fig:align_inter_production}(a), which shows the ratio
\begin{equation}\label{eq:ratios_prod}
  r^p(\Delta^\omega_2,\Delta^s_2) =
  \frac{\langle \widetilde{\omega}_i\widetilde{\omega}_j \widehat{s}_{ij} \rangle}
  {\sqrt{ \langle \widetilde{\omega}_i\widetilde{\omega}_j \widehat{s}_{ij} \rangle^2 + \langle \widetilde{\omega}_i\widetilde{\omega}_j s^r_{ij} \rangle^2 } },
\end{equation}
for all the possible combinations of filter widths. By definition,
the diagonal elements must be $1$ since $s^r_{ij}=0$ and $\langle
\widetilde{\omega}_i\widetilde{\omega}_j \widehat{s}_{ij}
\rangle>0$. Interestingly, the data reveal that $r^p$ reaches values
close to 1 for $\Delta^s_2>\Delta^\omega_2$, and the contribution of
$\langle \widetilde{\omega}_i \widetilde{\omega}_j
\widehat{s}_{ij}\rangle$ is large enough to support the idea of a
non-negligible role of vortex stretching in the energy
cascade. Surprisingly, $r^p$ is also large in magnitude but negative
for scales in the far lower part. The term (\ref{eq:DQw_inter})
appears in the dynamic equation for $\widehat{s}_{ij}\widehat{s}_{ij}$
acting as a source but with opposite sign. This suggests an
interesting connection between large-scale vorticity and small-scale
strain, closing the self-sustained cascade process.  However, we have
shown above that in that case there is no preferential alignment
between vorticity and strain eigenframe (figure
\ref{fig:align_inter_examples}b) and the physical relevance of this
result is unclear.  A simplified sketch of the different scenarios is
shown in figure \ref{fig:cas_ske}.
%
%================================================================
% /data4/adrian/Q1Q2R1R2/mfiles/cases/plot_Omega_S_scales.m
%================================================================
%
%================================================================
% Sketch from APS 2015 presentation
%================================================================

\begin{figure}
\vspace{0.5cm}
\centerline{
\psfrag{X}{ \raisebox{-0.2cm}{$\Delta^\omega_2$} } \psfrag{Y}{ $\Delta^s_2$ }
\includegraphics[width=0.47\textwidth]{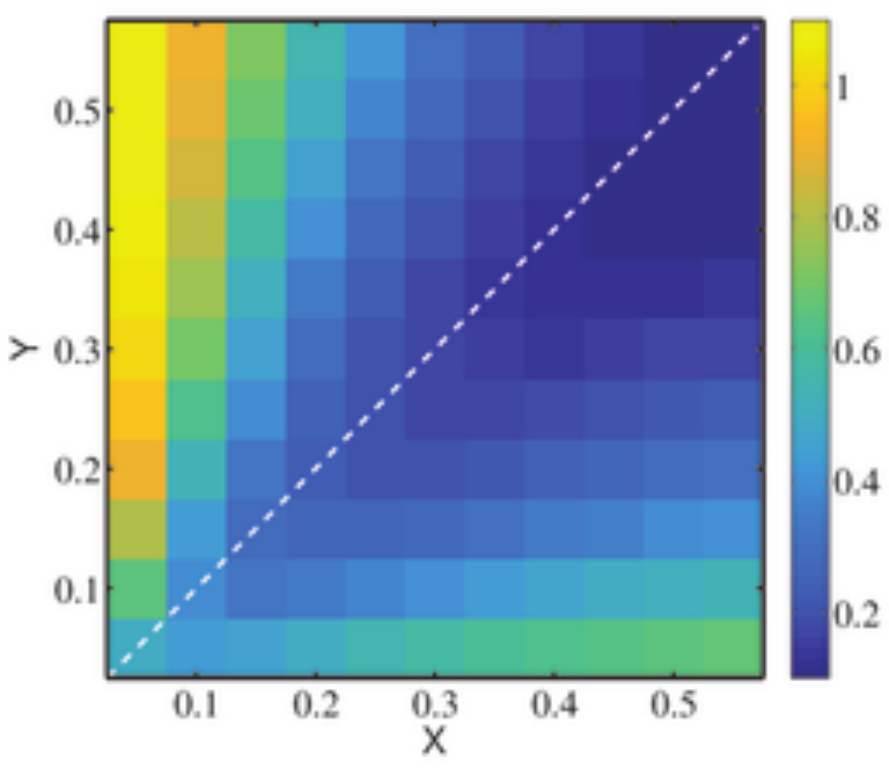}
\mylab{-5.3cm}{5.6cm}{(a)}
\psfrag{X}{ \raisebox{-0.2cm}{$\Delta^\omega_2$} } \psfrag{Y}{ }
\includegraphics[width=0.47\textwidth]{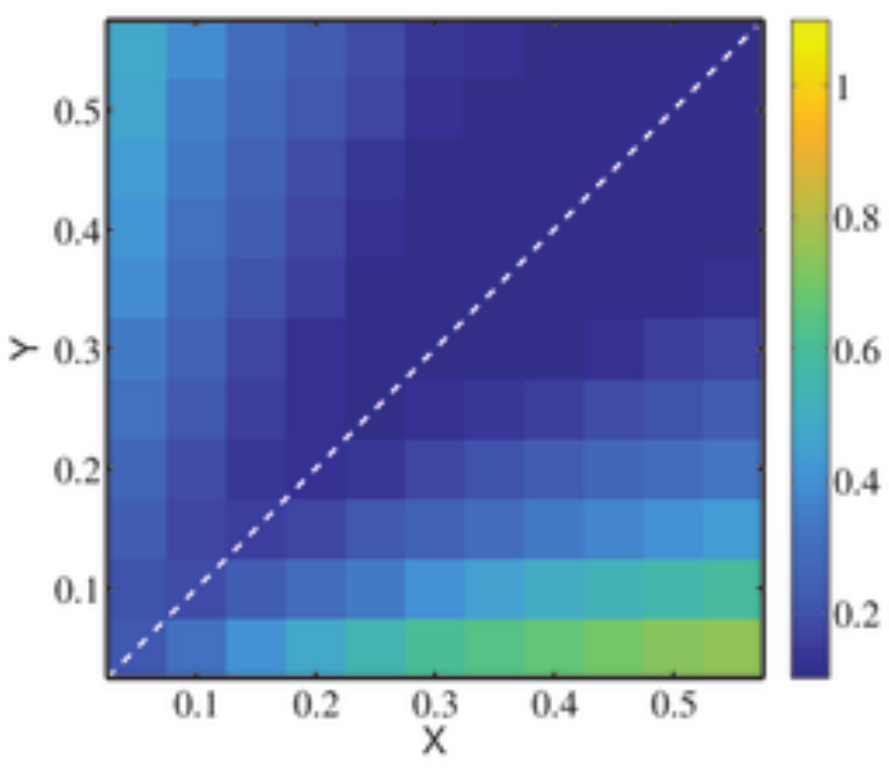}
\mylab{-5.3cm}{5.6cm}{(b)}
}
\caption{ (a) Ratio of the probability of
  $|\cos(\widetilde{\boldsymbol{\omega}},\widehat{\boldsymbol{\lambda}}_1)|>0.9$
  and
  $|\cos(\widetilde{\boldsymbol{\omega}},\widehat{\boldsymbol{\lambda}}_2)|>0.9$
  as a function of the vorticity and strain wall-normal filter widths,
  $\Delta^\omega_2$ and $\Delta^s_2$, respectively. Values equal to
  one imply equal probability of finding
  $|\cos(\widetilde{\boldsymbol{\omega}},\widehat{\boldsymbol{\lambda}}_1)|>0.9$
  and
  $|\cos(\widetilde{\boldsymbol{\omega}},\widehat{\boldsymbol{\lambda}}_2)|>0.9$. (b)
  Same as (a) but for the ratio of the probability of
  $|\cos(\widetilde{\boldsymbol{\omega}},\widehat{\boldsymbol{\lambda}}_3)|>0.9$
  and
  $|\cos(\widetilde{\boldsymbol{\omega}},\widehat{\boldsymbol{\lambda}}_2)|>0.9$. The
  dashed line is $\Delta^\omega_2=\Delta^s_2$. Results computed for
  the fluctuating velocity.
  \label{fig:align_inter_align_flu} }
\end{figure}

% Results for fluctuations
The results shown so far were calculated for the total velocity, and
following the structure of previous sections, they were repeated for
the fluctuations. Most of the conclusions discussed for the total
velocity apply to the fluctuating vorticity and rate-of-strain tensor.
The most remarkable difference is found in the ratio $r^p$ in figure
\ref{fig:align_inter_production}(b). The contribution of $\langle
\widetilde{\omega}_i \widetilde{\omega}_j \widehat{s}_{ij} \rangle$
decays with the distance to the diagonal and is around 20--30\% of
$\langle \widetilde{\omega}_i \widetilde{\omega}_j \widetilde{s}_{ij}
\rangle$ at those places where the alignment of
$\widetilde{\boldsymbol{\omega}}$ and
$\widehat{\boldsymbol{\lambda}}_1$ is of the same order as the one
from $\widetilde{\boldsymbol{\omega}}$ and
$\widehat{\boldsymbol{\lambda}}_2$ (figure
\ref{fig:align_inter_align_flu}a). Figure
\ref{fig:align_inter_align_flu} also shows that $r_1$ and $r_3$ attain
values below those obtained for the total velocity, which may be
connected to the counter-clockwise behavior in the
$R_\omega$--$Q_\omega$ plane in figure \ref{fig:fluctuations}(h) and
discussed in \S\ref{subsec:QsQwRsRw}.
%
%
%================================================================
% /data4/adrian/Q1Q2R1R2/mfiles/cases/plot_Omega_S_scales.m
%================================================================

%%%%%%%%%%%%%%%%%%%%%%%%%%%%%%%%%%%%%%%%%%%%%%%%%%%%%%%%%%%%%%%%%%%%%%%%%%%%
\section{Conclusions}\label{sec:conclusions}
%%%%%%%%%%%%%%%%%%%%%%%%%%%%%%%%%%%%%%%%%%%%%%%%%%%%%%%%%%%%%%%%%%%%%%%%%%%%

% Computation of RQ and filtering
We have studied the dynamics of the invariants of the filtered
velocity gradient tensor, $R$ and $Q$, in the logarithmic layer of an
incompressible turbulent channel flow. The invariants are gradients of
the velocities and hence, are dominated by the effect of the small
scales. By filtering the velocity field, we have applied the
topological and physical tools provided by the invariants to scales in
the inertial range.

% Numerical issues
We have paid special attention to the numerics involved in the
computation of the invariants in order to minimize numerical errors as
much as possible. The spatial derivatives were computed with spectral
methods, and the number of modes expanded by a factor of three in each
direction to reduce aliasing problems. The temporal derivatives were
computed with fourth-order finite differences using velocity fields
contiguous in time. Besides, all the calculations were performed in
double precision. More details about the numerical procedure can be
found in \cite{loz:hol:jim:2015}.

% Normalization
In order to compensate for the wall-normal inhomogeneity of the
channel, the invariants were scaled by the standard deviation of the
second invariant, $Q'$, as a function of the distance to the wall, and
their material derivatives were consistently computed to obtained
closed trajectories when the whole channel domain is considered as in
\citet{loz:hol:jim:2015}.

% Plane RQ: small change
The effect of filtering on the joint probability density function of
$R$ and $Q$, $J(R,Q)$, is found to be rather weak. The tear-drop shape
persists for larger scales, consistent with previous
findings \citep{Borue1998, VanDerBos2002, Luethi2007}, and the most
noteworthy change is the widening of $J(R,Q)$ along the $R$ axis for
the filtered cases.

% CMTs
The conditional mean trajectories in the normalized $R$--$Q$ plane
rotate clockwise for all of the cases. The CMTs describe almost closed
trajectories in the unfiltered case when normalized as $R/Q'^{3/2}$
and $Q/Q'$. However, they spiral outwards in the filtered cases, and
this effect intensifies with the filter width. The probability fluxes
show that the previous equilibrium is not achieved at the inertial
scales, in the sense that fluid from the outer region with associated
weak invariants enters to the logarithmic layer, and is later
intensified, that is, CMTs spiral outwards to larger values of the
normalized $R$ and $Q$.

Surprisingly, when the calculations were repeated for the invariants
of the fluctuating velocity gradient, the CMTs split into two families
for $Q<0$ and $Q>0$, with trajectories rotating clockwise and
counter-clockwise, respectively. The latter differs from the CMTs of
the invariants of the total velocity gradient tensor, and the cause
was traced back to the enstrophy/enstrophy-production cycle. It was
found that increasing enstrophy was on average associated with
contraction of vorticity, and therefore, the upper counter clockwise
cycle in the $R$--$Q$ plane can not be a consequence of the enstrophy
production (as it is for the total velocity) but the result of the
interaction of the fluctuations with the shear or with the interscale
transfer.  Nevertheless, it was also shown that they are the result of
an averaging process where the mean is 3-5 times smaller than the
corresponding standard deviation, and the CMTs represent broad trends
that may differ quite significantly from the instantaneous behavior of
the individual flow particles.

% Decomposition of RQ: important changes, shear responsible, fluctuations
Decomposing the invariants $R$ and $Q$ in their enstrophy ($Q_\omega$
and $R_\omega$) and strain ($Q_s$ and $R_s$) components reveals
substantial changes compared to those observed in the $R$--$Q$
plane. As the filter width increases, the
strain/strain-self-amplification and enstrophy/enstrophy-production
distributions become more symmetric. On the contrary, the joint
p.d.f.s of $Q_s$--$Q_\omega$ and $R_s$--$R_\omega$ become
progressively anti-correlated, i.e., $Q_s\approx -Q_\omega$ and
$R_s\approx -R_\omega$.  These results were explained considering that
the filter diminishes the effect of the small scales in favor of
larger wall-attached eddies, whose dynamics are controlled by the mean
shear \citep{Lozano2012, DelAlamo2006, Flores2010,
  Jimenez2012}. Interestingly, when the direct effect of the mean
shear is removed by computing the normalized invariants of the
fluctuating velocities, all the p.d.f.s collapse for filter widths
$\Delta_2>0.1h$, suggesting a self-similar multiscale behavior of the
fluctuating strain and enstrophy dynamical cycles.  Nevertheless, the
results obtained for the fluctuating velocities differ from those for
isotropic turbulence, which is an indication that some indirect effect
of the mean shear remains.

% Orbital periods: robust in all maps, linear with filter width
The orbital period $\tau$, i.e., the time employed by the CMTs to
complete one full revolution, computed in the $R$--$Q$,
$Q_\omega$--$R_\omega$ and $Q_s$--$R_s$ planes are all of the same
order, which is expected since they represent the same dynamical cycle
projected at different planes. Besides, the periods are independent of
the initial position of the CMTs, namely weak and strong events have
the same time-scale for a given filter width. This was explained by
noting that the trajectories associated with weak regions in a certain
space have shorter lengths but also slower conditional velocities. As
the CMTs move towards stronger events, they travel longer distances
but also move faster. These two effects compensate resulting in a
roughly constant $\tau$.  If we consider that strong and weak events
are respectively associated with small and large scales, that is not
strictly rigorous but reasonable on average, the previous results may
be related to the classical turbulent cascade where the energy is fed
into the largest scales and cascades downwards until is ultimately
dissipated at the smallest ones. In this scenario, the evolution of
the small scales is enslaved by the larger ones, and the orbital
periods of the strong small-scale events would simply reflect the
effect of the weaker larger ones \citep{jim:2013,cardesa2015}.

The orbital periods collapse for all the filter widths when scaled by
$\widetilde{Q}'^{-1/2}$, that is the natural eddy-turnover time of
eddies at scale $\Delta_2$. Also, when expressed as a function of the
filter width, they follow $\tau u_\tau/h \approx 8\Delta_2/h+0.08$. A
linear relation between lifetimes and scales has already been observed
in previous works \citep{DelAlamo2006, leh:gua:mck:2013, Lozano2014}
in the context of self-similar eddies in the logarithmic layer with
lifetimes proportional to their sizes, and dynamics controlled by the
mean shear. The periods obtained here are 8 times larger than the
lifetimes of individual eddies reported by \citet{Lozano2014} if
$\Delta_2$ is taken as the characteristic size of the wall-attached
motions. This disparity may be related to the large velocity
differences between CMTs and instantaneous trajectories mentioned
above. In this sense, the orbital periods are the average time
required by the fluid particles to undergo all the different
topologies, or from a dynamical point of view, to complete one cycle
(in the enstrophy/enstrophy-production,
strain/strain-self-amplification planes...) progressively.

% Alignment vorticity-strain: alignment with lambda_2
The angle between the vorticity and the eigenvectors of the
rate-of-strain tensor was also studied as a function of the filter
width. The results showed that the vorticity, $\boldsymbol{\omega}$,
tends to align with the second eigenvector of the rate-of-strain
tensor, $\boldsymbol{\lambda_2}$, which intensifies as the filter
width increases. Despite this, if the enstrophy production is
decomposed as the sum of $p_i=\lambda_i
\cos(\boldsymbol{\omega},\boldsymbol{\lambda_i})^2$ with $i=1,..,3$,
most of the contribution to its mean ($\langle \cdot \rangle$) is
caused by $\langle p_1\rangle$ in the unfiltered case, but $\langle
p_2 \rangle$ and $\langle p_3 \rangle$ steadily increase and decrease,
respectively, until they equal in magnitude the contribution of the
first for large filter widths. The changes in the ratio of the
standard deviations of $p_i$ as a function of the filter width is even
more pronounced. Again, this scale-dependent behavior was explained in
terms of the mean shear. When the calculations were repeated for the
fluctuating velocities, the results became scale-independent.  In this
case, there is still a preferential alignment of $\boldsymbol{\omega}$
and $\boldsymbol{\lambda_2}$, but a similar contribution of $\langle
p_1 \rangle$, $ \langle p_2 \rangle$ and $ \langle p_3 \rangle$ to the
mean enstrophy production at all the scales.  The previous results
reinforce the idea of self-similar dynamics in the inertial range when
the direct effect of the mean shear is removed.

% Vortex stretching cascade
Finally, we have investigated the energy cascade in terms of vortex
stretching where vortices at a given scale are stretched by the strain
at a larger one.  We have shown that the preferred alignment of the
vorticity and the intermediate eigenvector of the strain decreases
when vorticity and strain are each considered at a different scale. In
particular, the alignment of lower-scale vorticity and larger-scale
strain increases with the scale separation, and reaches values of the
same order or larger than those obtained at same scale.
Moreover, these interscale interactions between strain and vorticity
attain values between $0.5$-$1.6$ of those of the total enstrophy
production at the $\boldsymbol{\omega}$ scale. The scenario is
qualitatively similar for the fluctuating velocity but with a weaker
alignment of $\boldsymbol{\omega}$ and $\boldsymbol{\lambda_1}$, and
contributions to the total enstrophy production around
$0.2$-$0.3$. Although the results support a non-negligible role of the
phenomenological energy-cascade model formulated in terms of vortex
stretching, the details of such a cascade remain unknown, and
time-resolved data at higher Reynolds numbers is required to perform a
thorough analysis of the process.

%-----------------------------------------------------------%
\section*{Acknowledgments}
%-----------------------------------------------------------%
% Acknowlegments
The authors thank Beat L\"{u}thi for his contribution in the initial
phase of this project and Leander van Acker for his contribution in
the frame of an MSc thesis.  This work was supported in part by CICYT
under grant TRA2009-11498, and by the European Research Council under
grants ERC-2010.AdG-20100224 and
ERC-2014.AdG-669505. A. Lozano--Dur\'an was supported partially by an
FPI fellowship from the Spanish Ministry of Education and Science and
ERC. The computations were made possible by generous grants of
computer time from CeSViMa (Centro de Supercomputaci\'on y
Visualizaci\'on de Madrid) and from the Barcelona Supercomputing
Center.

%-----------------------------------------------------------%
\appendix
%-----------------------------------------------------------%

%-----------------------------------------------------------%
\section{Effects of the Reynolds number, computational domain and filter width aspect ratios}\label{sec:appendix}
%-----------------------------------------------------------%

\begin{figure}
\vspace{0.5cm}
\centerline{
\psfrag{X}{ \raisebox{-0.2cm}{$R/Q'^{3/2}$} }\psfrag{Y}{$Q/Q'$}
\includegraphics[width=0.45\textwidth]{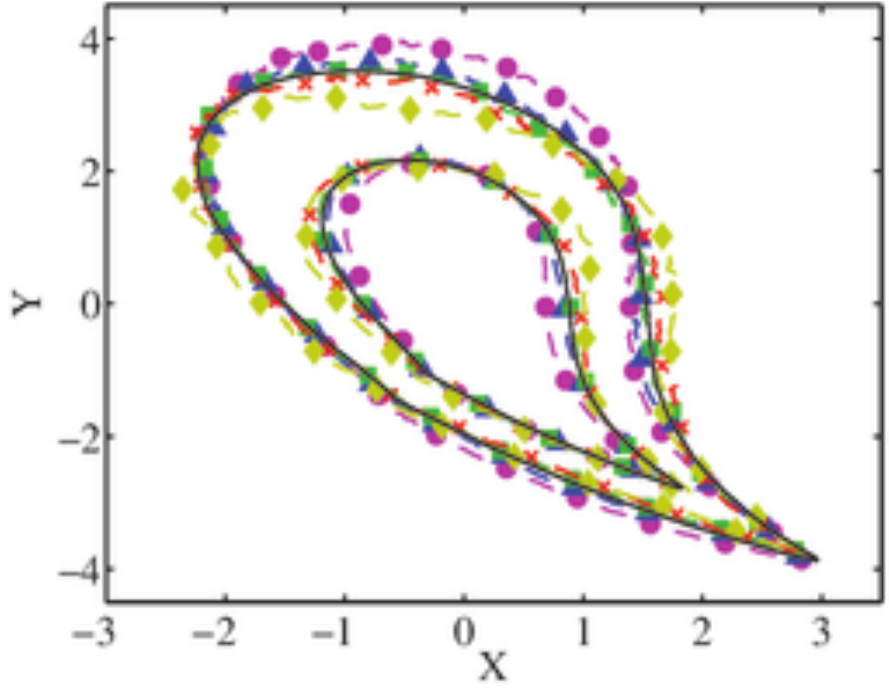}
\mylab{-5.3cm}{4.9cm}{(a)}
\psfrag{X}{ \raisebox{-0.2cm}{$Q_s/{Q'_s}$} }\psfrag{Y}{$Q_\omega/{Q'}_\omega$}
\includegraphics[width=0.45\textwidth]{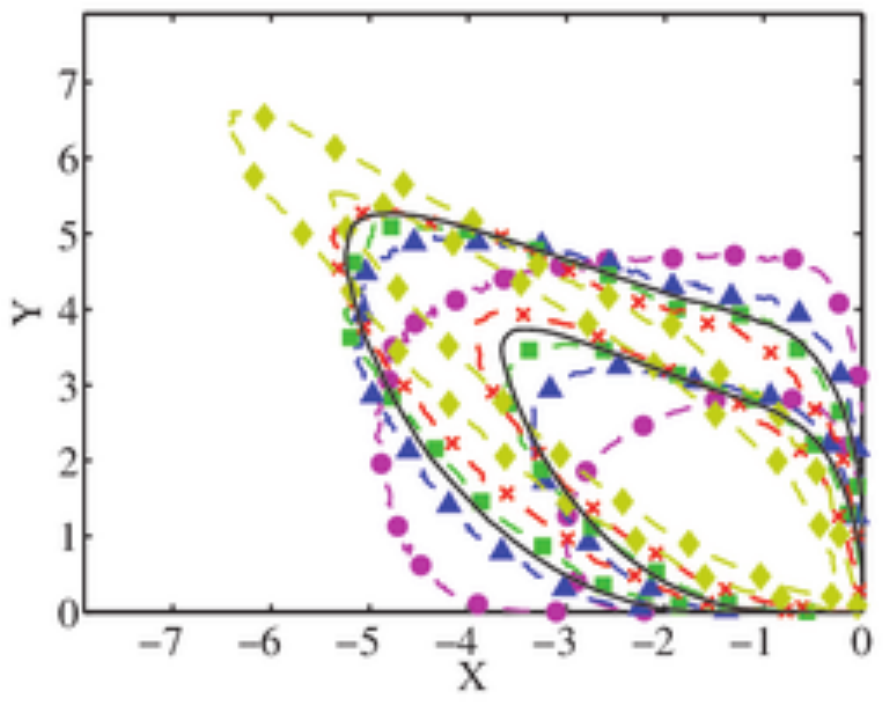}
\mylab{-5.3cm}{4.9cm}{(b)}
}
\vspace{0.5cm}
\centerline{
\psfrag{X}{ \raisebox{-0.2cm}{$R/Q'^{3/2}$} }\psfrag{Y}{$Q/Q'$}
\includegraphics[width=0.45\textwidth]{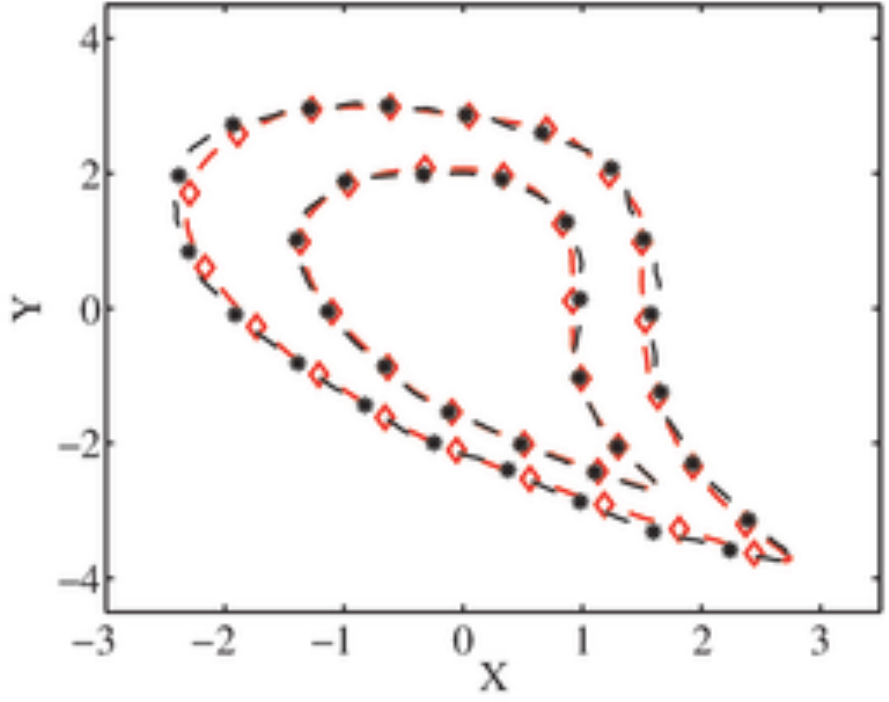}
\mylab{-5.3cm}{5cm}{(c)}
\psfrag{X}{ \raisebox{-0.2cm}{$Q_s/Q'_s$} }\psfrag{Y}{$Q_\omega/Q'_\omega$}
\includegraphics[width=0.45\textwidth]{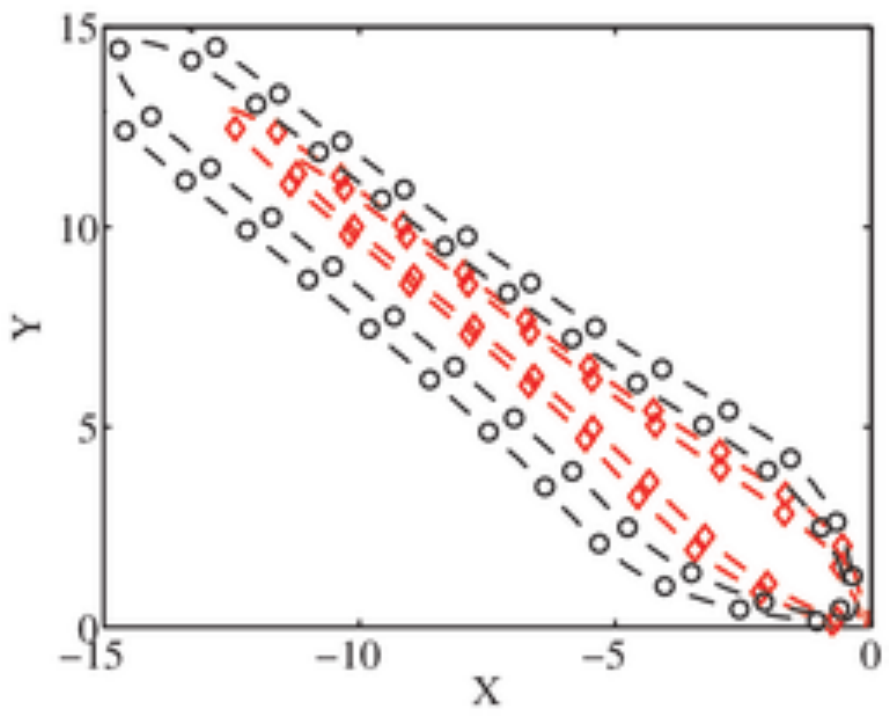}
\mylab{-5.3cm}{5cm}{(d)}
}
\caption{ (a) Joint probability density function of (a), $R$--$Q$ and
  (b), $Q_s$ and $Q_\omega$ at $Re_\tau=550$. Symbols and colors are
  as in table \ref{table:cases}.  The solid line is case F0.2. (c)
  Joint probability density function of $R$ and $Q$ computed for
  channels with streamwise and spanwise lengths of $\Diamond$,
  $L_1=2\pi h$ and $L_3=\pi h$; *, $L_1=8\pi h$ and $L_3=3\pi h$. In
  both cases the data was filtered with $\Delta_1=1.2h$,
  $\Delta_2=0.4h$ and $\Delta_2=0.6h$.  (d) Joint probability density
  functions of $Q_s$ and $Q_\omega$ filtered with $\Diamond$,
  $\Delta_1=0.9h$, $\Delta_2=0.3h$ and $\Delta_2=0.45h$; \circle,
  $\Delta_1=\Delta_2=\Delta_2=0.6h$.
\label{fig:appendix}
}
\end{figure}

% Reynolds number
The results presented in this paper were also computed for a DNS of a
turbulent channel at $Re_\tau=550$ with a resolution $\delta_1$,
$\delta_2$ and a numerical domain equal to the ones shown in table
\ref{table:DNS}.  The p.d.f.s at $Re_\tau=550$ collapse with those at
$Re_\tau=932$ in the $R$--$Q$ plane as shown in figure
\ref{fig:appendix}(a). However, the trends observed in the
distributions of the decomposed invariants at $Re_\tau=932$ are
qualitatively similar but less pronounced at $Re_\tau=550$. Figure
\ref{fig:appendix}(b) shows the joint p.d.f.s of $Q_s$ and $Q_\omega$
as an example.
%
%================================================================
% /data4/adrian/Q1Q2R1R2/mfiles/old/plotLifetimes.m
% /data4/adrian/Q1Q2R1R2/mfiles/old/plotQsQwRsRw_cte.m
%================================================================
%
%

% effect of the box size
We address next the effect of the computational domain in the results
presented above. The dataset were computed in boxes with streamwise
and spanwise dimensions of $L_1=2\pi h$ and $L_3=\pi h$, respectively.
\citet{Lozano2014} showed that these domains are large enough to
correctly capture the dynamics of the logarithmic layer. However, this
was done for the unfiltered case and it remains unclear whether it is
also valid for filtered fields.  The most restrictive case is the one
with the larger filter width, i.e., case F0.4 (see table
\ref{table:cases}). Figure \ref{fig:appendix}(b) compares the
iso-probability contours of $R$-$Q$ for case F0.4 with the results
obtained from a turbulent channel at the same Reynolds number,
filtered with the same filter width but with a much larger
computational domain, $L_1=8\pi h$ and $L_3=3\pi h$. The agreement
between the p.d.f.s computed in both domains is almost perfect and
suggests that the results in the previous sections are independent of
the size of the domain.

% effect of the filter width
The ratio of the filter widths $\Delta_1,\Delta_2$ and $\Delta_3$ was
chosen $\Delta_1/\Delta_2=3$, $\Delta_3/\Delta_2=1.5$ to match the
size of the eddies educed by \citet{Lozano2014}. This has a caveat,
because the effect of the filter is smaller in the wall-normal
derivatives than in the others. It was tested that modifying the
aspect ratio of the filter widths does not alter the dominant role of
$\partial u_1 /\partial x_2$ and qualitatively similar results to
those presented in the paper persist.  Figure \ref{fig:appendix}(d)
shows one example with a homogeneous filter
$\Delta_1=\Delta_2=\Delta_3=0.6h$ to illustrate that the strong
correlation between $Q_s$ and $Q_\omega$ remains.

%-----------------------------------------------------------%
\section{ Alternative filter }\label{sec:appendixFil}
%-----------------------------------------------------------%

\begin{figure}
\centerline{
\includegraphics[width=0.8\textwidth]{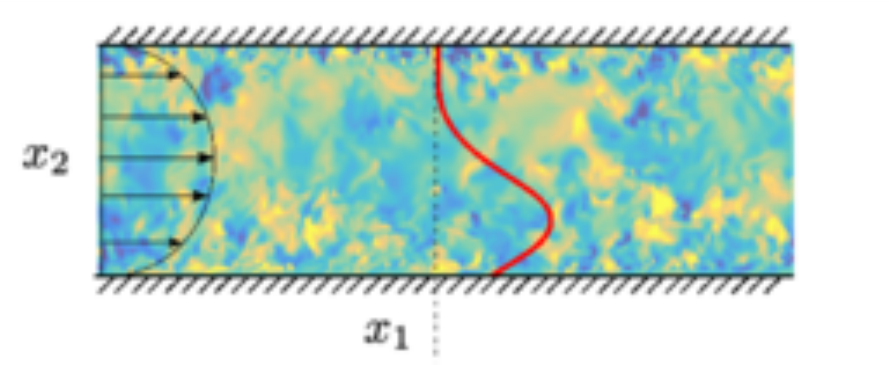}
}
\caption{ Alternative filter. Instantaneous wall-normal velocity used
  for filtering in the wall-normal direction. The red solid line
  represents the Gaussian kernel as a function of $x_2$ truncated at
  the wall. The colormap is the wall-normal
  velocity. \label{fig:filter_sketch_alter}}
\end{figure}

To assess the effect of (\ref{eq:filter}), all the results were
recomputed using the filter
\begin{equation}\label{eq:filterAlt}
\widetilde{u}_i(\boldsymbol{x})=
\iiint_{Vr} u_i(\boldsymbol{x}-\boldsymbol{x}') f(x_2) \exp
 \left( -\left(\frac{\pi x_1'}{\Delta_1}\right)^2
        -\left(\frac{\pi x_2'}{\Delta_2}\right)^2
        -\left(\frac{\pi x_3'}{\Delta_3}\right)^2
 \right) \mathrm{d}x_1'\mathrm{d}x_2'\mathrm{d}x_3',
\end{equation}
where $\Delta_1$, $\Delta_2$ and $\Delta_3$ are the filter widths in
the streamwise, wall-normal and spanwise directions, respectively, and
$V_r$ is the channel domain.  The wall-normal Gaussian shape of the
filter is maintained at all heights and truncated at the wall (see
figure \ref{fig:filter_sketch_alter}). The function $f(x_2)$ is a
normalization factor that accounts for the finite length of the domain
in $x_2$, and such that the integral of the filter kernel over $V_r$
is one.  Note that this makes the filtered velocity field slightly
compressible, particularly close to the wall and for large filter
widths. However, this effect is rather weak in the logarithmic layer,
and although not shown, the remaining compressible components of the
invariants, $Q_p=1/2 P_o^2$ and $R_p=-1/3 P_o^3 + P_oQ$, (where $P_o$
is the first invariant) computed for the filtered velocity are at
least $10^5$ times smaller than their $Q$ and $R$ counterparts from
(\ref{eq:invariants_Q}) and (\ref{eq:invariants_R}).
%
%================================================================
%
%================================================================
%

\begin{figure}%[h!tb]
\centerline{
\psfrag{X}{$R_\omega$}\psfrag{Y}{$Q_\omega$}
\includegraphics[width=0.432\textwidth]{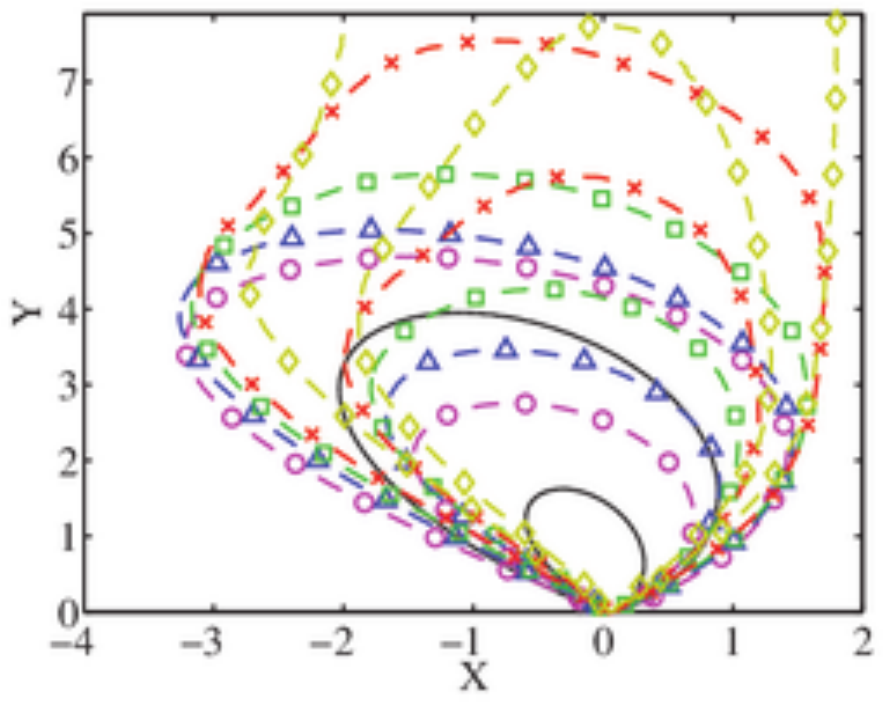}
\mylab{-1cm}{4.0cm}{(a)}
\psfrag{X}{$R_s$}\psfrag{Y}{$Q_s$}
\includegraphics[width=0.435\textwidth]{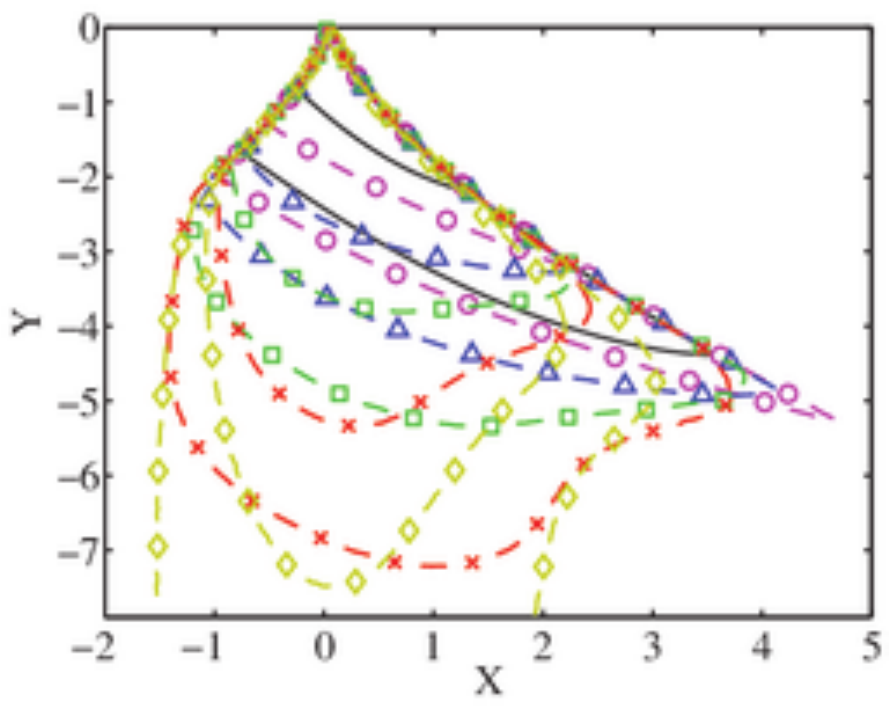}
\mylab{-1cm}{4.0cm}{(b)}
}
\centerline{
\psfrag{X}{$Q_s$}\psfrag{Y}{$Q_\omega$}
\includegraphics[width=0.435\textwidth]{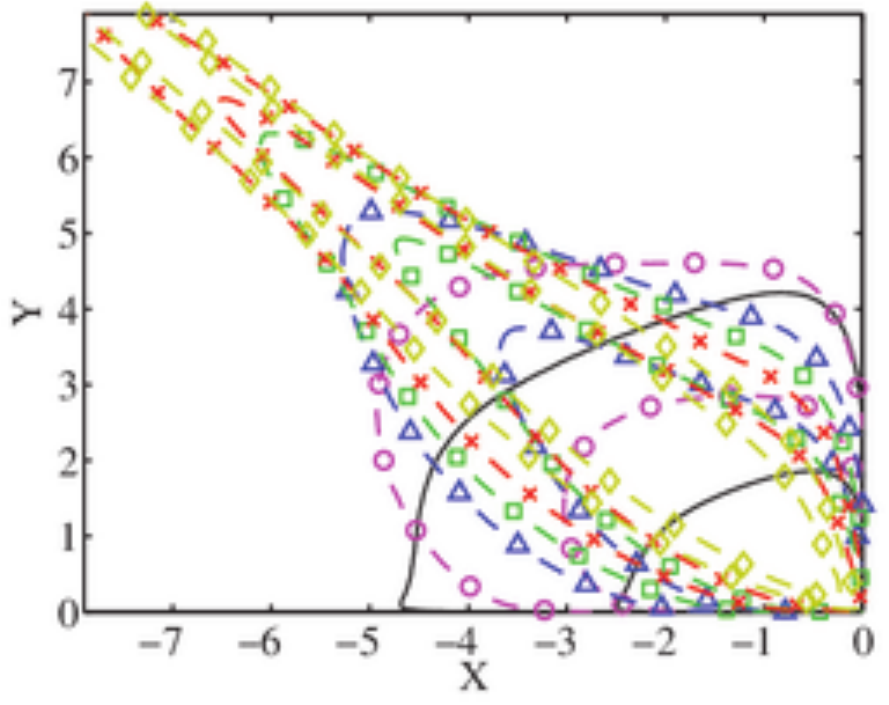}
\mylab{-1.0cm}{4.0cm}{(c)}
\psfrag{X}{$R_s$}\psfrag{Y}{$R_\omega$}
\includegraphics[width=0.435\textwidth]{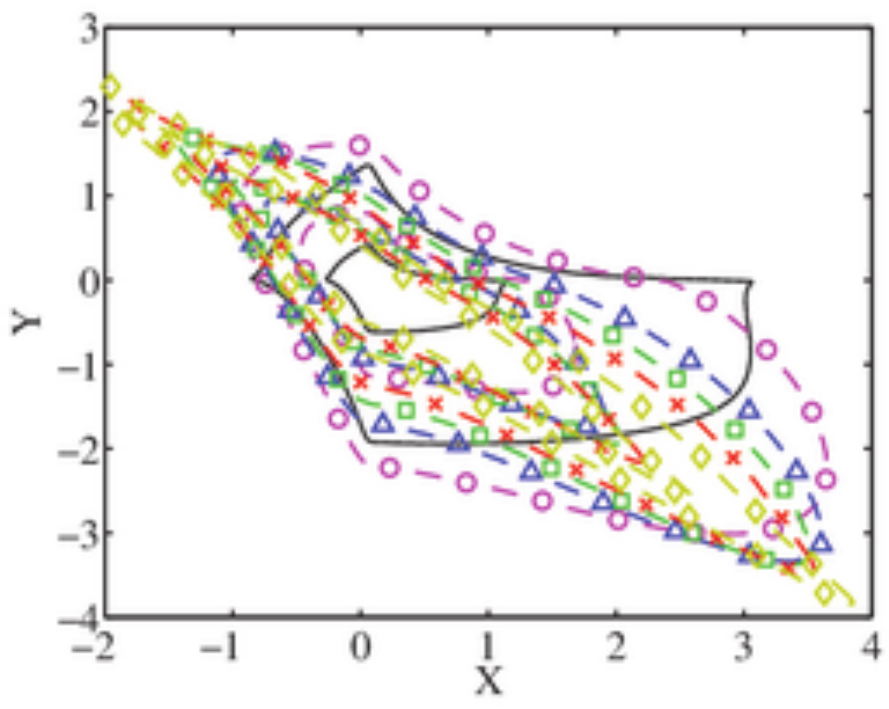}
\mylab{-1.0cm}{4.0cm}{(d)}
}
\caption{ Results filtering for the alternative filter. Joint
  probability density functions of (a), $R_\omega$--$Q_\omega$; (b),
  $R_s$--$Q_s$; (c), $Q_s$--$Q_\omega$; (d),
  $R_s$--$R_\omega$. Symbols and colors as in table
  \ref{table:cases}. The contours contain 90\% and 98\% of the data.
\label{fig:QsQwRsRw_Ft}}
\end{figure}

Figure \ref{fig:QsQwRsRw_Ft} shows some examples that are practically
identical to those from figure \ref{fig:QsQwRsRw}.
%
%================================================================
% /data4/adrian/Q1Q2R1R2/mfiles/plotQ1Q2R1R2_cte.m
%================================================================
%

%-----------------------------------------------------------%
\section{ Conservation of probability equation }\label{sec:appendixProb}
%-----------------------------------------------------------%

This Appendix provides some guidelines on how equation
(\ref{eq:prob_conser}) is obtained. For more details about the
procedure see chapter 2 in \cite{beck93}. Similar equations have also
been derived by Smoluchowski for the conservation of the particle
probability distribution function \citep{Doi88}.

Let's consider a number of experiments consisting of a tracer (fluid
particle) in a turbulent channel flow, and its associated invariants
and wall-normal position $(R,Q,x_2)$.  For simplicity, we will use $Q$
and $R$ but the following argument is also valid for $Q/Q'$ and
$R/Q'^{3/2}$.

Let's consider an initial condition for each experiment, $R$, $Q$ and
the initial position of the particle, and let the system evolve in
time. We will assume that the system is ``mixing'' (and hence
ergodic), so that every sufficiently smooth initial distribution
evolves to the ``natural invariant density'' \citep{beck93}. That is,
independently of the initial distribution of the test particles, they
eventually evolve in time in such a way that they are a fair
representation of the system.

Since $Q(x_2)$ and $R(x_2)$ are known, we can define a probability
density function $P = P(R,Q,x_2;t)$ at time $t$. The conservation of
the number of experiments (equivalently of $P$) is then given by
\begin{equation}\label{appendix:eq:prob}
\frac{\partial P}{\partial t} + \boldsymbol{\nabla}_{R,Q,x_2}\cdot \left( P \boldsymbol{w} \right) = 0, 
\end{equation}
where $\boldsymbol{\nabla}_{R,Q,x_2} = (\partial/\partial R, \partial/\partial
Q, \partial/\partial x_2)$ and $\boldsymbol{w}$ is the mean velocity vector
\begin{equation}
 \boldsymbol{w}=\left\langle \frac{\mathrm{D}}{\mathrm{D}t}(Q,R,x_2) \right\rangle_{R,Q,x_2},
\end{equation}
where $\langle \cdot \rangle_{R,Q,x_2}$ denotes conditional average at
point $(R,Q,x_2)$. 

Let's define a new probability 
\begin{equation}
J = a \int_{x_b}^{x_t} P \mathrm{d}x_2,
\end{equation}
with $a = 2/(x_t-x_b)$. Multiplying equation (\ref{appendix:eq:prob})
by $a$ and integrating from $x_2=x_b$ to $x_2=x_t$ yields to
\begin{equation}\label{appendix:eq:prob:final}
\frac{\partial J}{\partial t} + \boldsymbol{\nabla}_{R,Q}\cdot \left( J \boldsymbol{V} \right) = \psi_t + \psi_b, 
\end{equation}
where $\boldsymbol{\nabla}_{R,Q} = (\partial/\partial R,
\partial/\partial Q)$, $\boldsymbol{V}$ is as defined in (\ref{eq:v}),
$\psi_b = \alpha V_{b} J_b$, $\psi_t = -\alpha V_{t} J_t$ and
$\alpha=1/(x_t-x_b)$. $V_b$ and $V_t$ are the conditional wall-normal
velocities on the $R$--$Q$ plane at $x_2=x_{b}$ and $x_2=x_{t}$,
respectively, and
\begin{eqnarray}
 J_b &=& \frac{1}{a} P(R,Q,x_b),\\
 J_t &=& \frac{1}{a} P(R,Q,x_t),
\end{eqnarray}
are the probability density functions of $(R,Q)$ conditioned on $x_b$
and $x_t$, respectively.

%-----------------------------------------------------------%
\section{Two examples of the conditionally averaged velocity deviation}\label{sec:appendixB}
%-----------------------------------------------------------%

\begin{figure}
\vspace{0.5cm}
\centerline{
\psfrag{X}{ \raisebox{-0.2cm}{$R$} }\psfrag{Y}{$Q$}
\includegraphics[width=0.45\textwidth]{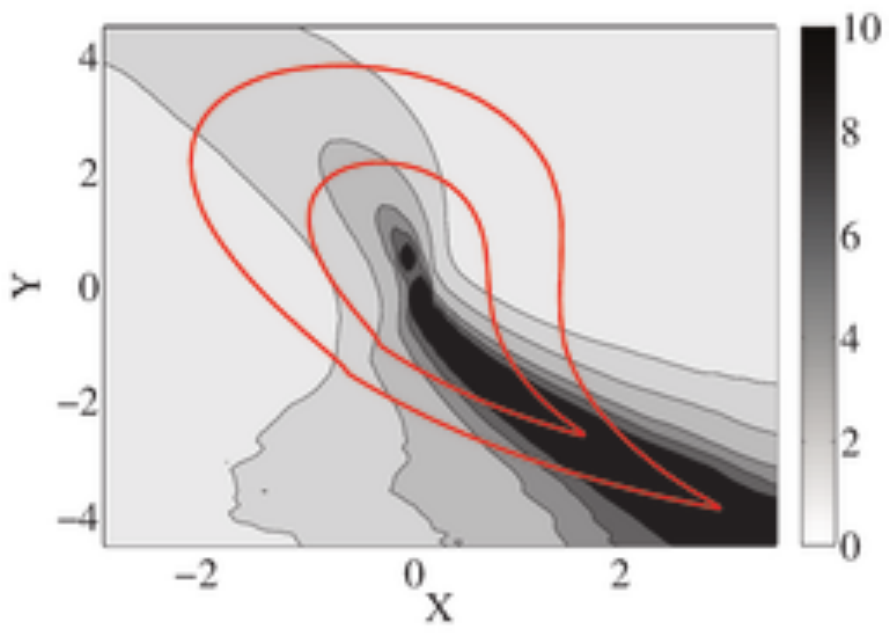}
\mylab{-5.3cm}{4.3cm}{(a)}
\psfrag{X}{ \raisebox{-0.2cm}{$R_\omega$} }\psfrag{Y}{$Q_\omega$}
\includegraphics[width=0.435\textwidth]{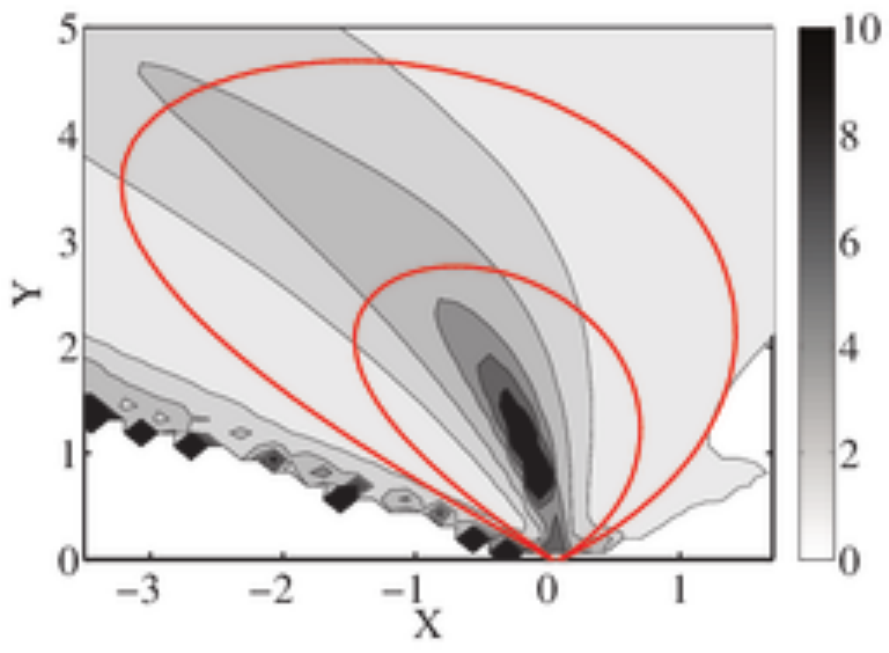}
\mylab{-5.3cm}{4.25cm}{(b)}
}
\caption{Ratio of the magnitude of the conditionally averaged velocity
  deviation $\boldsymbol{v}'$ and the mean $\boldsymbol{v}$,
  conditioned on (a), $R$--$Q$ and (b) $R_\omega$--$Q_\omega$ planes.
  Results for F0.1. Although the colorbar ranges from 0 to 10, values
  up to 100 are attained. The solid red lines in contain 90\% and 98\%
  of the data.
\label{fig:appendixB}
}
\end{figure}

This Appendix contains two more examples of the ratio of the magnitude
of the conditionally averaged velocity deviation $\boldsymbol{v}'$ and
the mean $\boldsymbol{v}$ discussed in \S\ref{subsec:RQ}. Figure
\ref{fig:appendixB} shows two cases conditioned on the $R$--$Q$ and
$R_\omega$--$Q_\omega$ planes, respectively, for case F0.1.
%
%================================================================
%================================================================
%
%

%-----------------------------------------------------------%
\section{ Results filtering in homogeneous directions }\label{sec:appendixFxz}
%-----------------------------------------------------------%

\begin{figure}%[h!tb]
\centerline{
\psfrag{X}{$R_\omega$}\psfrag{Y}{$Q_\omega$}
\includegraphics[width=0.432\textwidth]{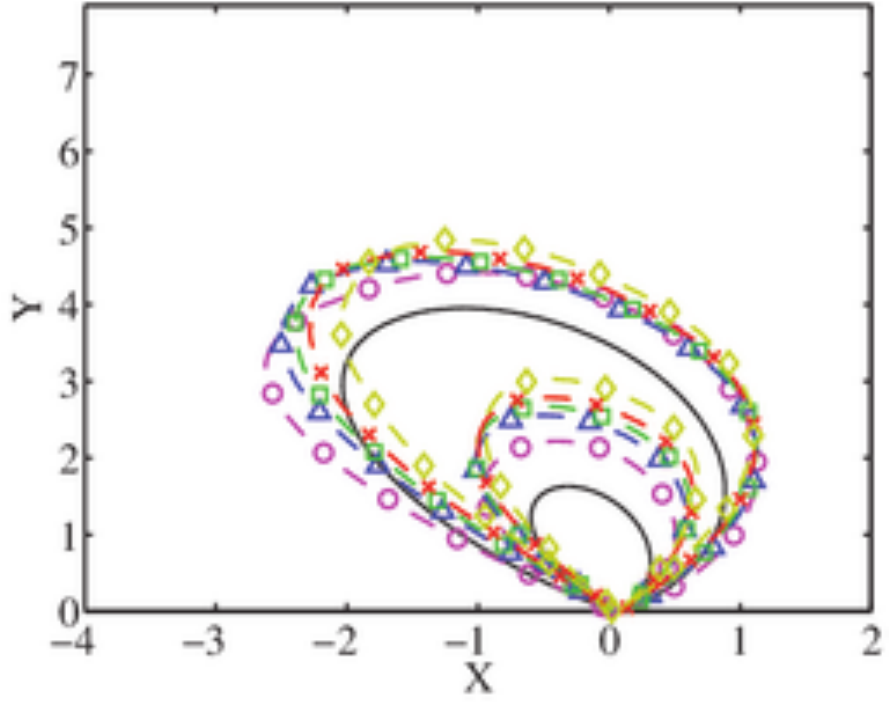}
\mylab{-1cm}{4.0cm}{(a)}
\psfrag{X}{$R_s$}\psfrag{Y}{$Q_s$}
\includegraphics[width=0.435\textwidth]{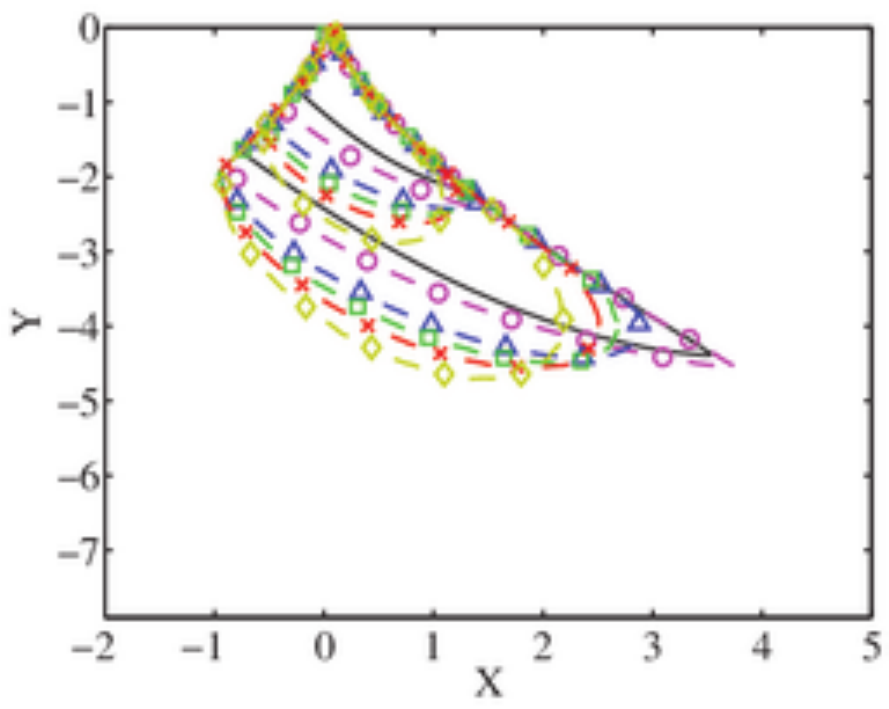}
\mylab{-1cm}{4.0cm}{(b)}
}
\centerline{
\psfrag{X}{$Q_s$}\psfrag{Y}{$Q_\omega$}
\includegraphics[width=0.435\textwidth]{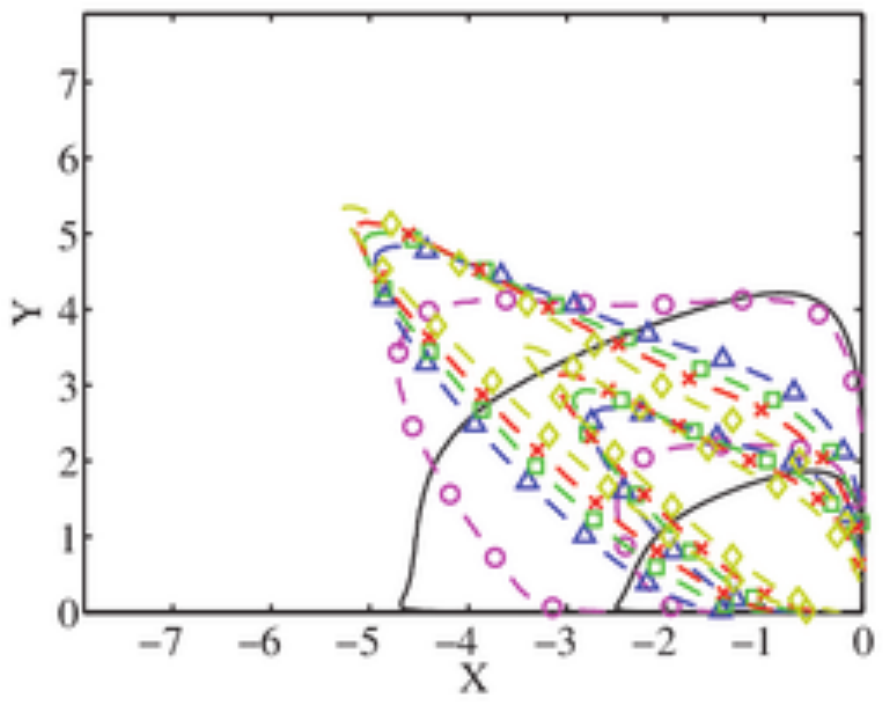}
\mylab{-1.0cm}{4.0cm}{(c)}
\psfrag{X}{$R_s$}\psfrag{Y}{$R_\omega$}
\includegraphics[width=0.435\textwidth]{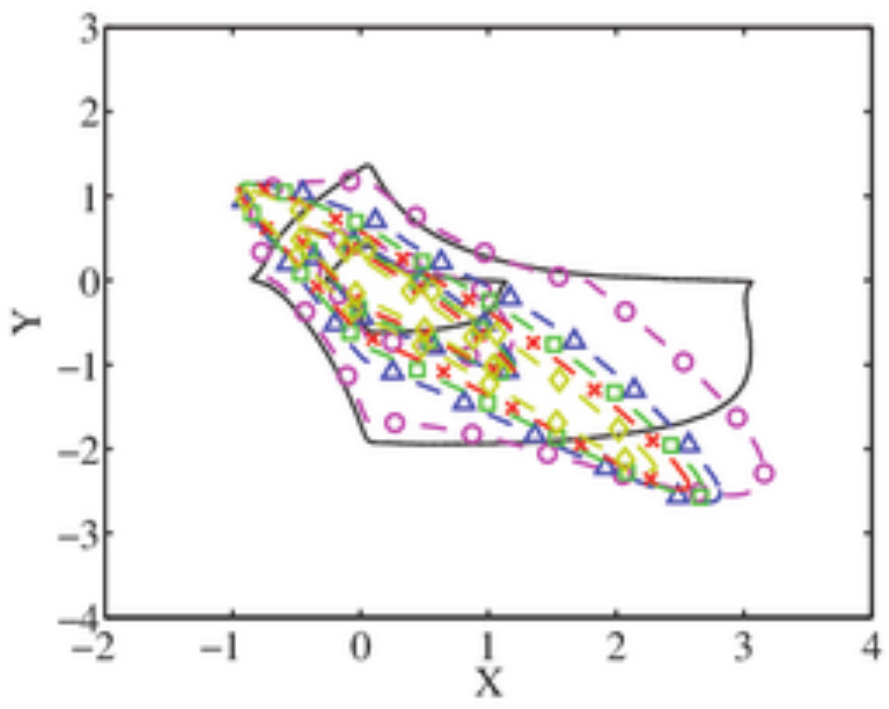}
\mylab{-1.0cm}{4.0cm}{(d)}
}
\caption{ Results filtering only in the $x_1$ and $x_3$
  directions. Joint probability density functions of (a),
  $R_\omega$--$Q_\omega$; (b), $R_s$--$Q_s$; (c), $Q_s$--$Q_\omega$;
  (d), $R_s$--$R_\omega$. Symbols and colors as in table
  \ref{table:cases}. The contours contain 90\% and 98\% of the data.
\label{fig:QsQwRsRw_Fxz}}
\end{figure}

The joint p.d.f.s from figure \ref{fig:QsQwRsRw} were recomputed
filtering the velocities with a Gaussian filter as the one in
(\ref{eq:filter}) but only applied in the two homogeneous directions.
The results, shown in figure \ref{fig:QsQwRsRw_Fxz}, are remarkable
different from those in figure \ref{fig:QsQwRsRw}, and the mean shear
never takes over as the primary effect.  This difference shows that
the wall-parallel filtering approach is not equivalent and justifies
the complication of filtering in the wall-normal direction.
%
%================================================================
% /data4/adrian/Q1Q2R1R2/mfiles/plotQ1Q2R1R2_cte.m
%================================================================
%

\bibliography{QR_filter}

\begin{thebibliography}{63}
\expandafter\ifx\csname natexlab\endcsname\relax\def\natexlab#1{#1}\fi
\def\au#1{#1} \def\ed#1{#1} \def\yr#1{#1}\def\at#1{#1}\def\jt#1{\textit{#1}}
  \def\bt#1{#1}\def\bvol#1{\textbf{#1}} \def\vol#1{#1} \def\pg#1{#1}
  \def\publ#1{#1}\def\arxiv#1{#1}\def\org#1{#1}\def\st#1{\textit{#1}}

\bibitem[del {\'A}lamo {\em et~al.\/}(2006)del {\'A}lamo, Jim{\'e}nez,
  Zandonade \& Moser]{DelAlamo2006}
{\sc \au{del {\'A}lamo, Juan~C.}, \au{Jim{\'e}nez, Javier}, \au{Zandonade,
  Paulo} \& \au{Moser, Robert~D.}} \yr{2006}  \at{Self-similar vortex clusters
  in the turbulent logarithmic region}.  \jt{J. Fluid Mech.}  \bvol{561},
  \pg{329--358}.

\bibitem[Ashurst {\em et~al.\/}(1987)Ashurst, Kerstein, Kerr \&
  Gibson]{Ashurst1987}
{\sc \au{Ashurst, Wm.~T.}, \au{Kerstein, A.~R.}, \au{Kerr, R.~M.} \&
  \au{Gibson, C.~H.}} \yr{1987}  \at{Alignment of vorticity and scalar gradient
  with strain rate in simulated {N}avier--{S}tokes turbulence}.  \jt{Phys.
  Fluids}  \bvol{30}~(8),  \pg{2343--2353}.

\bibitem[Atkinson {\em et~al.\/}(2012)Atkinson, Chumakov, Bermejo-Moreno \&
  Soria]{Atkinson2012}
{\sc \au{Atkinson, C.}, \au{Chumakov, S.}, \au{Bermejo-Moreno, I.} \&
  \au{Soria, J.}} \yr{2012}  \at{Lagrangian evolution of the invariants of the
  velocity gradient tensor in a turbulent boundary layer}.  \jt{Phys. Fluids}
  \bvol{24}~(10).

\bibitem[Batchelor \& Townsend(1949)]{bat:tow:1949}
{\sc \au{Batchelor, G.~K.} \& \au{Townsend, A.~A.}} \yr{1949}  \at{The nature
  of turbulent motion at large wave-numbers}.  \jt{Proc. Roy. Soc. London, A}
  \bvol{199}~(1057),  \pg{238--255}.

\bibitem[Beck \& Sch\"ogl(1993)]{beck93}
{\sc \au{Beck, Christian} \& \au{Sch\"ogl, Friedrich}} \yr{1993} {\em
  Thermodynamics of Chaotic Systems\/}.  \publ{Cambridge University Press},
  cambridge Books Online.

\bibitem[Betchov(1956)]{bet:1956}
{\sc \au{Betchov, R.}} \yr{1956}  \at{An inequality concerning the production
  of vorticity in isotropic turbulence}.  \jt{J. Fluid Mech.}  \bvol{1},
  \pg{497--504}.

\bibitem[Blackburn {\em et~al.\/}(1996)Blackburn, Mansour \&
  Cantwell]{Blackburn1996}
{\sc \au{Blackburn, Hugh~M.}, \au{Mansour, Nagi~N.} \& \au{Cantwell, Brian~J.}}
  \yr{1996}  \at{Topology of fine-scale motions in turbulent channel flow}.
  \jt{J. Fluid Mech.}  \bvol{310},  \pg{269--292}.

\bibitem[Borue \& Orszag(1998)]{Borue1998}
{\sc \au{Borue, Vadim} \& \au{Orszag, Steven~A.}} \yr{1998}  \at{Local energy
  flux and subgrid-scale statistics in three-dimensional turbulence}.  \jt{J.
  Fluid Mech.}  \bvol{366},  \pg{1--31}.

\bibitem[van~der Bos {\em et~al.\/}(2002)van~der Bos, Tao, Meneveau \&
  Katz]{VanDerBos2002}
{\sc \au{van~der Bos, Fedderik}, \au{Tao, Bo}, \au{Meneveau, Charles} \&
  \au{Katz, Joseph}} \yr{2002}  \at{Effects of small-scale turbulent motions on
  the filtered velocity gradient tensor as deduced from holographic particle
  image velocimetry measurements}.  \jt{Phys. Fluids}  \bvol{14}~(7),
  \pg{2456--2474}.

\bibitem[Cantwell(1992)]{Cantwell1992}
{\sc \au{Cantwell, Brian~J.}} \yr{1992}  \at{Exact solution of a restricted
  {E}uler equation for the velocity gradient tensor}.  \jt{Phys. Fluids}
  \bvol{4}~(4),  \pg{782--793}.

\bibitem[Cardesa {\em et~al.\/}(2013)Cardesa, Mistry, Gan \&
  Dawson]{Cardesa2013}
{\sc \au{Cardesa, J.~I.}, \au{Mistry, D.}, \au{Gan, L.} \& \au{Dawson, J.~R.}}
  \yr{2013}  \at{Invariants of the reduced velocity gradient tensor in
  turbulent flows}.  \jt{J. Fluid Mech.}  \bvol{716},  \pg{597--615}.

\bibitem[Cardesa {\em et~al.\/}(2015)Cardesa, Vela-Mart\'in, Dong \&
  Jim\'enez]{cardesa2015}
{\sc \au{Cardesa, Jos\'e~I.}, \au{Vela-Mart\'in, Alberto}, \au{Dong, Siwei} \&
  \au{Jim\'enez, Javier}} \yr{2015} {T}he propagation of kinetic energy across
  scales in turbulent flows. ArXiv:1505.00285v1,  \arxiv{arXiv: 1505.00285}.

\bibitem[Chacin \& Cantwell(2000)]{Chacin2000}
{\sc \au{Chacin, Juan~M.} \& \au{Cantwell, Brian~J.}} \yr{2000}  \at{Dynamics
  of a low reynolds number turbulent boundary layer}.  \jt{J. Fluid Mech.}
  \bvol{404},  \pg{87--115}.

\bibitem[Chertkov {\em et~al.\/}(1999)Chertkov, Pumir \&
  Shraiman]{Chertkov1999}
{\sc \au{Chertkov, M.}, \au{Pumir, A.} \& \au{Shraiman, B.I.}} \yr{1999}
  \at{Lagrangian tetrad dynamics and the phenomenology of turbulence}.
  \jt{Phys. Fluids}  \bvol{11}.

\bibitem[Chevillard {\em et~al.\/}(2011)Chevillard, Lévêque, Taddia,
  Meneveau, Yu \& Rosales]{Chevillard2011}
{\sc \au{Chevillard, Laurent}, \au{Lévêque, Emmanuel}, \au{Taddia,
  Francesco}, \au{Meneveau, Charles}, \au{Yu, Huidan} \& \au{Rosales, Carlos}}
  \yr{2011}  \at{Local and nonlocal pressure hessian effects in real and
  synthetic fluid turbulence}.  \jt{Phys. Fluids}  \bvol{23}~(9).

\bibitem[Chevillard \& Meneveau(2006)]{Chevillard2006}
{\sc \au{Chevillard, L.} \& \au{Meneveau, C.}} \yr{2006}  \at{Lagrangian
  dynamics and statistical geometric structure of turbulence}.  \jt{Phys. Rev.
  Lett.}  \bvol{97},  \pg{174501}.

\bibitem[Chevillard {\em et~al.\/}(2008)Chevillard, Meneveau, Biferale \&
  Toschi]{Chevillard2007}
{\sc \au{Chevillard, L.}, \au{Meneveau, C.}, \au{Biferale, L.} \& \au{Toschi,
  F.}} \yr{2008}  \at{Modeling the pressure hessian and viscous laplacian in
  turbulence: Comparisons with direct numerical simulation and implications on
  velocity gradient dynamics}.  \jt{Phys. Fluids}  \bvol{20}~(10).

\bibitem[Chong {\em et~al.\/}(1990)Chong, Perry \& Cantwell]{Chong1990}
{\sc \au{Chong, M.~S.}, \au{Perry, A.~E.} \& \au{Cantwell, B.~J.}} \yr{1990}
  \at{A general classification of three‐-dimensional flow fields}.  \jt{Phys.
  Fluids}  \bvol{2}~(5),  \pg{765--777}.

\bibitem[Chong {\em et~al.\/}(1998)Chong, Soria, Perry, Chacin, Cantwell \&
  Na]{Chong1998}
{\sc \au{Chong, M.~S.}, \au{Soria, J.}, \au{Perry, A.~E.}, \au{Chacin, J.},
  \au{Cantwell, B.~J.} \& \au{Na, Y.}} \yr{1998}  \at{Turbulence structures of
  wall-bounded shear flows found using {DNS} data}.  \jt{J. Fluid Mech.}
  \bvol{357},  \pg{225--247}.

\bibitem[Corrsin(1958)]{Corrsin1958}
{\sc \au{Corrsin, S.}} \yr{1958}  \bt{Local isotropy in turbulent shear flow}.
  Res. Memo 58B11.  \org{NACA}.

\bibitem[Davidson(2004)]{Davidson2004}
{\sc \au{Davidson, P.A.}} \yr{2004} {\em Turbulence: An Introduction for
  Scientists and Engineers\/}.  \publ{OUP Oxford}.

\bibitem[Doi \& Edwards(1988)]{Doi88}
{\sc \au{Doi, M.} \& \au{Edwards, S.F.}} \yr{1988} {\em The Theory of Polymer
  Dynamics\/}.  \publ{Clarendon Press}.

\bibitem[Elsinga \& Marusic(2010)]{Elsinga2010}
{\sc \au{Elsinga, G.~E.} \& \au{Marusic, I.}} \yr{2010}  \at{Evolution and
  lifetimes of flow topology in a turbulent boundary layer}.  \jt{Phys. Fluids}
   \bvol{22}~(1),  \pg{015102}.

\bibitem[Flores \& Jim{\'e}nez(2010)]{Flores2010}
{\sc \au{Flores, Oscar} \& \au{Jim{\'e}nez, Javier}} \yr{2010}  \at{Hierarchy
  of minimal flow units in the logarithmic layer}.  \jt{Phys. Fluids}
  \bvol{22}~(7),  \pg{071704}.

\bibitem[Gomes-Fernandes {\em et~al.\/}(2014)Gomes-Fernandes,
  Ganapathisubramani \& Vassilicos]{Gomes2014}
{\sc \au{Gomes-Fernandes, R.}, \au{Ganapathisubramani, B.} \& \au{Vassilicos,
  J.~C.}} \yr{2014}  \at{Evolution of the velocity-gradient tensor in a
  spatially developing turbulent flow}.  \jt{J. Fluid Mech.}  \bvol{756},
  \pg{252--292}.

\bibitem[Hamlington {\em et~al.\/}(2008)Hamlington, Schumacher \&
  Dahm]{Hamlington2008}
{\sc \au{Hamlington, Peter~E.}, \au{Schumacher, J\"org} \& \au{Dahm, Werner
  J.~A.}} \yr{2008}  \at{Local and nonlocal strain rate fields and vorticity
  alignment in turbulent flows}.  \jt{Phys. Rev. E}  \bvol{77},  \pg{026303}.

\bibitem[Jim\'enez(1992)]{jim:1992}
{\sc \au{Jim\'enez, Javier}} \yr{1992}  \at{Kinematic alignment effects in
  turbulent flows}.  \jt{Phys. Fluids}  \bvol{4}~(4).

\bibitem[Jim\'enez(2000)]{jim:2000}
{\sc \au{Jim\'enez, Javier}} \yr{2000}  \at{Intermittency and cascades}.
  \jt{J. Fluid Mech.}  \bvol{409},  \pg{99--120}.

\bibitem[Jim{\'e}nez(2012)]{Jimenez2012}
{\sc \au{Jim{\'e}nez, J.}} \yr{2012}  \at{Cascades in wall-bounded turbulence}.
   \jt{Ann. Rev. Fluid Mech.}  \bvol{44},  \pg{27--45}.

\bibitem[Jim{\'e}nez(2013)]{jim:2013}
{\sc \au{Jim{\'e}nez, J.}} \yr{2013}  \at{How linear is wall-bounded
  turbulence?}  \jt{Phys. Fluids}  \bvol{25},  \pg{110814}.

\bibitem[Jim{\'e}nez {\em et~al.\/}(1993)Jim{\'e}nez, Wray, Saffman \&
  Rogallo]{jim:wra:saf:rog:93}
{\sc \au{Jim{\'e}nez, Javier}, \au{Wray, Alan~A.}, \au{Saffman, Philip~G.} \&
  \au{Rogallo, Robert~S.}} \yr{1993}  \at{The structure of intense vorticity in
  isotropic turbulence}.  \jt{J. Fluid Mech.}  \bvol{255},  \pg{65--90}.

\bibitem[Kim {\em et~al.\/}(1987)Kim, Moin \& Moser]{Kim1987}
{\sc \au{Kim, John}, \au{Moin, Parviz} \& \au{Moser, Robert~D}} \yr{1987}
  \at{Turbulence statistics in fully developed channel flow at low {R}eynolds
  number}.  \jt{J. Fluid Mech}  \bvol{177},  \pg{133--166}.

\bibitem[LeHew {\em et~al.\/}(2013)LeHew, Guala \& McKeon]{leh:gua:mck:2013}
{\sc \au{LeHew, J.A.}, \au{Guala, M.} \& \au{McKeon, B.J.}} \yr{2013}
  \at{Time-resolved measurements of coherent structures in the turbulent
  boundary layer}.  \jt{Exp. Fluids}  \bvol{54}~(4),  \pg{1--16}.

\bibitem[Leung {\em et~al.\/}(2012)Leung, Swaminathan \& Davidson]{leung2012}
{\sc \au{Leung, T}, \au{Swaminathan, N} \& \au{Davidson, PA}} \yr{2012}
  \at{Geometry and interaction of structures in homogeneous isotropic
  turbulence}.  \jt{J. Fluid Mech.}  \bvol{710},  \pg{453--481}.

\bibitem[Li {\em et~al.\/}(2008)Li, Perlman, Wan, Yang, Meneveau, Burns, Chen,
  Szalay \& Eyink]{JHU2008}
{\sc \au{Li, Yi}, \au{Perlman, Eric}, \au{Wan, Minping}, \au{Yang, Yunke},
  \au{Meneveau, Charles}, \au{Burns, Randal}, \au{Chen, Shiyi}, \au{Szalay,
  Alexander} \& \au{Eyink, Gregory}} \yr{2008}  \at{A public turbulence
  database cluster and applications to study lagrangian evolution of velocity
  increments in turbulence}.  \jt{J. Turb.}  \pg{p. N31}.

\bibitem[Lozano-Dur{\'a}n {\em et~al.\/}(2012)Lozano-Dur{\'a}n, Flores \&
  Jim{\'e}nez]{Lozano2012}
{\sc \au{Lozano-Dur{\'a}n, Adri{\'a}n}, \au{Flores, Oscar} \& \au{Jim{\'e}nez,
  Javier}} \yr{2012}  \at{The three-dimensional structure of momentum transfer
  in turbulent channels}.  \jt{J. Fluid Mech.}  \bvol{694},  \pg{100--130}.

\bibitem[Lozano-Dur{\'a}n {\em et~al.\/}(2015)Lozano-Dur{\'a}n, Holzner \&
  Jim{\'e}nez]{loz:hol:jim:2015}
{\sc \au{Lozano-Dur{\'a}n, Adrian}, \au{Holzner, Markus} \& \au{Jim{\'e}nez,
  Javier}} \yr{2015}  \at{Numerically accurate computation of the conditional
  trajectories of the topological invariants in turbulent flows}.  \jt{J. Comp.
  Phys.}  \bvol{295},  \pg{805--814}.

\bibitem[Lozano-Dur\'an \& Jim\'enez(2014{\natexlab{{\em a\/}}})]{loz:jim:2014}
{\sc \au{Lozano-Dur\'an, Adri\'an} \& \au{Jim\'enez, Javier}}
  \yr{2014{\natexlab{{\em a\/}}}}  \at{Effect of the computational domain on
  direct simulations of turbulent channels up to ${Re}_\tau=4200$}.  \jt{Phys.
  Fluids}  \bvol{26}~(1),  \pg{011702}.

\bibitem[Lozano-Dur\'an \& Jim\'enez(2014{\natexlab{{\em b\/}}})]{Lozano2014}
{\sc \au{Lozano-Dur\'an, Adri\'an} \& \au{Jim\'enez, Javier}}
  \yr{2014{\natexlab{{\em b\/}}}}  \at{Time-resolved evolution of coherent
  structures in turbulent channels: characterization of eddies and cascades}.
  \jt{J. Fluid Mech.}  \bvol{759},  \pg{432--471}.

\bibitem[L\"{u}thi {\em et~al.\/}(2009)L\"{u}thi, Holzner \&
  Tsinober]{Luethi2009}
{\sc \au{L\"{u}thi, Beat}, \au{Holzner, Markus} \& \au{Tsinober, Arkady}}
  \yr{2009}  \at{Expanding the {Q}--{R} space to three dimensions}.  \jt{J.
  Fluid Mech.}  \bvol{641},  \pg{497--507}.

\bibitem[L\"{u}thi {\em et~al.\/}(2007)L\"{u}thi, Ott, Berg \&
  Mann]{Luethi2007}
{\sc \au{L\"{u}thi, Beat}, \au{Ott, S.}, \au{Berg, Jacob} \& \au{Mann, Jakob}}
  \yr{2007}  \at{Lagrangian multi-particle statistics}.  \jt{J. Turb.}
  \bvol{8},  \pg{N45},  \arxiv{arXiv:
  http://dx.doi.org/10.1080/14685240701522927}.

\bibitem[Mart{\'i}n {\em et~al.\/}(1998)Mart{\'i}n, Ooi, Chong \&
  Soria]{Martin1998}
{\sc \au{Mart{\'i}n, Jes{\'u}s}, \au{Ooi, Andrew}, \au{Chong, M.~S.} \&
  \au{Soria, Julio}} \yr{1998}  \at{Dynamics of the velocity gradient tensor
  invariants in isotropic turbulence}.  \jt{Phys. Fluids}  \bvol{10}~(9),
  \pg{2336--2346}.

\bibitem[Marusic {\em et~al.\/}(2013)Marusic, Monty, Hultmark \&
  Smits]{mar:mon:hul:smi:2013}
{\sc \au{Marusic, Ivan}, \au{Monty, Jason~P.}, \au{Hultmark, Marcus} \&
  \au{Smits, Alexander~J.}} \yr{2013}  \at{On the logarithmic region in wall
  turbulence}.  \jt{J. Fluid Mech.}  \bvol{716},  \pg{R3}.

\bibitem[Meneveau(2011)]{Meneveau2011}
{\sc \au{Meneveau, Charles}} \yr{2011}  \at{Lagrangian dynamics and models of
  the velocity gradient tensor in turbulent flows}.  \jt{Ann. Rev. Fluid Mech.}
   \bvol{43}~(1),  \pg{219--245}.

\bibitem[Moisy \& Jim\'enez(2004)]{Moisy2004}
{\sc \au{Moisy, F.} \& \au{Jim\'enez, J.}} \yr{2004}  \at{Geometry and
  clustering of intense structures in isotropic turbulence}.  \jt{J. Fluid
  Mech.}  \bvol{513},  \pg{111--133}.

\bibitem[Moser {\em et~al.\/}(1999)Moser, Kim \& Mansour]{Moser1999}
{\sc \au{Moser, Robert~D.}, \au{Kim, John} \& \au{Mansour, Nagi~N.}} \yr{1999}
  \at{Direct numerical simulation of turbulent channel flow up to ${Re}_\tau=
  590$}.  \jt{Phys. Fluids}  \bvol{11}~(4),  \pg{943--945}.

\bibitem[Mullin \& Dahm(2006)]{mul:2006}
{\sc \au{Mullin, John~A.} \& \au{Dahm, Werner J.~A.}} \yr{2006}  \at{Dual-plane
  stereo particle image velocimetry measurements of velocity gradient tensor
  fields in turbulent shear flow. {II}. experimental results}.  \jt{Phys.
  Fluids}  \bvol{18}~(3).

\bibitem[Naso \& Pumir(2005)]{Naso2005}
{\sc \au{Naso, Aurore} \& \au{Pumir, Alain}} \yr{2005}  \at{Scale dependence of
  the coarse-grained velocity derivative tensor structure in turbulence}.
  \jt{Phys. Rev. E}  \bvol{72},  \pg{056318}.

\bibitem[Naso {\em et~al.\/}(2006)Naso, Pumir \& Chertkov]{Naso2006}
{\sc \au{Naso, Aurore}, \au{Pumir, Alain} \& \au{Chertkov, Michael}} \yr{2006}
  \at{Scale dependence of the coarse-grained velocity derivative tensor:
  Influence of large-scale shear on small-scale turbulence}.  \jt{J. Turb.}
  \bvol{7},  \pg{N41}.

\bibitem[Naso {\em et~al.\/}(2007)Naso, Pumir \& Chertkov]{Naso2007}
{\sc \au{Naso, A.}, \au{Pumir, A.} \& \au{Chertkov, M.}} \yr{2007}
  \at{Statistical geometry in homogeneous and isotropic turbulence}.  \jt{J.
  Turb.}  \pg{p. N39}.

\bibitem[Ooi {\em et~al.\/}(1999)Ooi, Mart{\'i}n, Soria \& Chong]{Ooi1999}
{\sc \au{Ooi, Andrew}, \au{Mart{\'i}n, Jes{\'u}s}, \au{Soria, Julio} \&
  \au{Chong, M.~S.}} \yr{1999}  \at{A study of the evolution and
  characteristics of the invariants of the velocity-gradient tensor in
  isotropic turbulence}.  \jt{J. Fluid Mech.}  \bvol{381},  \pg{141--174}.

\bibitem[Pumir \& Naso(2010)]{Pumir2010}
{\sc \au{Pumir, Alain} \& \au{Naso, Aurore}} \yr{2010}  \at{Statistical
  properties of the coarse-grained velocity gradient tensor in turbulence:
  Monte-carlo simulations of the tetrad model}.  \jt{New J. Phys.}
  \bvol{12}~(12),  \pg{123024}.

\bibitem[She {\em et~al.\/}(1991)She, Jackson \& Orszag]{She1991}
{\sc \au{She, Zhen-Su}, \au{Jackson, Eric} \& \au{Orszag, Steven~A.}} \yr{1991}
   \at{Structure and dynamics of homogeneous turbulence: Models and
  simulations}.  \jt{Proc. Roy. Soc. London, A}  \bvol{434}~(1890),
  \pg{101--124}.

\bibitem[Soria {\em et~al.\/}(1994)Soria, Sondergaard, Cantwell, Chong \&
  Perry]{Soria1994}
{\sc \au{Soria, J.}, \au{Sondergaard, R.}, \au{Cantwell, B.~J.}, \au{Chong,
  M.~S.} \& \au{Perry, A.~E.}} \yr{1994}  \at{A study of the fine-‐scale
  motions of incompressible time-‐developing mixing layers}.  \jt{Phys.
  Fluids}  \bvol{6}~(2),  \pg{871--884}.

\bibitem[Tanahashi {\em et~al.\/}(2004)Tanahashi, Kang, Miyamoto \&
  Shiokawa]{Tanahashi2004}
{\sc \au{Tanahashi, M}, \au{Kang, S}, \au{Miyamoto, T} \& \au{Shiokawa, S}}
  \yr{2004}  \at{Scaling law of fine scale eddies in turbulent channel flows up
  to ${Re}_\tau= 800$}.  \jt{Int. J. Heat Fluid Flow}  \bvol{25},
  \pg{331--341}.

\bibitem[Tennekes \& Lumley(1972)]{Tennekes1972}
{\sc \au{Tennekes, H.} \& \au{Lumley, J.L.}} \yr{1972} {\em A First Course in
  Turbulence\/}.  \publ{MIT Press}.

\bibitem[Tsinober(1998)]{Tsinober1998}
{\sc \au{Tsinober, A.}} \yr{1998}  \at{Is concentrated vorticity that
  important?}  \jt{Eur. J. Mech. B-Fluid}  \bvol{17}~(4),  \pg{421--449}.

\bibitem[Tsinober {\em et~al.\/}(1997)Tsinober, Shtilman \& Vaisburd]{Tsi:1997}
{\sc \au{Tsinober, A}, \au{Shtilman, L} \& \au{Vaisburd, H}} \yr{1997}  \at{A
  study of properties of vortex stretching and enstrophy generation in
  numerical and laboratory turbulence}.  \jt{Fluid Dyn. Res.}  \bvol{21}~(6),
  \pg{477}.

\bibitem[Vieillefosse(1984)]{Vieillefosse1984}
{\sc \au{Vieillefosse, P.}} \yr{1984}  \at{Internal motion of a small element
  of fluid in an inviscid flow}.  \jt{Phys. Stat. Mech. Appl.}  \bvol{125}~(1),
   \pg{150--162}.

\bibitem[{Vieillefosse, P.}(1982)]{Vieillefosse1982}
{\sc \au{{Vieillefosse, P.}}} \yr{1982}  \at{Local interaction between
  vorticity and shear in a perfect incompressible fluid}.  \jt{J. Phys. France}
   \bvol{43}~(6),  \pg{837--842}.

\bibitem[Vincent \& Meneguzzi(1991)]{Vincent1991}
{\sc \au{Vincent, A.} \& \au{Meneguzzi, M.}} \yr{1991}  \at{The spatial
  structure and statistical properties of homogeneous turbulence}.  \jt{J.
  Fluid Mech.}  \bvol{225},  \pg{1--20}.

\bibitem[Vincent \& Meneguzzi(1994)]{Vincent1994}
{\sc \au{Vincent, A.} \& \au{Meneguzzi, M.}} \yr{1994}  \at{The dynamics of
  vorticity tubes in homogeneous turbulence}.  \jt{J. Fluid Mech.}  \bvol{258},
   \pg{245--254}.

\bibitem[Wang \& Lu(2012)]{Wang2012}
{\sc \au{Wang, Li} \& \au{Lu, Xi-Yun}} \yr{2012}  \at{Flow topology in
  compressible turbulent boundary layer}.  \jt{J. Fluid Mech.}  \bvol{703},
  \pg{255--278}.

\end{thebibliography}
\bibliographystyle{jfm}

\end{document}